 \documentclass[final,3p,times,sort&compress, 10pt]{elsarticle}

\usepackage{amsmath}
\usepackage{amssymb}
\usepackage{amsthm}
\usepackage{exscale}
\usepackage[mathscr]{eucal}
\usepackage{multirow}
\usepackage{longtable}
\usepackage{algorithm}
\usepackage{algorithmic}
\usepackage{ctable}
\usepackage{siunitx}
\usepackage{subfig}
\usepackage{placeins}
\usepackage{bm}

\def\longrightharpoonup{\relbar\joinrel\rightharpoonup}
\def\longleftharpoondown{\leftharpoondown\joinrel\relbar}

\def\longrightleftharpoons{
  \mathop{
    \vcenter{
      \hbox{
      \ooalign{
        \raise1pt\hbox{$\longrightharpoonup\joinrel$}\crcr
	  \lower1pt\hbox{$\longleftharpoondown\joinrel$}
	  }
      }
    }
  }
}

\journal{}

\begin{document}

\begin{frontmatter}



\title{A Technique for Computing Dense Granular Compressible Flows with Shock Waves}

\cortext[cor1]{Corresponding Author}
\author{Ryan W. Houim}
\ead{rhouim@umd.edu}
\author{Elaine S. Oran}
\address{Department of Aerospace Engineering, University of Maryland, College Park, MD 20742, USA}

\begin{abstract}
A numerical procedure was developed for solving equations for compressible granular multiphase flows in which the particle volume fraction can range dynamically from very dilute to very dense.  The procedure uses a low-dissipation and high-order numerical method that can describe shocks and incorporates a particulate model based on kinetic theory.
The algorithm separates edges of a computational cell into gas and solid sections where gas- and granular-phase Riemann problems are solved independently.  Solutions from these individual Riemann problems are combined to assemble the fully coupled convective fluxes and nonconservative terms for both phases.  The technique converges under grid refinement even with very high volume fraction granular interfaces.  The method can advect sharp granular material interfaces that coincide with multi-species gaseous contact surfaces without violating the pressure nondisturbing condition.  The procedure also reproduces known features from multiphase shock tube problems, granular shocks, transmission angles of compaction waves, and shock wave and dust-layer interactions.  This approach is relatively straightforward to use in an existing code based on Godunov's method and can be constructed from standard compressible solvers for the gas-phase and a modified AUSM$^+$-up scheme for the particle phase.  

\end{abstract}

\begin{keyword}
Multiphase flow
\sep
Compressible flow
\sep
Granular gas
\sep
nonconservative term
\sep
Riemann solver
\sep
Advection upstream splitting method 

\end{keyword}

\end{frontmatter}


\section{Introduction}

In the past, efforts were made to solve equations of granular multiphase flows because of their importance for coal combustors \cite{zhou2011two}, catalytic bed reactors \cite{zimmermann2005cfd}, biomass gasification \cite{gerber2010eulerian}, particle hoppers \cite{srivastava2003analysis}, scouring of sand underneath submerged pipes \cite{zhao2007numerical}, pyroclastic flows \cite{neri2003multiparticle}, etc. One common type of multiphase model for these situations solves a set of equations based on kinetic theory applied to solid particles \cite{GidaspowBook, Lun1984}.  These models are comprised of an inelastic granular gas with incompressible particles  \cite{GranularGasBook} coupled to a molecular gas.  Granular-gas models add a set of solid-phase conservation equations, which resemble the Navier-Stokes equations for a real gas with coupling terms between the gas and granular phases and can be considers as a variant of two-fluid models \cite{IshiiMultiphaseBook}.  There are  several existing codes, such as MFIX \cite{MFIX-Theory}, that solve such multiphase particle flow equations for low-speed gases and liquids.    Granular gases are characterized by the granular temperature ($\Theta_s$), representing  random translational kinetic energy of the particles, used in constitutive relations to compute a solids pressure ($p_s$) that describes an intergranular stress resulting from particle collisions that is analogous to gas-phase pressure.  Friction forces between colliding particles become substantial at high particle volume fraction and frictional-collisional pressure ($p_{\text{fric}}$) \cite{johnson1987friction} is often included.  

Despite emphasis on low-speed flows, dense granular multiphase flows are also important when the flows are high-speed and compressible, such as those that occur in explosions \cite{Sichel1995} in coal mines, sawmills, and flour mills, volcanic eruptions \cite{wilson1980relationships}, shock-induced lifting of dust layers \cite{chuanjie2012simulation,zydak2007modelling,wayne2013shock},  explosion suppression \cite{liu2013methane}, and interior ballistics of guns \cite{koo1977transient,Porterie1994,Nussbaum2006,markatos1983analysis,markatos1986modelling}.  Relatively little has been done to solve kinetic theory-based granular multiphase models suitable for the wide range of particle volume fractions and flow velocities that occur in compressible flows \cite{Fan2007LooseDust,chuanjie2012simulation}, which is in stark contrast to the plethora of methods for solving these equations for low-speed flows \cite{van2001comparative}.

Many other Eulerian compressible granular multiphase flows are modeled using the dust-gas approximation, which neglects particle-particle interactions and collision processes
\cite{collins1994DustyGas,Benkiewicz-2002-FluidDynamicsResearch-Al-Dust,saito2003numerical,fedorov2007reflection,PelantiDustyGas2006}.  These models have been used  extensively to study detonations of particle suspensions, explosion suppression from aerosolized particles, and lifting of dilute dust layers.  This type of model is not applicable when there are dense particle clusters, which could occur when dense dust layers are lifted by shocks \cite{Fan2007LooseDust}, or in coal mine explosions where settled layers of dust may accumulate to volume fractions on the order of 47\% \cite{BureauOfMines1988}.  In these cases, effects of particle-particle interactions must be included.

Other multiphase models for particulate flows applicable to high volume fractions include the Baer-Nunziato (BN) equations \cite{baer1986two}, the Nigmatulin equations \cite{NigmatulinMultiphaseMedia}, and many Eulerian interior gun ballistic models \cite{koo1977transient,Porterie1994,Nussbaum2006,markatos1983analysis,markatos1986modelling}. The BN equations were developed for modeling high-pressure combustion processes, such as detonation of high-explosive charges, where particle compression and distortion is significant.  The BN equations are a variant of the seven-equation model presented by Saurel and Abgrall \cite{saurel1999multiphase} with the interfacial velocity and pressure chosen to be the particle velocity and gas-phase pressure, respectively.  The BN model relaxes the particle incompressibility assumption used in the kinetic theory approach, but adds a nonconservative advection equation for the solid volume fraction.  The inclusion of this nonconservative advection equation changes the physical nature of compression waves in the solid phase with respect to kinetic theory models.  Compression waves for the solid phase BN equations represent pressure waves moving through the solid material, and, as a result, volume fraction changes only at solid contact surfaces.   Solid phase compression waves for kinetic theory-based models are produced by particle collisions which create compaction waves and granular shocks \cite{kamenetsky2000evolution} that directly change the volume fraction. 

 Interior gun ballistic models \cite{koo1977transient,Porterie1994,Nussbaum2006,markatos1983analysis,markatos1986modelling}  involve high particle loadings, relatively high pressures (on the order of 1000 atm or more), and high-speed flow.  Nevertheless, particle compression and deformation is insignificant at these high-pressure conditions and interior ballistic models often assume incompressible particles.  The intergranular stress is specified by equations that are similar to frictional-collisional pressure.  Many interior ballistic models can reduced from the kinetic theory-based multiphase equations by assuming zero granular temperature ($\Theta_s$) and assuming the intergranular stress is solely from frictional-collision pressure.  The multiphase approach used by Rogue \textit{et al.} \cite{Rogue1998ShockParticle} to simulate shocks interacting with dense particles curtains is similar to interior ballistic equations.  

Lagrangian methods are also used to model dense granular flow \cite{Helland2000210,LingBalachandar2013-Model-Experiment,nusca2004multidimensional} by tracking individual particles or groups of particles.  Such methods can be very computationally intensive when a large number of particle groups need to be tracked, which occurs when the granular phase approaches the packing limit or if simulations have large length scales.

In this paper we are interested in solving multiphase flows with large length scales where the particle packing can be relatively high and approach the packing limit, such as shock and dust layer interactions, but not so high that particle compressibility is significant.  Thus, we focus on kinetic theory-based granular models. 

Numerical solutions of dense granular flows have proved difficult for low-dissipation methods.  The origin of the problem is a combination of  nonconservative ``nozzling'' terms \cite{schwendeman2006riemann,karni2010hybrid,crochet2013numerical} arising from variations of the particle volume fraction within a control volume and  $pDV$ work done to the gas as particles enter and leave a control volume \cite{abgrall2010comment}. 
The gas-phase momentum equations reduce to the nondisturbing condition \cite{saurel1999multiphase} in the trivial case of uniform pressure and zero velocity,
\begin{equation*}
  \nabla \! \alpha_g p_g = p_g \nabla \! \alpha_g,
\end{equation*} 
where $\alpha_g$ is the gas-phase volume fraction, $p_g$ is the gas-phase pressure, and the right side is the nozzling term.   Computing the left side and nozzling terms independently   \cite{toroReactive2Phase,Nussbaum2006} can lead to situations where the left and right sides may not balance.  This induces flow and pressure oscillations that violate the nondisturbing condition and may cause the code to fail  \cite{liou2008TwoFluidSimple}.

Another complication is that the compaction wave speed  of the particles (analogous to the sound speed of a gas) ranges between zero, when $\Theta_s$ is zero or at low particle volume fractions, and diverges towards infinity as the packing limit is approached.  Zero compaction wave speed leads to a pressureless granular gas and hyperbolic degeneracy with linearly dependent flux eigenvectors.  Hyperbolic degeneracy rules out characteristic-based methods such as WENO \cite{Balsara-2000-JCompPhys-MPWENO} or PPM \cite{colella1984piecewise} and impedes the development of an exact Riemann solver for the granular phase.  Despite possible degeneracy, approximate Riemann solvers have been designed for pressureless dust \cite{collins1994DustyGas,PelantiDustyGas2006} and the Baer-Nunziato equations  \cite{schwendeman2006riemann,crochet2013numerical}, but not for kinetic theory-based granular models. 

Extreme sensitivity of the intergranular stress to minute fluctuations of solid volume fraction in dense regions approaching the packing limit makes it difficult to develop numerical methods that are both low-dissipation and robust.  Another difficulty is based on differences of how volume fraction influences the gas and granular phases: particles restrict the area where fluid can flow and so accelerate the gas by a nozzling term.  On the other hand, the volume fraction controls compressibility of the granular phase in the same way that density does for the gas phase.  

In place of exact solutions to kinetic theory or interior ballistic multiphase granular Riemann problems, the fully coupled system is sometimes solved using dissipative methods.  In such cases, refining the grid or using higher-order methods can trigger instabilities that can be masked by excessive dissipation \cite{Nussbaum2006}.  Recent methods seem to have overcome this for the BN equations \cite{karni2010hybrid,crochet2013numerical}. Another solution method, used in interior ballistics,  is to split the gas and granular conservation equations and solve a Riemann problem for each phase independently and to compute the nonconservative coupling terms separately \cite{toroReactive2Phase}.

Low-dissipation methods for solving the compressible Euler and Navier-Stokes equations are becoming widespread due, in part, to their excellent properties for compressible turbulence \cite{thornber2008improved, grinstein2007implicit, poludnenko2010interaction,poludnenko2011interaction}.  Fine grids are needed to resolve relevant turbulent features even with low-dissipation methods.  One pervasive problem for any multiphase model representing subgrid particles, whether Eulerian or Lagrangian, is that the computational cell size is limited by the particle diameter.  It is necessary to have enough particles in a computational cell so that the solid phase is comfortably a fluid.  The resulting grid may be too coarse to capture features of the gas-phase, such as the viscous sublayer of a turbulent boundary layer.  Models for high-speed, dynamic, multiphase, and non-Kolmogorov turbulence do not exist, so an eddy viscosity cannot be relied upon to artificially thicken turbulent boundary layers.  Instead, the only choice for the problems considered here is to use implicit large eddy simulation, which requires low dissipation numerical algorithms \cite{grinstein2007implicit,thornber2008improved}.

In the following sections we present a robust and low-dissipation numerical method for solving kinetic theory-based  multiphase equations in highly compressible flows.  The technique uses components from existing Riemann solvers and edge interpolation schemes, and incorporates a relatively small modification to the AUSM$^+$-up \cite{Liou-1996-JCompPhys-AUSM+, liou2006AUSM+UP} for the granular phase.  The fluxes and computed states obtained from solving the gas- and granular-phase Riemann problems are assembled to form the convective fluxes and nonconservative terms.  Numerical experiments are presented that show the ability of the method to preserve the nondisturbing condition, compute multiphase shock-tube problems, and simulate strong shocks interacting with dense layers of dust.

\section{Governing Equations for Multicomponent Granular Reacting Flows}

The kinetic theory-based granular multiphase flow equations describe a molecular gas coupled to an inelastic granular gas.    There are several available granular gas models that could be used \cite{GidaspowBook,GranularGasBook}, which differ mainly in how the intergranular stress and transport coefficients for the solid phase are computed. We choose the granular gas model presented in \cite{GidaspowBook} and neglect effects from the Basset (history) forces, virtual mass forces, and thermophoresis.    This paper is concerned with the solution process, not the model itself, exploring different granular gas models is left for future work.

The subscripts $g$ and $s$ refer to the gas and granular solid phases, respectively, in the discussion that follows.  Only one granular phase is considered, thus
\begin{equation}
\alpha_g + \alpha_s = 1,
\end{equation}
where $\alpha$ is the volume fraction.

\subsection{Governing Equations for the Gas Phase}
The governing equations for $N_g$ chemically reacting gaseous species, momentum, and energy for a chemically reacting multiphase flow \cite{Houim20118527}, and allowing for effects of phase change and heterogeneous reactions, are written as:
\begin{equation}
 \frac{\partial \alpha_g \rho_g Y_{g,j}}{\partial t} + \nabla \! \cdot \! [\alpha_g \rho_g Y_{g,j}(\mathbf{v}_g + \mathbf{V}_{g,j}^d)] = \alpha_g \dot{\omega}_{g,j} + \dot{M}_{g,j} 
\end{equation}
\begin{equation}
 \frac{\partial \alpha_g \rho_g \mathbf{v}_g}{\partial t} + \nabla \! \cdot \! (\alpha_g \rho_g \mathbf{v}_g  \mathbf{v}_g) + \nabla \! \alpha_g p_g = p_g \! \nabla \! \alpha_g +  \nabla \! \cdot \! (\alpha_g \sigma_g) - \mathbf{f}_{\text{Drag}} - \mathbf{f}_{\text{Lift}} + \mathbf{v}_{\text{int}} \dot{M} \ + \alpha_g \rho_g \mathbf{g} 
\end{equation}
\begin{eqnarray}
 \frac{\partial \alpha_g \rho_g E_g}{\partial t} + \nabla \! \cdot \! [\alpha_g \mathbf{v}_g (\rho_g E_g + p_g)] &=& -p_g \nabla \! \cdot \! (\alpha_s \mathbf{v}_s) + \nabla \! \cdot \! (\alpha_g \sigma_g \! \cdot \! \mathbf{v}_g)- \nabla \! \cdot \! (\alpha_g \mathbf{q}_g) - q_{\text{conv}} + \phi_{\text{visc}}  \\  \nonumber
 &&- (\mathbf{f}_{\text{Drag}} + \mathbf{f}_{\text{Lift}}) \! \cdot \! \mathbf{v}_s  +\alpha_g \rho_g \mathbf{g} \! \cdot \! \mathbf{v}_g - \phi_{\text{slip}} + E_{g,\text{int}} \dot{M}   
\end{eqnarray}
where $\alpha_g$, $Y_{g,j}$, $\rho_g$, $p_g$, $T_g$, $E_g$, $\mathbf{v}$, $\mathbf{V}_{g,j}^d$, $\mathbf{g}$, $\sigma$, and $\mathbf{q}_g$ are the volume fraction, mass fraction of species $j$, density, pressure, temperature, total energy, velocity components, mass diffusion velocity of species $j$, gravitational vector, deviatoric stress tensor, and heat diffusion vector, respectively. The homogeneous reaction rate due to chemical reactions is denoted by $\dot{\omega}_j$.  The net rate of phase change from the solid-phase to the gas-phase is represented by $\dot{M}$, and the mass production rate of species $j$ due to phase change is denoted by $\dot{M}_{g,j}$.  The interphase exchange terms $\mathbf{f}_{\text{Drag}}$, $\mathbf{f}_{\text{Lift}}$, $q_{\text{conv}}$, $\mathbf{v}_{\text{int}}$, and $E_{\text{int}}$ are the forces due to drag and lift, convective heat transfer, and interfacial velocity and energy transferred during phase change.   Dissipation of random granular translational kinetic energy ($E_s$) due to viscous effects and production of random granular kinetic energy due to relative velocity between gas particles are denoted by $\phi_{\text{visc}}$ and $\phi_{\text{slip}}$, respectively, and are discussed later.  The species mass equation for all $N_g$ species are solved, which implicitly satisfies mass conservation \cite{Houim20118527, billet2003adaptive}.  

The ideal gas equation-of-state is used to relate the pressure, chemical composition, temperature, and density of the gas phase,
\begin{equation}
  p_g = \rho_g R_u  T_g \sum_{j=1}^{N_g} \frac{Y_{g,j}}{M_j},
\end{equation}
where $M_j$ is the molecular weight of species $j$ and $R_u$ is the universal gas constant. The total energy of the gas phase, $E_g$, is given by
 \begin{equation}
  E_g = H_g -  \frac{p_g}{\rho_g} = \sum_{j=1}^{N_g} Y_{g,j} \Bigl(h_{fj}^0 + \int_{T_0}^{T_g} C_{Pj}(s) ds \Bigr) - \frac{p_g}{\rho_g} + 
       \frac{\mathbf{v}_g \! \cdot \! \mathbf{v}_g}{2},
\end{equation}
where $H_g$ is the total enthalpy, $h_{fj}^0$ and $C_{Pj}$ are the enthalpy of formation at reference temperature $T_0$ and constant-pressure specific heat of species $j$.  Thermodynamic data is taken from Goos \textit{et al.} \cite{Burcat}.  The sound speed is 
\begin{equation}
   c_g^2 = \gamma \frac{p_g}{\rho_g},
\end{equation}
where $\gamma$ is the ratio of specific heats.  

\subsection{Governing Equations for the Particulate Phase}

The governing equations for the particle phase are similar to the Navier-Stokes equations for the gas phase.  The main difference is that compressibility is introduced by changes in solid volume fraction rather than the material density of the particles, so that compaction waves (analogous to acoustic waves, rarefactions, and shocks in a fluid) traveling through granular media directly change the particle volume fraction.       The solid-phase governing equations for mass with $N_s$ species within each particle, momentum, pseudo-thermal energy, and internal energy are:
\begin{equation}
 \frac{\partial \alpha_s \rho_s}{\partial t} + \nabla \! \cdot \! (\alpha_s \rho_s \mathbf{v}_s) = -\dot{M} 
\end{equation} 
\begin{equation}
 \frac{\partial \alpha_s \rho_s Y_{s,j}}{\partial t} + \nabla \! \cdot \! (\alpha_s \rho_s Y_{s,j} \mathbf{v}_s) = \alpha_s \dot{\omega}_{s,j} + \dot{M}_{s,j}
\end{equation}
\begin{equation}
 \frac{\partial \alpha_s \rho_s \mathbf{v}_s}{\partial t} + \nabla \! \cdot \! (\alpha_s \rho_s \mathbf{v}_s  \mathbf{v}_s)  + \nabla \! p_s + \nabla \! p_{\text{fric}} = -\alpha_s \nabla \! p_g + \nabla \! \cdot \! (\alpha_s \sigma_s) + \mathbf{f}_{\text{Drag}} + \mathbf{f}_{\text{Lift}} - \mathbf{v}_{\text{int}} \dot{M} \ + \alpha_s \rho_s \mathbf{g} 
\end{equation}
\begin{equation}
\label{eqn:PTE}
\frac{\partial \alpha_s \rho_s E_s}{\partial t} + \nabla \! \cdot \! (\alpha_s \rho_s E_s \mathbf{v}_s)= -p_s \! \nabla \! \cdot \! \mathbf{v}_s +  \alpha_s \sigma_s : \! \nabla \mathbf{v}_s + \nabla \! \cdot \! (\alpha_s \lambda_s \nabla \Theta_s) -\dot{\gamma}  - \phi_{\text{visc}} + \phi_{\text{slip}} - E_{s,\text{int}} \dot{M}
\end{equation}
\begin{eqnarray}
\label{eqn:granThermEn}
\frac{\partial \alpha_s \rho_s e_s }{\partial t} + \nabla \! \cdot \! (\alpha_s \rho_s e_s \mathbf{v}_s) = q_{\text{conv}} - e_{s,\text{int}} \dot{M} + \dot{\gamma},
\end{eqnarray}
where $p_s$, $p_{\text{fric}}$, $e_s$, $E_s$, and $\dot{\gamma}$ are the solids pressure derived from kinetic theory, friction-collisional pressure, internal energy, pseudo-thermal energy, and dissipation of $E_s$ due to inelastic particle collisions.  The homogeneous reaction rate of granular species $i$ in the granular phase is denoted by $\dot{\omega}_{s,i}$. Pseudo-thermal energy, $E_s$, represents the energy due to random translational motion of the particles
\begin{equation}
E_s = \frac{3}{2} \alpha_s \rho_s \Theta_s.
\label{eqn:PTE_Def}
\end{equation} 
$E_s$ is described by a granular temperature, $\Theta_s$, defined as the mean-square of the particle velocity fluctuations.

The mean kinetic energy of the solid phase is not included in Eqns.~\eqref{eqn:PTE} and \eqref{eqn:PTE_Def} to avoid small truncation errors of kinetic energy that can lead to unphysical values of $E_s$.  Dissipation of $E_s$ due to inelastic particle collisions ($\dot{\gamma}$) and viscous damping ($\phi_{\text{visc}}$) act as sinks and often reduce it to the point where it is small compared to the mean kinetic energy, $ \alpha_s \rho_s \mathbf{v}_s \! \cdot \! \mathbf{v}_s/2$.    Small truncation errors in the mean kinetic energy cause small fluctuations of $\Theta_s$, which in dense regimes produces severe oscillations of intergranular stress that can degrade calculations to the point of failure.  The Flash astrophysics code \cite{Flash} uses a similar approach and removes kinetic energy from the total energy in calculations where the internal energy is four orders of magnitude lower than the kinetic energy.

The granular temperature, $\Theta_s$, is not related to the solid temperature of the particles, $T_s$, which is determined from the solid internal energy, $e_s$,
\begin{equation}
   e_s = \sum_{j=1}^{N_s} Y_{s,j} \left(e_{fj}^0 + \int_{T_0}^{T_s} C_{V,s,j}(s) ds \right).
\end{equation}
where $N_s$ is the number of species in the solid phase, $e_{fj}^0$ is an internal energy of formation, and $C_{V,s,j}$ is the constant-volume specific heat of species $j$ in the particle phase.  
This equation assumes a small enough Biot modulus that the temperature distribution within each particle is uniform.  More complex equations that include the effect of nonuniform temperature distributions inside the particles  could be used when the Biot number is large \cite{koo1977transient,markatos1983analysis,Nussbaum2006}.  

The density of the solid particles is given by
\begin{equation}
  \rho_s = \sum_{j=1}^{N_s} Y_{s,j} \rho_{s,j}(T_s),
\end{equation}
where $\rho_{s,j}$ is the density of species $j$. In this work, we take $\rho_s$ to be constant.  Relaxing this assumption should be straightforward and is a topic of future work.  

The solids pressure, $p_s$, is given by an equation of state for a granular gas \cite{GidaspowBook},
\begin{equation}
p_s = \rho_s \Theta_s [ \alpha_s(1+2(1+e)\alpha_s g_0)],
\end{equation}
where $e$ is the coefficient of restitution.  The radial distribution function, $g_0$, is defined by \cite{GidaspowBook},
\begin{equation}
  \frac{1}{g_0} = 1 - \left( \frac{\alpha_s}{\alpha_{s,\text{max}}} \right) ^{1/3},
  \label{eqn:radDist}
\end{equation}
where $\alpha_{s,\text{max}}$ is the packing limit, which is an input parameter commonly set to 0.65 \cite{Agrawal2001role,igci2008filtered}. Other expressions for $g_0$ and $p_s$ could be used \cite{kamenetsky2000evolution, van2001comparative}. Frictional-collisional pressure is used in highly packed granular regions \cite{johnson1987friction},
\begin{equation}
   p_{\text{fric}}[\text{Pa}] = \begin{cases}
      0 &\mbox{if } \alpha_s < \alpha_{s,\text{crit}} \\
      0.1 \alpha_s \displaystyle{\frac{(\alpha_s - \alpha_{s,\text{crit}})^2}{(\alpha_{s,\text{max}} - \alpha_s)^5}} &\mbox{if } \alpha_s \geq \alpha_{s,\text{crit}}
    \end{cases}
    \label{eqn:pFric}
\end{equation}
where $\alpha_{s,\text{crit}}$ is 0.5 unless otherwise noted and $p_{\text{fric}}$ is in units of Pa.   Other expressions for friction pressure could be used as well \cite{GidaspowBook,koo1977transient,markatos1983analysis,Nussbaum2006,LingBalachandar2013-Model-Experiment,jenike1987theory,van2001comparative,saurel1999multiphase}. 
(Friction pressure is often called intergranular stress in internal ballistic calculations.)  In this paper we call the sum of the solids and friction pressures the total intergranular stress,
\begin{equation}
  p_{s,\text{tot}}=p_s + p_{\text{fric}},
\end{equation}
to avoid ambiguity

Friction pressure is a necessary addition to $p_s$ when the particles occupy a high volume fraction.  The binary collision assumption in the Boltzmann equation, which is used to derive the granular-phase equation state, becomes invalid at high volume fractions when particles are in contact with several other particles. 
 In high-volume fraction regions, high rates of granular cooling from $\dot{\gamma}$ often reduce $\Theta_s$ to absolute zero rapidly and result in zero solids pressure.  The addition of friction pressure is one way to add intergranular stress in dense regions needed to limit compaction of the solid phase.  Another way, which is not used in this paper, is to limit the minimum granular temperature  near the packing limit \cite{kamenetsky2000evolution}.  This approach has a physical basis.  The coefficient of restitution, $e$, is not a constant, as assumed in many kinetic theory-based granular multiphase models, but is a function of impact velocity.  Higher impact velocities (higher $\Theta_s$) convert more pseudo-thermal energy, $E_s$, into internal energy, $e_s$, through viscoelastic deformation \cite{GranularGasBook}.  Collisions become increasingly elastic ($e \rightarrow 1$) as the granular temperature decreases.  This effect limits both the rate at which $\Theta_s$ can decrease and its lower value.  Recent granular gas models account for this to some extent, but have not been used in multiphase flow calculations.  

The pressure term in the $E_s$ equation, Eqn.~\eqref{eqn:PTE}, does not include flow work from frictional pressure.  This is consistent from a thermodynamics point-of-view.  The friction pressure models used in this work are  analogous to a barotropic equation of state for a fluid.  It is possible to show, using thermodynamic arguments, that the total energy equation for a barotropic fluid simplifies to the mechanical energy equation.  Performing a similar analysis for granular mixtures reveals that only the solids pressure, $p_s$, contributes to changes in $E_s$ from compression and expansion (changes in volume fraction) of the particles.  If the mean kinetic energy were included in the definition of $E_s$ the friction pressure would enter the total pseudo-thermal energy equation, Eqn.~\eqref{eqn:PTE_Def}, as $\mathbf{v}_s \! \cdot \! \nabla p_{\text{fric}}$, while the solids pressure would be present in the usual manner, $\nabla \! \cdot \! (\mathbf{v}_s p_s)$.

The granular-compaction wave speed can be derived for the particulate phase in a manner identical to that of a real gas \cite{serna2005capturing}. The compaction wave speed for a general granular gas with both solids and friction pressure that is a function of $\alpha_s$ is 
\begin{equation}
   c_s^2 = \frac{1}{\rho_s} \left[\frac{ \partial (p_s + p_\text{fric})}{\partial \alpha_s} \Biggr|_{\Theta_s} + 
                            \frac{2}{3}\frac{\Theta_s \left(  \frac{\partial p_s}{\partial \Theta_s} \bigr|_{\alpha_s} \right)^2} {\rho_s \alpha_s^2} \right].
\end{equation}
Using the $p_s$ and $p_{\text{fric}}$ definitions above gives  
\begin{equation}
  c_s^2 = \Theta_s \left( A + \frac{2}{3} A^2 + \alpha_s B \right) + c_{\text{fric}}^2
\end{equation}
where
\begin{equation}
   A = 1 + 2(1+e)\alpha_s g_0, \ \ \ \
   B = 2(1 + e)(g_0 + \alpha_s g_0'), \ \ \ \ 
   g_0' = \frac{g_0^2}{3 \alpha_{s,\text{max}}} \left(\frac{\alpha_{s,\text{max}}}{\alpha_s} \right)^{2/3}.
\end{equation}
The frictional contribution to the compaction wave speed is
\begin{equation}
  c_{\text{fric}}^2 [\text{m}^2/\text{s}^2] = \begin{cases}
        0 &\mbox{if } \alpha_s < \alpha_{s,\text{crit}} \\
        \displaystyle{\frac{1}{\rho_s[\text{kg/m}^3]} \frac{(\alpha_s - \alpha_{s,\text{crit}})}{(\alpha_{s,\text{max}} - \alpha_s)^5} \left[ \alpha_s \left( \frac{1}{5} + \frac{1}{2} \frac{\alpha_s - \alpha_{s,\text{crit}}}{\alpha_{s,\text{max}} - \alpha_s} \right) + \frac{\alpha_s - \alpha_{s,\text{crit}}}{10} \right]} &\mbox{if } \alpha_s \geq \alpha_{s,\text{crit}}.
      \end{cases}
\end{equation}

\subsection{Solution Process for the Governing Equations}

The purpose of this paper is to present a numerical method for solving the equations for dense granular multiphase flow with a focus on hyperbolic and parabolic terms.  Therefore, phase changes and reactions are not  considered in the remainder of this paper.  
 The solution process uses a Strang operator splitting scheme,
\begin{equation}
  \mathbf{U}^{t+2 \Delta t}=\mathcal{H}_{xy}^{\Delta t} \ \mathcal{P}_{xy}^{\Delta t} \ \mathcal{S}^{2\Delta t} \ \mathcal{P}_{xy}^{\Delta t} \ \mathcal{H}_{xy}^{\Delta t}(\mathbf{U}^t),
\end{equation}
where $\mathbf{U}$ is the vector of conserved variables, $\mathcal{H}_{xy}^{\Delta t}$ represents the directionally unsplit integration of the convective terms for a time-step size of $\Delta t$, $\mathcal{P}_{xy}^{\Delta t}$ represents integration of the parabolic terms for a time-step size of $\Delta t$, and $\mathcal{S}^{2\Delta t}$ represents integration of the source term operator for a time-step of $2 \Delta t$.  The method of solution for each of these operators will be discussed in turn.  

\section{Solution of the Hyperbolic Terms, $\mathcal{H}_{xy}^{\Delta t}$}

For brevity, only the  discretization of the one-dimensional equations in the x-direction is discussed, noting that discretization in y- and z-directions is similar.  Superscripts $L$ and $R$ on a variable refer to left- and right-biased reconstructions on the edge of a computational cell in the following discussion. 

The convective terms for the granular phase are:
\begin{eqnarray}
     \frac{\partial \alpha_s \rho_s}{\partial t} + \nabla \! \cdot \! (\alpha_s \rho_s  u_s) &=& 0 \\
    \frac{\partial \alpha_s \rho_s Y_{s,j}}{\partial t} +\nabla \! \cdot \! (\alpha_s \rho_s Y_{s,j} u_s) &=& 0 \\
    \frac{\partial \alpha_s \rho_s \mathbf{v}_s}{\partial t} + \nabla \! \cdot \! (\alpha_s \rho_s \mathbf{v}_s \mathbf{v}_s) + \nabla \! p_s + \nabla \! p_{\text{fric}} &=& p_g \nabla \! \alpha_g + 
      \mathbf{f}_{\text{Lift}} \\
   \frac{\partial \alpha_s \rho_s E_s}{\partial t} + \nabla \! \cdot \! (\alpha_s \rho_s E_s u_s) &=& -p_s \nabla \! \cdot \! \mathbf{v}_s \\
   \frac{\partial \alpha_s \rho_s e_s}{\partial t} + \nabla \! \cdot \! (\alpha_s \rho_s e_s u_s) &=& 0. 
\end{eqnarray}
The Magnus lift force is defined by
\begin{equation}
 \mathbf{f}_{\text{lift}} = C_l \alpha_s \rho_g (\mathbf{v}_s - \mathbf{v}_g) \! \times \! (\nabla \! \times \! \mathbf{v}_g),
 \label{Eqn:fLift}
\end{equation}
where the lift coefficient, $C_l$, has a value of 0.5 unless otherwise noted \cite{Drew1987113}.  The lift force is discretized using standard second-order differencing if $\min(\alpha_s) > \alpha_{s,\text{min}}$ for all points in the stencil where $\alpha_{\text{min}}=10^{-10}$ is the minimum particle volume fraction.

The gas-phase convective terms are
\begin{eqnarray}
     \frac{\partial \alpha_g \rho_g Y_{g,j}} {\partial t} + \nabla \! \cdot \! (\alpha_g \rho_g Y_{g,j} \mathbf{v}_g) &=& 0 \\
    \frac{\partial \alpha_g \rho_g \mathbf{v}_g}{\partial t} + \nabla \! \cdot \! (\alpha_g \rho_g \mathbf{v}_g \mathbf{v}_g) + \nabla \alpha_g p_g &=& p_g \! \nabla \! \alpha_g -
      \mathbf{f}_{\text{Lift}} \\
   \frac{\partial \alpha_g \rho_g E_g}{\partial t} + \nabla \! \cdot \! [\mathbf{v}_g(\alpha_g \rho_g E_g + \alpha_g p_g)] &=& -p_g \nabla \! \cdot \! \alpha_s \mathbf{v}_s -
     \mathbf{f}_{\text{Lift}} \! \cdot \! \mathbf{v}_s.
\end{eqnarray}
The left-hand side is in a form that is solvable by many standard numerical algorithms as long as the nozzling term ($p_g \! \nabla \! \alpha_g$) in the momentum equation and the $pDV$ work term ($p_g \! \nabla \! \cdot \! \alpha_s \mathbf{v}_s$) for the energy equation are treated as independent source terms.   Nevertheless, the pressure nondisturbing condition \cite{saurel1999multiphase} is usually \textit{not} satisfied with such an approach, as discussed in the introduction.

\begin{figure}
  \centering
  \subfloat[Actual control volume.]{\includegraphics[width=0.49\textwidth]{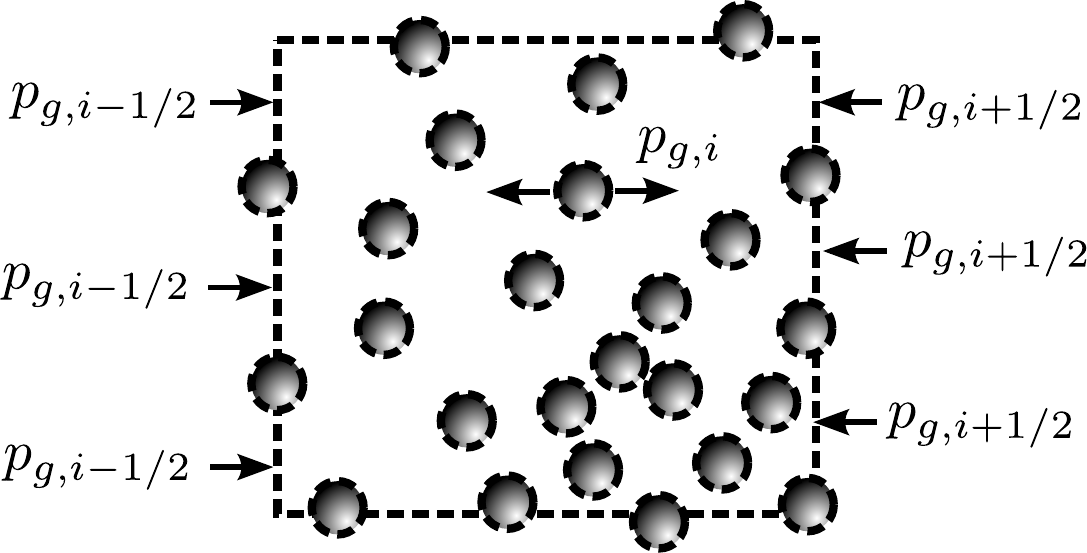}} \
  \subfloat[Control volume with particles collapsed]{\includegraphics[width=0.49\textwidth]{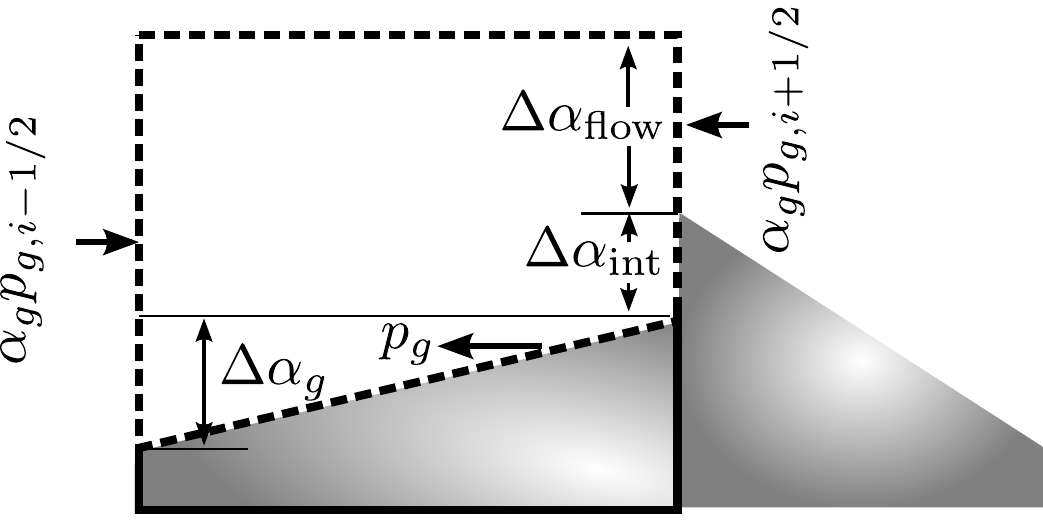}}
\caption{Application of the concept of two-fluid approach of Chang and Liou \cite{Chang-Two-Fluid-2007} to granular flows.}
\label{fig:ChangMultifluid} 
\end{figure}

An approach that has been shown to work for coupling compressible gases to compressible liquids with two-fluid multiphase models is that of Chang and Liou \cite{Chang-Two-Fluid-2007}.  They consider control volumes for the gas and liquid phases separately. Each computational cell edge is divided into gas-gas, liquid-liquid, and gas-liquid sections.  The hyberbolic fluxes in each section are computed independently with an appropriate Riemann solver.  Application of their concept to granular multiphase flows is illustrated in Fig.~\ref{fig:ChangMultifluid}.
Each face of a computational cell edge for the gas phase is divided into two sections: gas phase on both sides, with a fractional area of
\begin{equation} 
  \Delta \alpha_{g,\text{flow}} = \min(\alpha_{g,i+\frac{1}{2}}^L, \alpha_{g,i+\frac{1}{2}}^R),
\end{equation}
and gas on one side and particles on the other, with fractional areas of
\begin{equation} 
  \Delta \alpha_{g,\text{int},i+\frac{1}{2}}^L = \max(0,\alpha_{g,i+\frac{1}{2}}^R - \alpha_{g,i+\frac{1}{2}}^L), \quad
  \Delta \alpha_{g,\text{int},i+\frac{1}{2}}^R = \max(0,\alpha_{g,i+\frac{1}{2}}^L - \alpha_{g,i+\frac{1}{2}}^R)
\end{equation}
on the right and left faces of computational cell $i$, respectively.  The total gas-phase flux vector at  edge $i+1/2$, $\mathbf{F}_{g,i+\frac{1}{2}} $ of the control volume is \cite{Chang-Two-Fluid-2007}:
\begin{equation}
 \mathbf{F}_{g,i+\frac{1}{2}} = \Delta \alpha_{g,\text{flow},i+\frac{1}{2}} \mathbf{F}_{g-g} + \Delta \alpha_{g,\text{int},i+\frac{1}{2}}^L \mathbf{F}_{g-s} + \Delta \alpha_{g,\text{int},i+\frac{1}{2}}^R \mathbf{F}_{s-g} 
\end{equation}
where $\mathbf{F}_{g-g}$ is the flux resulting from the solution to a gas-gas Riemann problem, $\mathbf{F}_{g-s}$ is the gas-phase flux resulting from a gas-particle Riemann problem with the gas phase on the left and particles on the right, and $\mathbf{F}_{s-g}$ is the gas-phase flux from a gas-particle Riemann problem with the gas phase on the right and particles on the left. Chang and Liou \cite{Chang-Two-Fluid-2007} discretize the nonconservative nozzling term  within each computational cell as
\begin{equation}
  p_g \frac{\partial \alpha_g}{\partial x} \approx p_{g,i} \frac{\Delta \alpha_g}{\Delta x} =  p_{g,i} \frac{\alpha_{g,i+\frac{1}{2}}^L - \alpha_{g,i-\frac{1}{2}}^R}{\Delta x},
\end{equation} 
where $\alpha^L$ and $\alpha^R$ refer to the left- and right-biased interpolation of the volume fraction to the cell face.  

The Riemann problem for the gas-gas section can be solved by any standard method (such as HLLC \cite{harten1983upstream,ToroHLLC}). If the particles were assumed to be compressible, an interfacial gas-solid Riemann problem would have to be solved in the gas-solid section.  Instead, the particles are assumed to be incompressible, and a two-shock or two-rarefaction solution could be used based on relative velocity between the gas and particles.   Because the granular Riemann problem takes jumps of volume fraction into account when computing the granular fluxes (discussed below), there is not any need to modify the granular fluxes in the solid-gas section when computing the solid fluxes for the cell on the right in Fig.~\ref{fig:ChangMultifluid}.  

The procedure discussed above can be simplified substantially by using the incompressibility assumption of the particles.  The volume fraction (while integrating the hyperbolic terms) is controlled by motion of the solid particles.  This implies that volume fraction is not a function of the gas-phase state.  Thus, the product rule can be applied to the pressure gradient term ($\nabla \alpha_g p_g$) which simplifies the momentum equation to
\begin{equation}
      \frac{\partial \alpha_g \rho_g \mathbf{v}_g}{\partial t} + \nabla \! \cdot \! (\alpha_g \rho_g \mathbf{v}_g \mathbf{v}_g)  = -\alpha_g \! \nabla \! p_g.
\end{equation}
It is easy to satisfy the pressure nondisturbing condition with this simplification.  

The above a simplification is not rigorously possible if particle incompressibility is not assumed.  The solid material density would be a function of  the total isotropic stress acting on the particle: $\rho_s = \rho_s(T_s, p_g, p_{s,\text{tot}}$) as well as $T_s$.  Conservation of particle mass would then dictate that solid volume fraction is implicitly a function of $\rho_s$. The high pressure behind the shock would increase $\rho_s$ with a resulting decrease in particle diameter. Thus, a gas-phase shock would cause a discontinuous decrease of  $\alpha_s$ in addition to a pressure jump.  As a result, $\nabla \! \alpha_g p_g \neq \alpha_g \! \nabla \! p_g + p_g \! \nabla \! \alpha_g$ across a shock.  If the particle incompressibility is assumed then $\rho_s = \rho_s(T_s)$ and $\nabla \! \alpha_g p_g = \alpha_g \! \nabla \! p_g + p_g \! \nabla \! \alpha_g$  since $\alpha_s$ would not be a function of the gas-phase pressure and, as a result, would not change discontinuously with a shock.

One problem that merging the pressure gradient and nozzling terms introduces is that there are no portions of the gas-phase hyperbolic terms that resemble the Euler equations.   Despite this, the control volume shown in Fig.~\ref{fig:ChangMultifluid} is still valid and the solution to the Riemann problem on the gas-gas fraction of the cell face is physically correct.  The pressure from the gas-gas Riemann problem is stored and used to compute $\nabla \! p_g$ of the nonconservative pressure term $\alpha_g \nabla \! p_g$.  The only requirement for evaluating the gas-phase flux is that pressure is explicitly calculated. Not all Riemann solvers do this.  For example HLLC \cite{harten1983upstream,ToroHLLC} and AUSM \cite{liou2006AUSM+UP} explicitly calculate pressure while Roe's Riemann solver \cite{Roe-1981-JCompPhys}, HLL \cite{harten1983upstream}, and Rusanov \cite{Toro-book} do not.

The semi-discrete form of the convective terms in one dimension, based on the above simplification, at computational cell $i$ with grid size $\Delta x$ and cell edges located at $i+1/2$ and $i-1/2$ are:
\begin{eqnarray}
     \frac{d (\alpha_g \rho_g Y_{g,j})_i}{d t} &=& -\frac{\alpha_{g,i+\frac{1}{2}}\rho_{g,i+\frac{1}{2}}^- u_{g,i+\frac{1}{2}}^- Y_{g,j,i+\frac{1}{2}}^- - \alpha_{g,i-\frac{1}{2}}\rho_{g,i-\frac{1}{2}}^+ u_{g,i-\frac{1}{2}}^+Y_{g,j,i-\frac{1}{2}}^+}{\Delta x} \\
    \frac{d (\alpha_g \rho_g u_g)_i}{d t} &=& -\frac{\alpha_{g,i+\frac{1}{2}}\rho_{g,i+\frac{1}{2}}^- u_{g,i+\frac{1}{2}}^-u_{g,i+\frac{1}{2}}^- - \alpha_{g,i-\frac{1}{2}}\rho_{g,i-\frac{1}{2}}^+ u_{g,i-\frac{1}{2}}^+ u_{g,i-\frac{1}{2}}^+}{\Delta x} -\alpha_{g,i}\frac{p_{g,i+\frac{1}{2}}^- - p_{g,i-\frac{1}{2}}^+}{\Delta x} \\
   \frac{d (\alpha_g \rho_g E_g)_i}{d t} &=& -\frac{ \alpha_{g,i+\frac{1}{2}}u_{g,i+\frac{1}{2}}^-(\rho_{g,i+\frac{1}{2}}^-E_{g,i+\frac{1}{2}}^- +  p_{g,i+\frac{1}{2}}^-) -  \alpha_{g,i-\frac{1}{2}}u_{g,i-\frac{1}{2}}^+(\rho_{g,i-\frac{1}{2}}^+ E_{g,i-\frac{1}{2}}^+ +  p_{g,i-\frac{1}{2}}^+)}{\Delta x} \\
       &&-p_{g,i} \frac{\alpha_{s,i+\frac{1}{2}}u_{s,i+\frac{1}{2}} - \alpha_{s,i-\frac{1}{2}}u_{s,i-\frac{1}{2}}}{\Delta x} \nonumber \\
     \frac{d (\alpha_s \rho_s)_i}{d t} &=& -\frac{\dot{m}_{s,i+\frac{1}{2}} - \dot{m}_{s,i-\frac{1}{2}}}{\Delta x} \\
     \frac{d (\alpha_s \rho_s Y_{s,j})_i}{d t} &=& -\frac{\dot{m}_{s,i+\frac{1}{2}} Y_{s,j,i+\frac{1}{2}} - \dot{m}_{s,j,i-\frac{1}{2}}Y_{s,j,i-\frac{1}{2}}}{\Delta x} \\
    \frac{d (\alpha_s \rho_s u_s)_i}{d t} &=& -\frac{\dot{m}_{s,i+\frac{1}{2}}u_{s,i+\frac{1}{2}} + p_{s,\text{tot},i+\frac{1}{2}} -  \dot{m}_{s,i-\frac{1}{2}}u_{s,i-\frac{1}{2}} - p_{s,\text{tot},i-\frac{1}{2}}}{\Delta x} -\alpha_{s,i}\frac{p_{g,i+\frac{1}{2}}^- - p_{g,i-\frac{1}{2}}^+}{\Delta x} \\
   \frac{d (\alpha_s \rho_s E_s)_i}{d t} &=& -\frac{\dot{m}_{s,i+\frac{1}{2}}  E_{s,i+\frac{1}{2}} - \dot{m}_{s,i-\frac{1}{2}}E_{s,i-\frac{1}{2}}}{\Delta x} - p_{s,i+\frac{1}{2}} \frac{u_{s,i+\frac{1}{2}} - u_{s,i-\frac{1}{2}}}{\Delta x} \\
   \frac{d (\alpha_s \rho_s e_s)_i}{d t} &=& -\frac{\dot{m}_{s,i+\frac{1}{2}}  e_{s,i+\frac{1}{2}} - \dot{m}_{s,i-\frac{1}{2}}e_{s,i-\frac{1}{2}}}{\Delta x}, 
\end{eqnarray}
where subscript $j$ is the species index and $\dot{m}_s=\alpha_s \rho_s u_s$ is the mass flux for the solid phase.  Gas-phase variables with $+$ and $-$ superscripts  (e.g., $\rho_{g,i+\frac{1}{2}}^+$ and $\rho_{g,i-\frac{1}{2}}^-$) are needed for the double-flux model \cite{billet2003adaptive} discussed below. Examination of the semi-discrete equations shows that the gas and granular phases are coupled directly through the volume fraction, gas-phase pressure gradient, and $pDV$ work from particles entering and leaving the control volume.  The edge-centered fluxes, pressure, velocities, volume fraction, etc. needed to assemble the conservative and nonconservative convective terms are calculated with approximate solutions to separate gas and granular Riemann problems discussed below.

\subsection{Approximate Solution of the Gas-Phase Riemann Problem}

\begin{figure}
  \centering
  \subfloat[$\rho_g$, $Y_{He}$, $Y_{N_2}$, and $1/(\gamma-1)$]{\includegraphics[width=0.49\textwidth]{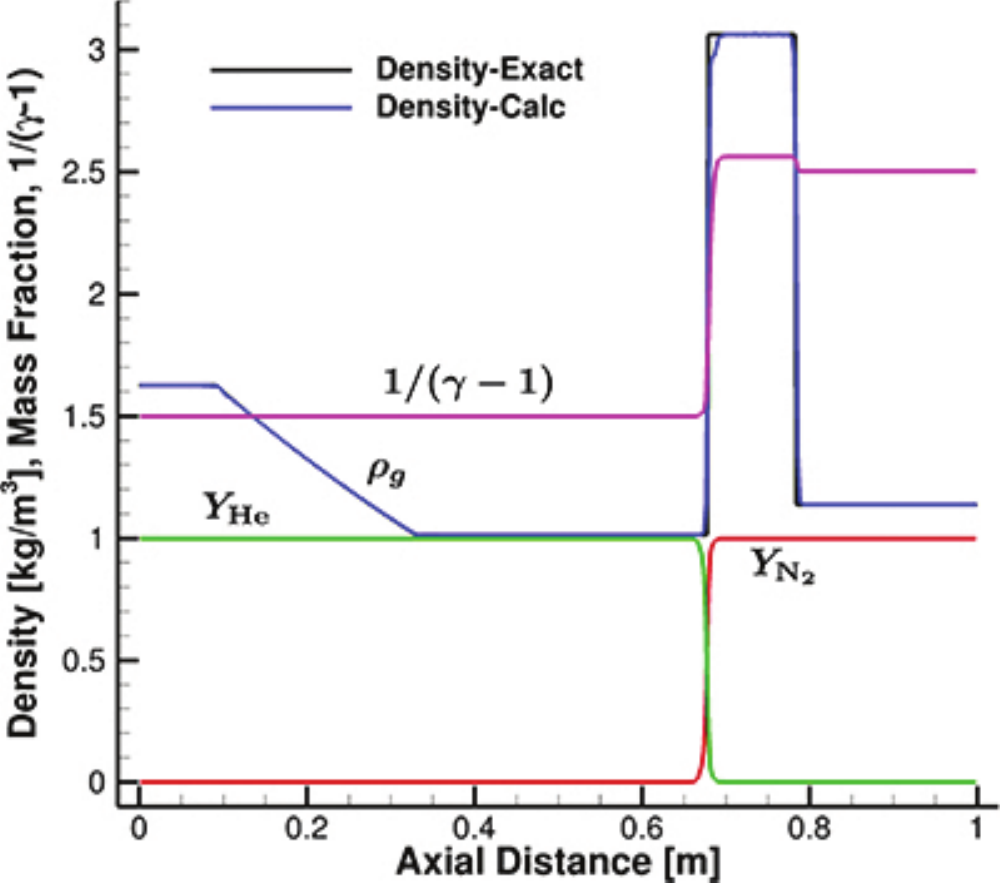}}
  \subfloat[Temperature]{\includegraphics[width=0.49\textwidth]{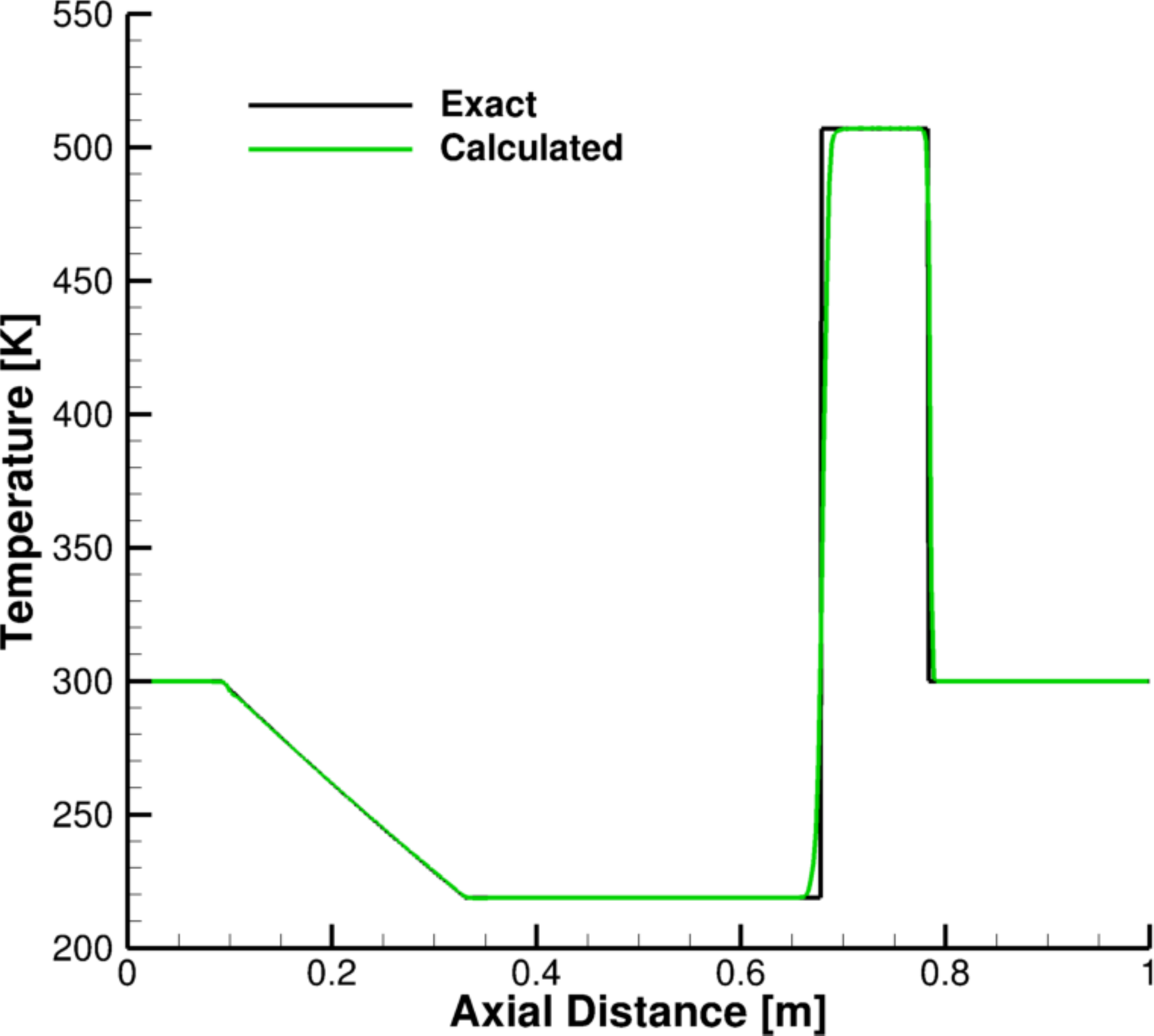}}
\caption{Comparison of the exact and computed solution to the Helium-Nitrogen Riemann problem using the double flux model with 400 points at a time of 400 $\mu$s. The initial discontinuity was placed at $x=0.5$ m and the pressure of the pure He on the left was 10 atm and the pressure of the pure N$_2$ on the right was 1 atm.  The temperature of both fluids was initially  300 K.}
\label{fig:HEN2Riemann} 
\end{figure}

The ratio of specific heats is, in general, a function of temperature and chemical composition.  Advecting a multicomponent gaseous interface can produce severe pressure oscillations \cite{billet2003adaptive, Houim20118527} similar to those produced by advecting multiphase interfaces.  The quasi-conservative double-flux model \cite{billet2003adaptive}  that prevents these oscillations works well for a variety of complex reacting flows such as premixed flames, cellular structure of detonations, and shock waves interacting with and diffusion flames~\cite{Houim20118527}.  It is currently one of the only methods that can converge to the correct weak solution of the multicomponent Riemann problem as shown in Fig.~\ref{fig:HEN2Riemann}.  Fully conservative methods for multicomponent gaseous flows  have recently been developed \cite{johnsen2012preventing}, but these have not been extended to situations where  specific heat is a function of temperature.    

Details of the double-flux model \cite{billet2003adaptive, Houim20118527} are briefly summarized here for completeness.  First, at the start of the calculation, specific heats for each gaseous species are stored in tables for linear interpolation with uniform temperature increments of $\Delta T = 1$ K.  If $T_m < T_g < T_{m+1}$ the constant-pressure specific heat of species $j$ is
\begin{equation}
  C_{Pj}(T_g) = a_j^m T_g + b_j^m,
  \label{eqn:cpTabular}
\end{equation}
where $m$ is the temperature interval that $T_g$ resides in the table.  The specific total energy of the mixture is then calculated in a form similar to a constant-property single-component ideal gas,
\begin{equation}
   E_g = h_0^m + \frac{p_g}{\rho_g(\gamma - 1)} + \frac{\mathbf{v}_g \! \cdot \! \mathbf{v}_g}{2},
\end{equation}
where
\begin{equation}
  h_0^m = \sum_{j=1}^{N_g}Y_{g,j}(\bar{h}_{j0}^m - \bar{b}_j^m T_m), \ \ 
  \bar{h}_{j0}^m = h_{fi}^0 + \sum_{k=1}^m \left[ \int_{T_{k-1}}^{T_k} \left(a_j^k s + b_j^k \right) ds \right], \ \ 
  \bar{b}_{j}^m = \frac{a_j^m}{2}(T_g + T_m) + b_j^m.
\end{equation}
An extension of the analysis performed by Billet and Abgrall \cite{billet2003adaptive} to multiphase mixtures shows that if  $u_g = u_s = u$ and $\nabla \! p_g=\nabla \! u = 0$,  the uniform pressure and velocity in a multicomponent granular flow is preserved if $\gamma$ and $\alpha_g \rho_g h_0^m$ are frozen in each cell for the entire time step. Thus, the gas mixture in each computational cell is treated as its own constant-property single-component ideal gas.  The flux at each cell face must be calculated twice as a consequence: once for the cell on the left side of the face using $\gamma_i$ and $\alpha_{g,i} \rho_{g,i} h_{0,i}^m$, and then again for the cell on the right using $\gamma_{i+1}$ and $\alpha_{g,i+1} \rho_{g,i+1} h_{0,i+1}^m$.  The flux and pressure gradients are then calculated for cell $i$ using the solution to the Riemann problem on each cell face that used $\gamma_i$.  Since $\alpha_g \rho_g h_{0}^m$ is frozen, the quantity $\alpha_g \rho_g u_g h_{0}^m$ is not added to the energy flux.   

Any Riemann solver can be used for the gas-phase provided that the pressure and velocity can are explicitly computed, as discussed above.  An additional requirement for the double-flux model is that the flux evaluation method preserves a stationary contact.  HLLC and AUSM meet both of these requirements.  In this paper we use the HLLC solver \cite{harten1983upstream, ToroHLLC} to compute the approximate solution of the gas-phase Riemann problem, which is rotated near shocks to avoid carbuncle-related anomalies \cite{pandolfi2001Carbuncle, huang2010cures,Houim20118527}.   

HLLC  has been modified to return $\mathbf{P}_g = (\rho_g, Y_{g,j}, u_g, p_g, E_g)^T$ rather than the fluxes:
\begin{equation}
   \mathbf{P}_{g,i+\frac{1}{2}} = \begin{cases}
        \mathbf{P}^L_{g,i+\frac{1}{2}} & \ \ \mbox{if } 0 \leq S^L  \\
        \mathbf{P}_g^{L*}              & \ \ \mbox{if } S^L \leq 0 \leq S^* \\
        \mathbf{P}_g^{R*}              & \ \ \mbox{if } S^* \leq 0 \leq S^R \\
        \mathbf{P}^R_{g,i+\frac{1}{2}} & \ \ \mbox{if } S^R \leq 0, 
   \end{cases}
\end{equation}
where
\begin{equation}
  \mathbf{P}_{g}^K = \begin{bmatrix}
     \rho_{g}^K \\
     Y_{g,j}^K \\
     u_{g}^K \\
     p_{g}^K \\
     E_{g}^K \\ 
  \end{bmatrix}, 
  \quad
  \mathbf{P}_g^{K*} = \begin{bmatrix}
      \rho_{g}^{K*} \\
      Y_{g,j}^K \\
      S^* \\
      p^* \\
      E_g^{K*} 
  \end{bmatrix},  
\end{equation}
the superscript $K$ refers to either $L$ or $R$,
\begin{equation}
  \rho_g^{K*} = \rho_g^K \frac{S^K - u_g^K}{S^K - S^*}, \quad
  E_g^{K*} = E_g^K + \frac{p^* S^* - p_g^K u_g^K}{\rho_g^K (S^K - u_g^K)},
\end{equation}
and $\rho_g^K$ is computed from the interpolated pressure, temperature, species mass fractions, and the equation of state.  The velocity of gas-phase contact surface ($S^*$) is
\begin{equation}
  S^* = \frac{p_g^R - p_g^L + \rho_g^L u_g^L (S^L - u_g^L) - \rho_g^R u_g^R (S^R - u_g^R)}
             {\rho_g^L (S^L - u_g^L) - \rho_g^R (S^R - u_g^R)}
\end{equation}
and the pressure at the contact surface ($p^*$) is
\begin{equation}
  p^* = p_g^L + \rho_g^L (S^L - u_g^L) (S^* - u_g^L).
\end{equation}
The left and right wave speeds are estimated using a Roe-averaged approach \cite{Einfeldt1991273}
\begin{equation}
   S^L = \min(u_g^L - c_g^L, \tilde{u}_g - \tilde{c}_g), \quad 
   S^R = \max(u_g^R + c_g^R, \tilde{u}_g + \tilde{c}_g),
\end{equation}
where
\begin{equation}
  \tilde{\mathbf{v}}_g = \frac{\sqrt{\rho_g^L}\mathbf{v}_g^L + \sqrt{\rho_g^R}\mathbf{v}_g^R} {\sqrt{\rho_g^L}\sqrt{\rho_g^R}}, \quad
  \tilde{H}_g = \frac{\sqrt{\rho_g^L}H_g^L + \sqrt{\rho_g^R}H_g^R} {\sqrt{\rho_g^L}\sqrt{\rho_g^R}},
  \quad
  \tilde{c}_g = \left[(\gamma - 1) \left(\tilde{H}_g - \frac{1}{2} \tilde{\mathbf{v}}_g \! \cdot \tilde{\mathbf{v}}_g \right) \right]^{1/2},
\end{equation}
and, for the double-flux model or a constant-$\gamma$ gas,
\begin{equation}
   H_g^K = \frac{\gamma}{\gamma - 1}\frac{p_g^K}{\rho_g^K} + \frac{1}{2}\mathbf{v}_g^K \! \cdot \! \mathbf{v}_g^K.
\end{equation}

The double-flux model requires that the Riemann problem be calculated twice at each cell edge to give $\mathbf{P}_{g,i+\frac{1}{2}}^-$ and $\mathbf{P}_{g,i+\frac{1}{2}}^+$.  This is the reason for separating the edge-centered variables with superscripts + and - in Eqns.~(39)-(41) and (44).  $\mathbf{P}_{g,i+\frac{1}{2}}^-$ is calculated from the interpolated primitive variables and $\gamma_{i}$.  $\mathbf{P}_{g,i+\frac{1}{2}}^+$ is calculated from the same interpolated primitive variables, but using $\gamma_{i+1}$ rather than $\gamma_i$.

The primitive variables ($Y_{g,i}$, $p_g$, $T_g$, $u_g$, and $v_g$) needed for the gas-phase Riemann solver are interpolated using a low-dissipation method \cite{Houim20118527}.  The density, $\rho_g$, of the interpolated states is computed from the interpolated mass fractions, pressure, and temperature.  The interpolation method uses fifth-order symmetric bandwidth-optimized WENO \cite{martin2006bandwidth} with nonlinear error controls \cite{taylor2007NonlinearWeightedWenoSYMBO} and an adaptive TVD slope limiter to interpolate primitive variables to the cell edges. The low-Mach number velocity adjustment procedure of Thornber \textit{et al.} \cite{Thornber-2007-JCompPhys-MILES} is used to reduce numerical dissipation for the gas-phase.

\subsection{Approximate Solution of the Granular-Phase Riemann Problem}

Calculation of the granular flux has its own set of complications.  One of the most restrictive issues is that the compaction wave speed and solid pressure become zero when the granular temperature is zero or at low volume fractions resulting in hyperbolic degeneracy, as discussed previously.  One approximate Riemann solver used for the particulate phase for a pressureless ($p_{s,\text{tot}}=0$) dust model  is from Collins \textit{et al.} \cite{collins1994DustyGas}:
\begin{equation}
  \label{eqn:pressurelessRiemann}
  \mathbf{F}_{s,i+\frac{1}{2}} = \begin{cases}
               \mathbf{F}_s(\mathbf{U}^{L}_s)   & \text{   if   }  \ u^L_{s} \geq 0 \ \text{and} 
                                                                  \ u_{s}^R > 0 \\
               \mathbf{F}_s(\mathbf{U}^R_s)   & \text{   if   } \ u_{s}^L < 0 \ \text{and} 
                                                                  \ u_{s}^R \leq 0 \\
               \mathbf{F}_s(\mathbf{U}^L_s) +
               \mathbf{F}_s(\mathbf{U}^R_s)   & \text{   if   } \ u_{s}^L > 0 \ \text{and} 
                                                                  \ u_{s}^R < 0 \\
               0                               & \text{   if   } \ u_{s}^L < 0 \ \text{and} 
                                                                  \ u_{s}^R > 0,
                 \end{cases}
\end{equation}
where $\mathbf{U}_s$ is the conserved variable vector for the solid phase and $\mathbf{F}_s$ is the flux vector of the particle phase, excluding solids pressure and the nonconservative terms.  In other locations, where the granular phase is not pressureless, Eqn.~\eqref{eqn:pressurelessRiemann} is incorrect and a more usual Riemann solver will work.  Ideally, however, the same Riemann solver would handle both cases without any \textit{ad hoc} switches, which is the case for the AUSM$^+$-up solver \cite{liou2006AUSM+UP}.  Many traditional Riemann solvers, such as HLLC \cite{harten1983upstream, ToroHLLC}, do not reduce to a form similar to Eqn.~\eqref{eqn:pressurelessRiemann} for a pressureless gas.  

The AUSM$^+$-up flux is
\begin{equation}
  \mathbf{F}_{s,i+\frac{1}{2}} = \mathbf{p}_{\frac{1}{2}} + \dot{m}_{s,i+\frac{1}{2}} \begin{cases}
                           \psi^L & \text{ if   } \dot{m}_{s,i+\frac{1}{2}} \geq 0 \\
                           \psi^R & \text{ if   } \dot{m}_{s,i+\frac{1}{2}}  <   0, \\
                                     \end{cases}
\end{equation}
where $\mathbf{p}_{\frac{1}{2}}$ is a pressure flux vector that is nonzero only for the momentum equation, $\mathbf{p}_{\frac{1}{2}} = (0, 0, p_{\frac{1}{2}},0,0)^T$), and $\psi$ is the vector of the passively advected scalars, $\psi = (1, Y_{s,j}, u_s, E_s, e_s)^T$.  The mass flux at the cell interface, modified slightly to avoid problems with hyperbolic degeneracy and to add dissipation as the packing limit ($\alpha_{s,\text{max}}$) is approached, is defined by
\begin{equation}
  \label{eqn:granAusm}
  \dot{m}_{s,i+\frac{1}{2}} = \mathcal{F} + (c_{\frac{1}{2}} + \epsilon) M_{\frac{1}{2}} \begin{cases}
                         \alpha_s^L \rho_s^L & \text{   if   } M_{\frac{1}{2}} \geq 0 \\
                         \alpha_s^R \rho_s^R & \text{   if   } M_{\frac{1}{2}} < 0.
                                    \end{cases}
\end{equation}
where $\epsilon$ is a small number (10$^{-10}$)  to avoid division by zero when the compaction wave speed is zero and $M_{1/2}$ is the Mach number of the particles based on the compaction wave speed at the cell edge, defined below.  $\mathcal{F}$ is an extra dissipation term developed to stabilize calculations that approach the packing limit where friction pressure and compaction wave speed become extremely sensitive to minute volume fraction fluctuations. It was found through numerical experimentation that $\mathcal{F}$ with a functional form similar to the dissipation term for the Rusanov flux \cite{Toro-book} works well to suppress oscillations as the packing limit is approached,
\begin{equation}
  \label{eqn:granF}
  \mathcal{F} = \frac{c_{\frac{1}{2}} + u_{\frac{1}{2}}\mathcal{G}}{2} \frac{\max(\alpha_s^L, \alpha_s^R)}{\alpha_{s,\text{max}}} \left[ \alpha_s^L \rho_s^L - \alpha_s^R \rho_s^R \right],
\end{equation}
where $\alpha_{s,\text{max}}$ is the packing limit and $\mathcal{G}$ is a dissipation-controlling parameter determined while interpolating the granular primitive variables ($\alpha_s$, $Y_{s,i}$, $\mathbf{v}_s$, $\Theta_s$, $T_s$) at edge $i+1/2$ discussed below.  

The primitive variables are initially reconstructed using fifth-order symmetric bandwidth-optimized WENO \cite{martin2006bandwidth} with nonlinear error controls \cite{taylor2007NonlinearWeightedWenoSYMBO}.  The WENO-interpolated variables are then processed through a TVD slope limiter.  The interpolation for a slope-limited variable $Q$ is
\begin{equation}
  \label{eqn:slopeLimiter}
  Q_{i+\frac{1}{2}}^L = Q_i + 0.5(Q_i - Q_{i-1}) \phi_{\text{TVD}},
\end{equation}
where $\phi_{\text{TVD}}$ is the slope limiter.  The TVD slope limiter used in this work is based on \cite{Houim20118527,kim2005MLP5-II}
\begin{equation}
  \phi_{\text{TVD}} = \max\biggl[0, \min\biggl(\mathcal{G}, \mathcal{G} \frac{Q_{i+1} - Q_{i}}{Q_{i} - Q_{i-1}}, 
                    2 \frac{\hat{Q}_{i+1/2}^L - Q_{i}}{Q_{i} - Q_{i-1}} \biggr) \biggl],
\end{equation}
where $\hat{Q}_{i+1/2}^L$ is the original left-biased interpolated variable using WENO and   
\begin{equation}
   \label{eqn:granG}
   \mathcal{G} = \max[2(1-\mathcal{D} \zeta^2),0],
\end{equation}
where $\mathcal{D} \geq 0$ and is a user defined constant that is set to 0 unless otherwise noted, $\alpha_s^M$ is the maximum solid volume fraction in the entire stencil used by WENO for both left- and right-biased interpolations, and
\begin{equation}
   \zeta = \begin{cases}
     \displaystyle{\frac{\alpha_s^M - \alpha_{s,\text{crit}}}{\alpha_{s,\text{max}} - \alpha_{s,\text{crit}}}} \quad &\text{if} \quad \alpha_s^M > \alpha_{s,\text{crit}} \\
     0     \quad &\text{if} \quad \alpha_s^M < \alpha_{s,\text{crit}}.   
   \end{cases} 
\end{equation}
where $\alpha_{s,\text{crit}}$ and $\alpha_{s,\text{max}}$ are the critical volume fraction and packing limit [see Eqns.~\eqref{eqn:radDist} and \eqref{eqn:pFric}].  The form of $\mathcal{G}$ and $\mathcal{F}$ work well in dense granular regions to suppress oscillations, has relatively little influence on regions with low particle concentration, and does not interfere with the ability of AUSM$^+$-up to capture stationary granular contact surfaces.  

The effect of $\mathcal{D}$ is twofold.  First, it degrades the edge reconstruction scheme for the granular phase to first-order in dense regions.  Second, $\mathcal{D}$  explicitly increases the dissipation of AUSM$^+$-up. The particular form of $\mathcal{G}$ is chosen such that the transition is smooth between low-volume fraction regions (where $\mathcal{G}=2$) and high-volume fraction regions ($\mathcal{G}=0$.)  Increasing $\mathcal{D}$ increases the rate at with $\mathcal{G}$ transitions from 2 to 0 as a function of $\alpha_s$.  A value of $\mathcal{D}=1$ seems to work well in calculations where the solid volume fraction approaches the packing limit.  

 The compaction wave speed at the cell edge for Eqn.~\eqref{eqn:granAusm} is
\begin{equation}
  c_{\frac{1}{2}} = \sqrt{\frac{\alpha_s^L \rho_s^L (c_{s}^L)^2 + \alpha_s^R \rho_s^R (c_{s}^R)^2}{\alpha_s^L \rho_s^L +\alpha_s^R \rho_s^R}},
\end{equation}

 The granular Mach number is
\begin{equation}
  M_{\frac{1}{2}} = \mathcal{M}_4^+(M^L) + \mathcal{M}_4^-(M^R) - 
              2 \frac{K_p}{f_a} \max(1 - \sigma \bar{M}^2,0)
               \frac{p_{s,\text{tot}}^R - p_{s,\text{tot}}^L}{(\alpha_s^L \rho_s^L + \alpha_s^R \rho_s^R) (c_{\frac{1}{2}} + \epsilon)^2},
  \label{eqn:AUSMgranMach}
\end{equation}
where and $f_a = 1$. $K_p$ and $\sigma$ are AUSM dissipation parameters discussed below and
\begin{equation}
 M^L = \frac{u_{s}^L}{c_{\frac{1}{2}}  + \epsilon }, \quad  M^R = \frac{u_{s}^R}{c_{\frac{1}{2}}  + \epsilon }, \quad \bar{M}^2 = \frac{(u_s^L)^2 + (u_s^R)^2}{2 (c_{\frac{1}{2}}  + \epsilon)^2}.
 \label{eqn:AUSM_ML}
\end{equation}

The split pressure at the cell face is:
\begin{equation}
  p_{\frac{1}{2}} = \mathcal{P}_5^+(M^L)p_{s,\text{tot}}^L + \mathcal{P}_5^-(M^R)p_{s,\text{tot}}^R - 
              K_u f_a c_{\frac{1}{2}} \mathcal{P}_5^+(M^L) \mathcal{P}_5^-(M^R)(\alpha_s^L \rho_s^L + \alpha_s^R \rho_s^R)
                                                                        (u_s^R - u_s^L),
\end{equation}
where $K_u$ is another AUSM dissipation parameter.   

The pressure and Mach number splitting polynomials needed to complete the AUSM flux are
\begin{eqnarray}
  \mathcal{M}_1^{\pm}(M) &=& \frac{1}{2} (M \pm \bigl| M \bigr |), \\
  \mathcal{M}_2^{\pm}(M) &=& \pm\frac{1}{4}( M \pm 1)^2,  \\
  \mathcal{M}_4^{\pm}(M) &=& \begin{cases}
                             \mathcal{M}_1^{\pm}(M)  & \text{   if   }\bigl | M \bigr| \geq 1 \\
                             \mathcal{M}_2^{\pm}(M)[ 1 \mp 16 \beta \mathcal{M}_{2}^{\mp}(M)] &
                                        \text{  if   } \bigl |M \bigr | < 1,
                             \end{cases} \\
  \mathcal{P}_5^{\pm}(M) &=& \begin{cases}
                             \frac{\mathcal{M}_1^{\pm}(M)}{M}  & 
                                       \text{   if   }\bigl | M \bigr| \geq 1 \\
                             \mathcal{M}_2^{\pm}(M)[ (\pm 2 -M) \mp 16 \xi
                                                     M \mathcal{M}_{2}^{\mp}(M)] &
                                        \text{  if   } \bigl |M \bigr | < 1,
                             \end{cases}
\end{eqnarray}
where $\beta = 0.125$ and $\xi$ is defined by:
\begin{equation}
  \xi = \frac{3}{16}(-4 + 5 f_a^2).
\end{equation}

The AUSM dissipation parameters are determined from $\mathcal{G}$ by
\begin{align}
   K_p &= 0.25 + 0.75(1 - \mathcal{G}/2) \\
   K_u &= 0.75 + 0.25(1 - \mathcal{G}/2) \\
   \sigma &= 0.75 + 0.25(1 - \mathcal{G}/2)
\end{align}
This modified AUSM$^+$-up scheme for the granular phase reduces to the original if $\mathcal{F}=0$, $\mathcal{G}=2$, and $\epsilon = 0$.  The small number, $\epsilon$, is inserted into Eqns.~\eqref{eqn:granAusm}, \eqref{eqn:AUSMgranMach}, and \eqref{eqn:AUSM_ML}, but not the others to produce a properly upwinded flux when $c_{\frac{1}{2}}=0$ by canceling the $\epsilon$ used to compute $c_{\frac{1}{2}}$ and $M_{\frac{1}{2}}$.   In the case of zero solids pressure, the quantity $(c_{\frac{1}{2}} + \epsilon) M_{\frac{1}{2}}$ used to compute the $\dot{m}_{s,i+\frac{1}{2}}$ reduces to
\begin{eqnarray*}
   (c_{\frac{1}{2}} + \epsilon) M_{\frac{1}{2}} &=& \frac{\epsilon}{2}\left[ \frac{ u_s^L + |u_s^L| }{\epsilon} + \frac{ u_s^R - |u_s^R| }{\epsilon}\right] \\
                               &=& \begin{cases}
                                              {u}^{L}_s  & \text{   if   }  \ u^L_{s} \geq 0 \ \text{and} 
                                                                                                 \ u_{s}^R > 0 \\
                                              {u}^R_s   & \text{   if   } \ u_{s}^L < 0 \ \text{and} 
                                                                                                 \ u_{s}^R \leq 0 \\
                                              {u}^L_s +
                                              {u}^R_s   & \text{   if   } \ u_{s}^L > 0 \ \text{and} 
                                                                                                 \ u_{s}^R < 0 \\
                                              0                               & \text{   if   } \ u_{s}^L < 0 \ \text{and} 
                                                                                                 \ u_{s}^R > 0,
                                                \end{cases}
\end{eqnarray*}
which is similar to the pressureless Riemann solver of Collins \textit{et al.} \cite{collins1994DustyGas} and indicates that the modification to AUSM provides a correct upwinded flux even when $c_{\frac{1}{2}}=0$.

\begin{figure}
\centering
\includegraphics[width=0.7\textwidth]{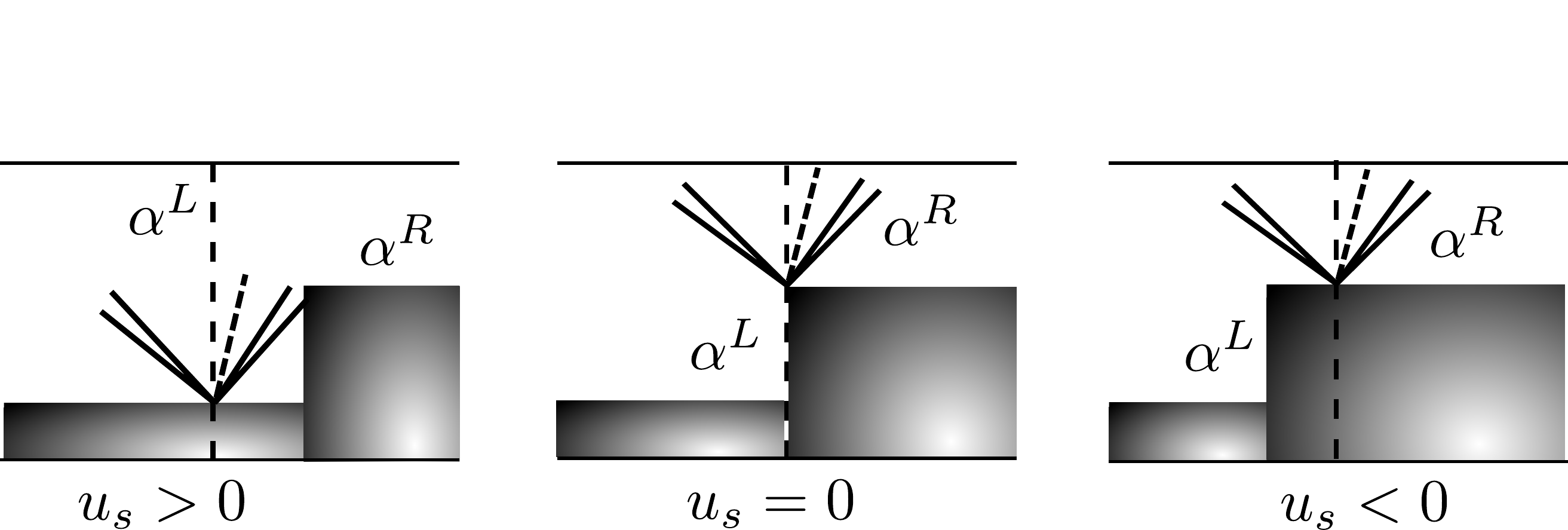}
\caption{Schematic of the gas-phase volume fraction at a cell face at $t=0^+$.}
\label{fig:Hyp_Blockage_2}
\end{figure}

In addition to the granular-phase fluxes, AUSM$^+$-up is also modified to return $u_{s,i+\frac{1}{2}}$ needed for $p_s \nabla \! \cdot \! \mathbf{v}_s$,  the quantity $ \alpha_{s,i+\frac{1}{2}} u_{s,i+\frac{1}{2}}$ for the $pDV$ work term ($p_g \! \nabla \! \cdot \! \alpha_s \mathbf{v}_s$),  and $\alpha_{g,i+\frac{1}{2}}$ needed to complete the gas-phase fluxes. 
Computation of the gas-phase edge volume fraction is based on where the particles have moved at $t=0^+$, as shown in Fig.~\ref{fig:Hyp_Blockage_2},
and is computed by
\begin{equation}
   \alpha_{g,i+1/2} = \begin{cases}
         \alpha_{g,i+\frac{1}{2}}^L & \mbox{if } \dot{m}_{s,i+\frac{1}{2}} > 0 \\
         \alpha_{g,i+\frac{1}{2}}^R & \mbox{if } \dot{m}_{s,i+\frac{1}{2}} \leq 0.
   \end{cases}
   \label{eqn:edgeVolFrac}
\end{equation}
Similarly,
\begin{equation}
   \alpha_{s,i+\frac{1}{2}}u_{s,i+\frac{1}{2}} = \begin{cases}
         \dot{m}_{s,i+\frac{1}{2}}/\rho_{s,i+\frac{1}{2}}^L & \mbox{if } \dot{m}_{s,i+\frac{1}{2}} > 0 \\
         \dot{m}_{s,i+\frac{1}{2}}/\rho_{s,i+\frac{1}{2}}^R & \mbox{if } \dot{m}_{s,i+\frac{1}{2}} \leq 0
   \end{cases},
   \quad
    u_{s,i+\frac{1}{2}} = \begin{cases}
        \alpha_{s,i+\frac{1}{2}}u_{s,i+\frac{1}{2}}/\alpha_{s,i+\frac{1}{2}}^L & \mbox{if } \dot{m}_{s,i+\frac{1}{2}} > 0 \\
        \alpha_{s,i+\frac{1}{2}}u_{s,i+\frac{1}{2}}/\alpha_{s,i+\frac{1}{2}}^R & \mbox{if } \dot{m}_{s,i+\frac{1}{2}} \leq 0. 
   \end{cases}
\end{equation}

\subsection{Solution Algorithm for the Hyperbolic Terms}

The procedure to integrate the convective terms of the conserved variables from time level $n$ to $n+1$ for a multistage explicit integration method using the double-flux model for multicomponent mixtures is:
\begin{enumerate}
  \item Calculate and store $\gamma$ and $\alpha_g \rho_g h_0^m$ at each cell.
  \item Compute the solutions to the gas and granular Riemann problems at each cell edge for each direction.
        \begin{enumerate}
        \item Interpolate the granular phase primitive variables ($\alpha_s$, $Y_{s,j}$, $u_s$, $\Theta_s$, and $T_s$) from the left and right.
            \begin{enumerate}
               \item Compute $\zeta$ from Eqn. (65).
               \item Compute $\mathcal{G}$ from Eqn. (64).
               \item Interpolate the primitive variables.  We use the six-point bandwidth-optimized WENO method 
                  \cite{martin2006bandwidth,taylor2007NonlinearWeightedWenoSYMBO}.
               \item Apply the slope limiter to the WENO-interpolated granular primitive variables using Eqns. (62) and (63).
            \end{enumerate}
        \item Interpolate the gas-phase primitive variables ($Y_{g,j}$, $p_g$, $T_g$, and $u_g$) to the cell face from the left and right.  In this work we use the scheme given in \cite{Houim20118527}.  
        \item Solve the granular-phase Riemann problem using the modified AUSM$^+$-up scheme detailed in Sec. 3.2.  Store the $u_{s,i+\frac{1}{2}}$, $\alpha_{s,i+\frac{1}{2}} u_{s,i+\frac{1}{2}}$, and $\alpha_{g,i+\frac{1}{2}}$ in addition to the granular fluxes.
        \item Solve the gas-phase Riemann problem at each cell edge twice using the HLLC method detailed in Sec. 3.1 to get $\mathbf{P}_{g,i+\frac{1}{2}}^+$ and $\mathbf{P}_{g,i+\frac{1}{2}}^-$.  Rotate the HLLC solver near shocks using the method outlined in \cite{Houim20118527} to avoid shock anomalies. 
        \end{enumerate}
   \item Compute the lift force ($\mathbf{f}_{\text{lift}}$) from Eqn.~\eqref{Eqn:fLift} at each cell center using second order finite-difference if $\alpha_s > \alpha_{s,\text{min}}$ for all points in the stencil where $\alpha_{s,\text{min}}=10^{-10}$.   
  \item Assemble the right-hand-side to discretize the convective terms, Eqns.~(39)-(46).   
  \item Update the conserved variables, $\mathbf{U}_g$ and $\mathbf{U}_s$, using the fully-assembled right-hand-side of the hyperbolic operator and the chosen time-marching method.  We use the third-order strong-stability-preserving Runge-Kutta \cite{spiteri2003new} scheme.
  \item Update the gas-phase variables in a manner consistent with the double flux model via:
       \begin{equation} 
       \begin{split}
          Y_{g,j} &= \frac{\alpha_g \rho_g Y_{g,j}}{\sum \alpha_g \rho_g Y_{g,j}}, \quad  u_g = \frac{\alpha_g \rho_g u_g}{\sum \alpha_g \rho_g Y_{g,j}}, \quad  T_g   = \frac{p_g}{R_u \sum \frac{\rho_g Y_{g,j}}{M_j}}, \\
          p_g &= \frac{\gamma^n -1}{\alpha_g} \biggl[\alpha_g \rho_g E_g - (\alpha_g \rho_g h_0^m)^n - \alpha_g \rho_g \frac{\mathbf{v}_g \! \cdot \! \mathbf{v}_g}{2} \biggr]. 
       \end{split}
       \end{equation}
  \item Repeat steps 2-6 for each stage of the time-marching algorithm.  
  \item Use the granular-phase conserved variables, gas-phase primitive variables, gaseous species densities ($ \alpha_g \rho_g Y_{g,j}$), and momenta calculated from the final stage of the time-stepping algorithm as values for the next time step, $n\!+\!1$, and use them to calculate $\gamma^{n+1}$ and $(\alpha_g \rho_g h_0^m)^{n+1}$.  
  \item Perform the final step of the double-flux method to update the total gas-phase energy for time step $n+1$,
      \begin{equation}
        (\alpha_g \rho_g E_g)^{n+1} = \frac{\alpha_g^{n+1} p_g^{n+1}}{\gamma^{n+1} - 1} + (\alpha_g \rho_g h_0^m)^{n+1} + \alpha_g^{n+1} \rho_g^{n+1} \frac{\mathbf{v}_g^{n+1} \! \cdot \! \mathbf{v}_g^{n+1}}{2}.
       \end{equation}
  \item Check for computational cells with very low particle volume fractions.  If $\alpha_s < \alpha_{s,\text{min}}$, where $\alpha_{s,\text{min}} = 10^{-10}$, remove the granular phase and scale the gas-phase conserved variables to account for gas-phase volume gained by removal of the particulate phase:   $\mathbf{U}_g^{n+1} = \mathbf{U}_g^{n+1}/(1 - \alpha_s^{n+1})$ and $\mathbf{U}_s^{n+1} = 0$ if $\alpha_s^{n+1} < \alpha_{s,\text{min}}$.
\end{enumerate}

If the double-flux model is not used then steps 1 and 9 are unnecessary, step 2(d) would be completed using a single solution to the Riemann problem, and step 6 would be completed using the actual equation of state for the gas.

\section{Solution of the Parabolic Terms, $\mathcal{P}_{xy}^{\Delta t}$}

The parabolic portions of the governing equations are:
\begin{equation}
 \frac{\partial \alpha_g \rho_g Y_{g,j}}{\partial t} = - \nabla \! \cdot \! \alpha_g \rho_g Y_{g,i}\mathbf{V}_{g,j}^d
\end{equation}
\begin{equation}
 \frac{\partial \alpha_g \rho_g \mathbf{v}_g}{\partial t}  =  \nabla \! \cdot \! (\alpha_g \sigma_g) 
\end{equation}
\begin{equation}
 \frac{\partial \alpha_g \rho_g E_g}{\partial t} = - \nabla \! \cdot \! (\alpha_g \mathbf{q}_g) 
\end{equation}
\begin{equation}
 \frac{\partial \alpha_s \rho_s \mathbf{v}_s}{\partial t} = \nabla \! \cdot \! (\alpha_s \sigma_s)
\end{equation}
\begin{equation}
   \label{eqn:EsPara}
\frac{\partial \alpha_s \rho_s E_s}{\partial t}= \nabla \! \cdot \! (\alpha_s \lambda_s \nabla \Theta_s) + \nabla \! \cdot \! (\alpha_s \mathbf{v}_s \! \cdot \! \sigma_s) - \mathbf{v}_s \! \cdot \! \nabla \! \cdot \! (\alpha_s \sigma_s)
\end{equation}
where the deviatoric stress tensor, $\sigma$, is defined as:
\begin{equation}
   \label{eqn:StressTensor}
   \sigma = \mu \left( \nabla \mathbf{v} + \nabla \mathbf{v}^T \right) + \left( \kappa - \frac{2}{3} \mu \right) \nabla \! \cdot \! \mathbf{v} \mathbf{I},
\end{equation}
and $\mu$ and $\kappa$ are the shear and bulk viscosities. The granular viscous dissipation term in Eqn.~\eqref{eqn:PTE}, $\alpha_s \sigma_s \! \! : \! \! \nabla \mathbf{v}_s$, is rewritten as $\nabla \! \cdot \! (\alpha_s \mathbf{v}_s \! \cdot \! \sigma_s) - \mathbf{v}_s \! \cdot \! \nabla \! \cdot \! (\alpha_s \sigma_s)$ in Eqn.~\eqref{eqn:EsPara}. This ensures that calculation of the particle deviatoric stresses and viscous dissipation are consistent at computational cell edges.

The model for the gas-phase transport coefficients, diffusion velocity, $\mathbf{V}_{g,j}^d$, and heat diffusion vector, $\mathbf{q}_g$, used in this work can be found in  \cite{Houim20118527}.

\subsection{Granular-Phase Transport Coefficients}

The total granular deviatoric stress and diffusion of $E_s$ are written as $\alpha_s \sigma_s$ and $\alpha_s \lambda_s \nabla \Theta$, respectively, rather than using their original definitions without the linear factor of $\alpha_s$.  Thus, the original expressions for the granular transport coefficient are divided by a linear factor of $\alpha_s$. (The reason for doing this is discussed later.)   The granular shear viscosity is divided into collisional, kinetic, and frictional components \cite{Lun1984},
\begin{equation}
   \mu_s = \rho_s d_s \sqrt{\Theta_s}[f_{\text{coll}}(\alpha_s) + f_{\text{kin}}(\alpha_s)] + \mu_{\text{fric}}
\end{equation}
where $d_s$ is the particle diameters, $e$ is the coefficient of restitution,
\begin{equation}
 f_{\text{coll}}(\alpha_s)= \frac{4}{5 \sqrt{\pi}} \alpha_s g_0 (1 + e),
\end{equation}
\begin{equation}
  f_{\text{kin}}(\alpha_s) = \frac{\sqrt{\pi}}{6 ( 3 - e)}\left[1 + \frac{2}{5}(1+e)(3 e - 1)\alpha_s g_0 \right],
\end{equation}
and $\mu_{\text{fric}}$ is a simplified expression for the frictional viscosity,
\begin{equation}
 \mu_{\text{fric}} = \frac{p_{\text{fric}}}{\alpha_s + \epsilon} \sin{\psi},
\end{equation}
where $\psi$ is the angle of internal friction that is set to $\pi/6$ unless specified otherwise and $\epsilon$ is a small number to avoid division by zero.  Other formulations for $\mu_{\text{fric}}$ could be used as well \cite{schneiderbauer2012comprehensive}. 
The granular bulk viscosity is \cite{Lun1984}
\begin{equation}
 \kappa_s = \rho_s d_s \sqrt{\Theta_s} f_{\kappa}(\alpha_s)
\end{equation}
where
\begin{equation}
  f_{\kappa}(\alpha_s) = \frac{4}{3 \sqrt{\pi}}g_0 (1+e)
\end{equation}
and 
\begin{equation}
\eta = \frac{1 + e}{2}
\end{equation}

The granular thermal conductivity is \cite{Lun1984}
\begin{equation}
   \lambda_s = \rho_s d_s \sqrt{\Theta_s} f_{\lambda}
\end{equation}
where
\begin{equation}
\frac{15 \sqrt{\pi}}{4(41-33 \eta)} \left[1 + \frac{12}{5} \eta^2 (4 \eta - 3) \alpha_s g_0 + \frac{16}{15 \pi} (41 - 33 \eta) \eta \alpha_s g_0 \right]
\end{equation}
The values of the granular transport coefficients are limited to 100 $\text{kg}/\text{m} \, \text{s}$ for the viscosities and 100 $\text{kg}/\text{m}^3 \text{s}$ for the thermal conductivity to prevent excessively small time step sizes when the packing limit is approached.

\subsection{Discretization of the Parabolic Terms}
\begin{figure}
  \centering
  \subfloat[]{\includegraphics[width=0.25\textwidth]{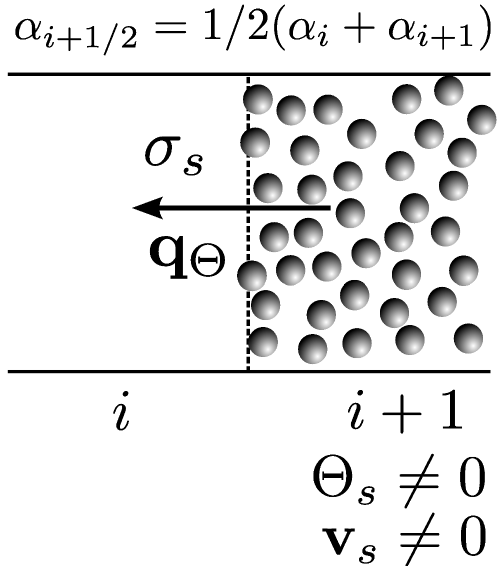}} \quad \quad \quad
  \subfloat[]{\includegraphics[width=0.25\textwidth]{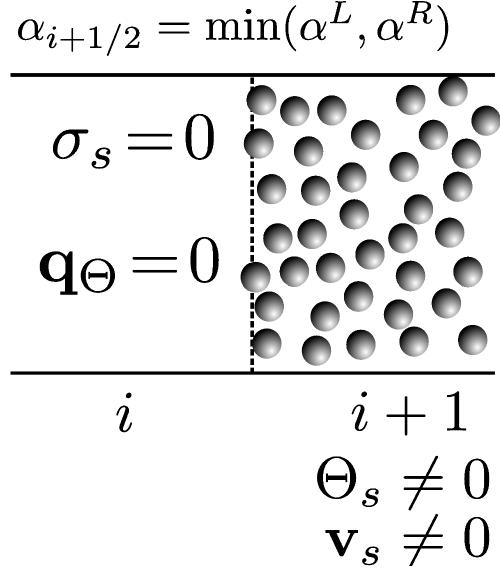}}
\caption{Illustration of unphysical granular pseudo-thermal energy diffusion flux ($\mathbf{q}_{\Theta} = - \alpha_s \lambda_s \! \nabla \Theta_s$) and viscous stress computed (a) with $\alpha_{s,i+\frac{1}{2}} = 1/2(\alpha_{s,i} + \alpha_{s,i+1})$ and (b) a fix to the problem by defining $\alpha_{s,i+\frac{1}{2}} = \min(\alpha^L, \alpha^R)$.}
\label{fig:GranularParabolicProblem} 
\end{figure}

Discretization of the parabolic terms is more straight forward than the hyperbolic terms.  There is, however, a question about which volume fraction to use at a cell edge, as shown in Fig.~\ref{fig:GranularParabolicProblem}. 
If a simple average or a centered interpolation is used to estimate the volume fractions at the cell faces, the result may be unphysical. For example, if there is a sharp granular interface with no particles in one cell and many particles in the neighboring cell, an average may result in a finite granular diffusion flux into the cell with no particles.
This behavior is unphysical. There are not any particles in the neighboring cell and, by definition, there cannot be any granular viscous stress or diffusion of $E_s$ at that interface.  

\begin{figure} 
  \centering
  \subfloat[]{\includegraphics[width=0.25\textwidth]{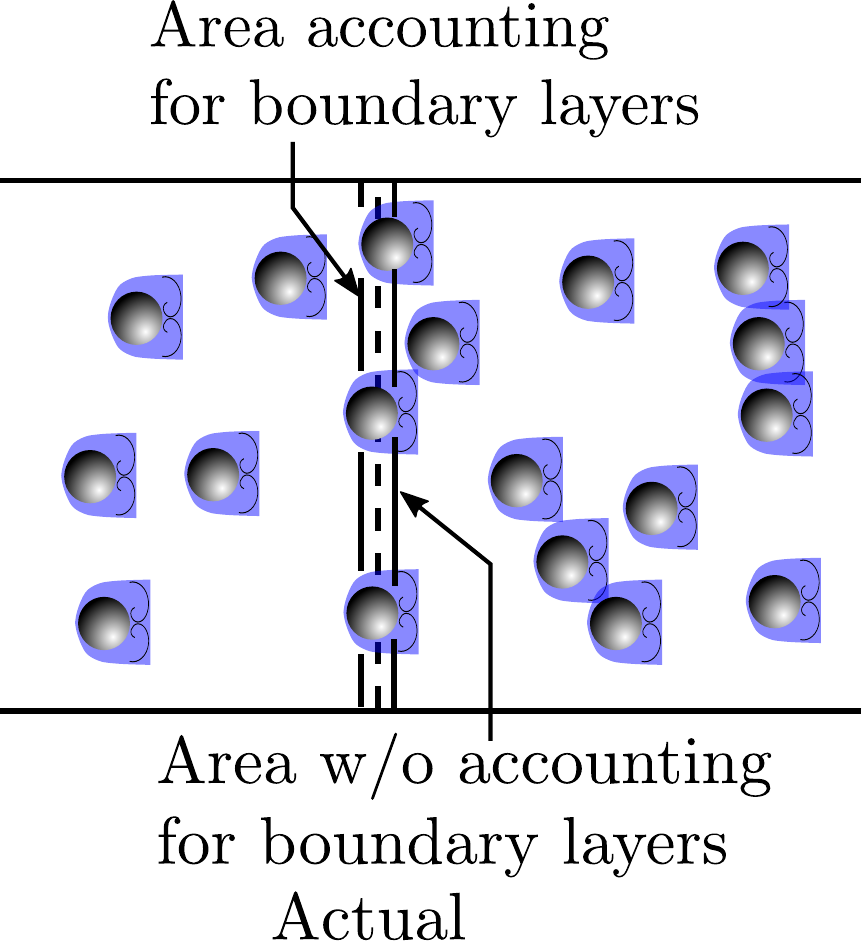}} \quad \quad
  \subfloat[]{\includegraphics[width=0.26\textwidth]{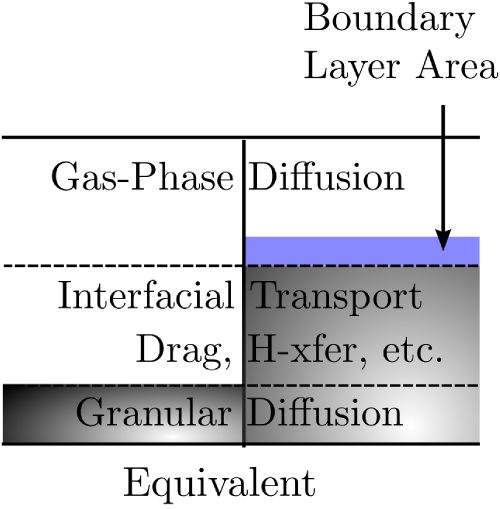}} \quad \quad
  \subfloat[]{\includegraphics[width=0.25\textwidth]{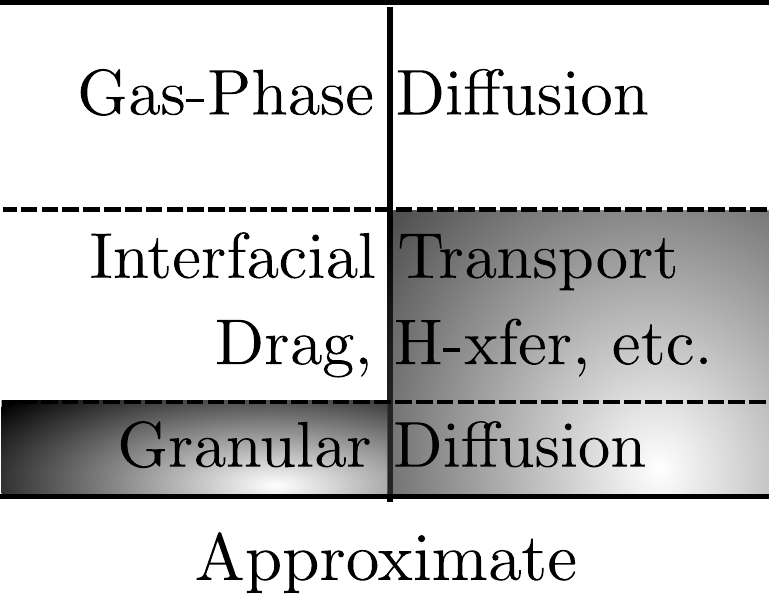}} 
\caption{Illustration of granular material interfaces for the parabolic terms: (a) actual granular system with small boundary layers on the individual particles, (b) equivalent system with all the particles compressed, and (c) approximate system neglecting the boundary layer effect on the particles.}
\label{fig:ParabolicCollapsedParticles} 
\end{figure}

For example, consider a sharp granular interface between cells $i$ and $i+1$ with volume fractions $\alpha_{s,i+\frac{1}{2}}^L$ and $\alpha_{s,i+\frac{1}{2}}^R$ on either side [see Fig.~\ref{fig:ParabolicCollapsedParticles}(c)].  A stability analysis on Eqn.~\eqref{eqn:EsPara} shows that the time step is
\begin{equation}
  \Delta t \leq \Delta x^2 \frac{3}{4}  \frac{\rho_{s,i+\frac{1}{2}}}{\lambda_{s,i+\frac{1}{2}}} \frac{\alpha_{s,i+\frac{1}{2}}^L}{\alpha_{s,i+\frac{1}{2}}}.
\end{equation}
If the volume fraction in cell $i$ is very dilute ($\alpha_{s,i+\frac{1}{2}}^L = 10^{-6}$), there are many particles in cell $i+1$ ($\alpha_{s,i+\frac{1}{2}}^R = 0.4$), and we define $\alpha_{s,i+\frac{1}{2}} = (\alpha_{s,i+\frac{1}{2}}^L + \alpha_{s,i+\frac{1}{2}}^R)/2$, the time-step restriction is on the order of $\Delta t = 1$ ps for a grid size of $\Delta x = 200$ $\mu$m.  This is unrealistic as particle collisions simply do not occur on that time scale. Worse yet, if there are no particles to the left of interface, the time-step size is zero.  The result is physically correct if the volume fraction at the cell face is defined using the minimum value of $\alpha$ on either side. If we define $\alpha_{s,i+\frac{1}{2}} = \min(\alpha_{s,i+\frac{1}{2}}^L, \alpha_{s,i+\frac{1}{2}}^R)$, the time-step size is $\Delta t = 1$ $\mu$s, which is much more reasonable.    

Given the above arguments we define,
\begin{equation}
   \label{eqn:alphaFace}
   \alpha_{s,i+\frac{1}{2}}=\min(\alpha_{s,i+\frac{1}{2}}^L, \alpha_{s,i+\frac{1}{2}}^R), \ \ \ \  \alpha_{g,i+\frac{1}{2}}=\min(\alpha_{g,i+\frac{1}{2}}^L, \alpha_{g,i+\frac{1}{2}}^R)
\end{equation}
for the factor of $\alpha_{s,i+\frac{1}{2}}$ and $\alpha_{g,i+\frac{1}{2}}$ multiplying diffusion fluxes, conduction, and viscous terms for the gas and granular phases.  First-order interpolation is used for $\alpha_{s,g,i+1/2}^{L,R}$ and a simple arithmetic average for the transport coefficients themselves is used to estimate their edge values.  A linear dependence of $\alpha_s$ was factored out of the granular transport coefficients to make this step more convenient.  Equation~\eqref{eqn:alphaFace} for $\alpha_{s,i+\frac{1}{2}}$ is justified physically, as shown in Fig.~\ref{fig:ParabolicCollapsedParticles}.

Diffusion processes only occur with molecular collisions in the gas-gas section or particle collisions in the solid-solid section.  Neglecting boundary layer effects, only the gas-gas section  [$\min(\alpha_{g,,i+\frac{1}{2}}^L, \alpha_{g,i+\frac{1}{2}}^R)$] is available for gas-phase diffusion and the solid-solid section [$\min(\alpha_{s,i+\frac{1}{2}}^L, \alpha_{s,i+\frac{1}{2}}^R)$] is available for granular diffusion in Fig.~\ref{fig:ParabolicCollapsedParticles}(c).  The gas-solid section represents effects from interphase drag, heat transfer, mass transfer, etc., which are already taken into account via inhomogeneous source terms. Thus, there is no need to model interphase diffusion processes in the gas-solid section.

After the volume fraction for the gas and granular phases is found at the cell edges, the diffusion fluxes and viscous stresses are computed using a second-order accurate conservative approach.  The gradient of generic variable $Q$ normal to a cell face is
\begin{equation}
  \frac{d Q}{dx} \biggr|_{i+1/2} \approx \frac{Q_{i+1} - Q_{i}}{\Delta x},
\end{equation}
and gradients tangential to a cell face are approximated using
\begin{equation}
  \frac{d Q}{dy} \biggr|_{i+1/2,j} \approx \frac{Q_{i+1,j+1} + Q_{i,j+1} - Q_{i+1,j-1} - Q_{i,j-1}}{4 \Delta y}.
\end{equation}

An explicit Runge-Kutta-Chebyshev (RKC) scheme \cite{verwer2004rkc} is used to advance the parabolic-split equations in time, which allows an arbitrary number of stages to increase stability.  Details of applying RKC to compressible reacting flow and how it was used in this work can be found in \cite{Houim20118527}, with the exception that the number of RKC stages was forced to be an odd number.  RKC with an even number of stages has recently been found to produce instabilities \cite{meyer2012second,meyer2014stabilized}.

\subsection{Solution Algorithm for the Parabolic Terms}
The procedure for integrating the parabolic terms of the conserved variables from time level $n$ to $n+1$ for using RKC is:
\begin{enumerate}
\item Estimate the number of RKC stages \cite{Houim20118527} and force the number to be odd.
\item Compute the primitive variables at each grid point.
\item Compute the transport properties at each grid point. 
\item At each cell face compute the diffusion fluxes and viscous stresses:
   \begin{enumerate}
    \item Determine $\alpha_{g,i+\frac{1}{2}}$ and $\alpha_{s,i+\frac{1}{2}}$ at the cell faces using Eqn.~\ref{eqn:alphaFace} with first-order interpolation.
    \item Average the transport properties to the cell face via: $\xi_{i+\frac{1}{2}} = 1/2(\xi_{i} + \xi_{i+1})$.
    \item Form all of the gradients needed at the cell face.
    \item Calculate the viscous stresses and diffusion fluxes.
   \end{enumerate}
\item Assemble the right-hand-side of the parabolic terms.
\item Update the conserved variables based on the RKC algorithm.
\item Repeat steps 2-6 for each RKC stage.  
\end{enumerate}

\section{Solution of the Source Terms, $\mathcal{S}^{2 \Delta t}$}
Neglecting phase change and chemical reaction, the remaining nonzero inhomogeneous source terms are:
\begin{equation}
 \frac{d \alpha_g \rho_g \mathbf{v}_g}{d t}  = -  \mathbf{f}_{\text{Drag}}  + \alpha_g \rho_g \mathbf{g}
\end{equation}
\begin{equation}
 \frac{d \alpha_g \rho_g E_g}{d t} = - q_{\text{conv}} -\mathbf{f}_{\text{Drag}} \! \cdot \! \mathbf{v}_s + \phi_{\text{visc}}   - \phi_{\text{slip}} + \alpha_g \rho_g \mathbf{g} \! \cdot \! \mathbf{v}_g
\end{equation}
\begin{equation}
 \frac{d \alpha_s \rho_s \mathbf{v}_s}{d t} =  \mathbf{f}_{\text{Drag}}   + \alpha_s \rho_s \mathbf{g}
\end{equation}
\begin{equation}
\frac{d \alpha_s \rho_s E_s}{d t} = -\dot{\gamma}  - \phi_{\text{visc}} + \phi_{\text{slip}} 
\end{equation}
\begin{equation}
\frac{d \alpha_s \rho_s e_s }{d t}  = q_{\text{conv}}  + \dot{\gamma}
\end{equation}
For each equation, the source terms are split into several sub steps due to the wide variety of time scales between convection, drag, granular cooling ($\dot{\gamma}$), etc.  A Strang-splitting method is used
\begin{equation}
  \mathcal{S}^{2\Delta t} = \mathcal{S}_{qD}^{\Delta t} \ 
                           \mathcal{S}_{\Theta}^{\Delta t} \
                           \mathcal{S}_{\Theta}^{\Delta t} \
                           \mathcal{S}_{qD}^{\Delta t} 
\end{equation}
where $\mathcal{S}_{qD}^{\Delta t}$ is the advancement of drag, convective heat transfer, and gravity ($\mathbf{g}$), $\mathcal{S}_{\Theta}^{\Delta t}$ is the advancement of  $\phi_{\text{slip}}$, $\phi_{\text{visc}}$, and $\dot{\gamma}$.  

\subsection{Model and Solution for Drag and Heat Transfer, $\mathcal{S}_{qD}^{\Delta t}$}

The equations considering only drag and heat transfer are
\begin{equation*}
 \frac{d \alpha_g \rho_g \mathbf{v}_g}{d t}  = -  \mathbf{f}_{\text{Drag}} + \alpha_g \rho_g \mathbf{g}, \quad 
 \frac{d \alpha_s \rho_s \mathbf{v}_s}{d t} =  \mathbf{f}_{\text{Drag}} + \alpha_s \rho_s \mathbf{g},
 \end{equation*}
 \begin{equation*}
 \frac{d \alpha_g \rho_g E_g}{d t} = - q_{\text{conv}} -\mathbf{f}_{\text{Drag}} \! \cdot \! \mathbf{v}_g +  \alpha_g \rho_g \mathbf{g} \! \cdot \! \mathbf{v}_g, \quad
\frac{d \alpha_s \rho_s e_s }{d t}  = q_{\text{conv}}.
\end{equation*}
The drag force in this work is given by the Gidaspow correlation \cite{GidaspowBook}, which is valid for particle volume factions ranging from dilute to the packing limit,
\begin{equation}
 \mathbf{f}_{\text{drag}} = K_{sg} (\mathbf{v}_g - \mathbf{v}_s),
\end{equation}
where
\begin{equation}
K_{sg} = \begin{cases}
   0.75 C_d \displaystyle{\frac{\rho_g \alpha_g \alpha_s | \mathbf{v}_g - \mathbf{v}_s |}{d_s \alpha_g^{2.65}}} &\mbox{if } \alpha_g \geq 0.8 \\
   \displaystyle{ 150 \frac{\alpha_s^2 \mu_g}{\alpha_g d_s^2} - 1.75 \frac{\rho_g \alpha_s | \mathbf{v}_g - \mathbf{v}_s |}{d_s}} &\mbox{if } \alpha_g < 0.8.
   \end{cases}
\end{equation}
The drag coefficient over a single sphere, $C_d$, is
\begin{equation}
   C_d = \begin{cases} 
      \displaystyle 24 ( \alpha_g Re)^{-1}[1 + 0.15 (\alpha_g Re)^{0.687}] &\mbox{if } \alpha_g Re < 1000 \\
      0.44 &\mbox{if } \alpha_g Re \geq 1000,
   \end{cases}
\end{equation}
and the Reynolds number is defined by
\begin{equation}
  Re = \displaystyle{\frac{\rho_g | \mathbf{v}_g - \mathbf{v}_s | d_s}{\mu_g}}.
\end{equation}
Other forms of the drag coefficient are applicable \cite{van2001comparative,MFIX-Theory}.

The convective heat flux between the gas and particles is
\begin{equation}
   q_{\text{conv}} = h_{sg}(T_g - T_s),
\end{equation}
where the heat transfer coefficient, $h_{sg}$ between the gas and solid phases is
\begin{equation}
   h_{sg} = 6 \frac{\alpha_s \lambda_g N\!u }{d_s^2},
\end{equation}
the Nusselt number correlation of Gunn \cite{Gunn1978}, valid from $0 < \alpha_s \leq 0.65$ and $Re < 10^5$, is used
\begin{equation}
   N\!u = (7-10 \alpha_s + 5 \alpha_s^2)(1 + 0.7 Re^{0.2}Pr_g^{1/3}) + (1.33-2.4 \alpha_s + 1.2 \alpha_s^2)Re^{0.7}Pr_g^{1/3}
\end{equation}
and $Pr_g$ is the gas-phase Prandtl number.

If the drag and heat transfer coefficients and specific heats are evaluated using the initial conditions (parameters denoted with a superscript $0$) and frozen during integration, the drag and heat transfer terms can be computed analytically \cite{PelantiDustyGas2006}.  Then the change in momentum from drag, $\Delta \mathbf{M}$, is
\begin{equation}
   \Delta \mathbf{M} = \frac{\mathbf{v}_g^0 - \mathbf{v}_s^0}{\xi_D}\left[ \frac{1}{K_{sg} \xi_D \Delta t + 1} - 1 \right],
\end{equation}
where 
\begin{equation}
   \xi_D = \frac{1}{\alpha_g \rho_g} + \frac{1}{\alpha_s \rho_s}.
\end{equation}
The change in internal energy from convection, $\Delta e$, is
\begin{equation}
   \Delta e = \frac{T_g^0 - T_s^0}{\xi_e}\left[ e^{-h_{sg} \xi_e \Delta t} - 1 \right],
\end{equation}
where
\begin{equation}
   \xi_e = \frac{1}{\alpha_g \rho_g C_{V,g}^0} + \frac{1}{\alpha_s \rho_s C_{V,s}^0}.
\end{equation}

Then the momentum and energies at $t=\Delta t$ are
\begin{eqnarray}
  (\alpha_g \rho_g \mathbf{v}_g)^{\Delta t} &=& (\alpha_g \rho_g \mathbf{v}_g)^{0} + \Delta \mathbf{M} + \alpha_g \rho_g \mathbf{g} \Delta t \\
  (\alpha_s \rho_s \mathbf{v}_s)^{\Delta t} &=& (\alpha_s \rho_s \mathbf{v}_s)^{0} - \Delta \mathbf{M} + \alpha_s \rho_s \mathbf{g} \Delta t\\
    (\alpha_g \rho_g E_g)^{\Delta t} &=& (\alpha_s \rho_s E_g)^0 + \Delta e + \frac{\alpha_g \rho_g}{2} \left[ \mathbf{v}_g^{\Delta t} \! \cdot \mathbf{v}_g^{\Delta t} - \mathbf{v}_g^{0} \! \cdot \mathbf{v}_g^{0}  \right] \\
  (\alpha_s \rho_s e_s)^{\Delta t} &=& (\alpha_s \rho_s e_s)^0 - \Delta e.
\end{eqnarray}

\subsection{Model and Solution for Pseudo-Thermal Energy Production and Dissipation, $\mathcal{S}_{\Theta}^{\Delta t}$}

The equations for  source terms considering only sources and sinks of pseudo-thermal energy ($E_s$) are
\begin{equation}
 \frac{d \alpha_g \rho_g E_g}{d t} =  \phi_{\text{visc}}   - \phi_{\text{slip}} 
\end{equation}
\begin{equation}
\frac{d \alpha_s \rho_s E_s}{d t} = -\dot{\gamma}  - \phi_{\text{visc}} + \phi_{\text{slip}}
\end{equation}
\begin{equation}
\frac{d \alpha_s \rho_s e_s }{d t}  = \dot{\gamma}.
\end{equation}

The model for viscous damping of $E_s$, $\phi_{\text{visc}}$, is adopted from Gidaspow \cite{GidaspowBook}
\begin{equation}
  \phi_{\text{visc}} = 3 K_{sg} \Theta_s.
\end{equation}
Alternative models for $\phi_{\text{slip}}$ could also be used \cite{Koch1999}.

Production of pseudo-thermal energy due to velocity slip between the gas and solid phases, $\phi_{\text{slip}}$, is given by Koch and Sangani \cite{Koch1999} with their correction factor fixed at unity,
\begin{equation}
   \phi_{\text{slip}} = f_{\text{slip}} \frac{| \mathbf{v}_g - \mathbf{v}_s |^2}{\sqrt{\Theta_s}},
\end{equation}
where
\begin{equation}
 f_{\text{slip}} = \frac{81 \alpha_s \mu_g^2}{g_{0} d_s^3 \rho_s \sqrt{\pi}}.
\end{equation}
The granular dissipation term, $\dot{\gamma}$, converts $E_s$ into $e_s$ from inelastic collisions between particles.  The granular dissipation model adopted for this study \cite{Lun1984} is a variant of Haff's cooling law \cite{Haff1983,GranularGasBook}
\begin{equation}
   \label{eqn:granGamma}
   \dot{\gamma} = \displaystyle{\frac{\rho_s \Theta_s^{3/2}}{d_s} f_{\gamma}},
\end{equation}
where
\begin{equation} 
   f_{\gamma} = \frac{12(1 - e^2)g_0 \alpha_s^2}{\sqrt{\pi}  }.
\end{equation}

A predictor-corrector approach is used to integrate these terms independently so that analytic solutions can be used.  First the dissipation of $E_s$ from viscous damping is computed
\begin{equation}
   \Theta_s^* = \Theta_s^0 \exp \left[\frac{-2 K_{sg}\Delta t}{\alpha_s \rho_s} \right].
\end{equation}
Next the production of PTE from velocity slip is integrated
\begin{equation}
   \Theta_s^{**} = \left[\frac{\xi_{\text{slip}}}{\alpha_s \rho_s} \Delta t + \frac{3}{2} \Theta_s^* \right]^{2/3},
\end{equation}
where 
\begin{equation}
   \label{eqn:xiSlip}
   \xi_{\text{slip}} = f_{\text{slip}} \frac{| \mathbf{v}_g^0 - \mathbf{v}_s^0 |^2}{K_{sg} \xi_D \Delta t + 1}.
\end{equation}
The factor $1/(K_{sg} \xi_D \Delta t + 1)$ in Eqn.~\eqref{eqn:xiSlip} results from averaging $| \mathbf{v}_g^0 - \mathbf{v}_s^0 |^2$ during particle acceleration from drag.   Finally, dissipation of $E_s$ due to inelastic granular collisions is integrated
\begin{equation}
  \Theta_s^{\Delta t} = \Theta_s^{**} \frac{9 (\alpha_s \rho_s)^2}{3 \alpha_s \rho_s + \Delta t f_{\gamma} \sqrt{\Theta_s^{**}}}.
\end{equation}
With the final and intermediate granular temperatures known, the gas- and granular energies are
\begin{eqnarray}
  (\alpha_s \rho_s E_s)^{\Delta t} &=& \frac{3}{2} \alpha_s \rho_s \Theta_s^{\Delta t} \\
  (\alpha_s \rho_s e_s)^{\Delta t} &=& (\alpha_s \rho_s e_s)^0 - \frac{3}{2} \alpha_s \rho_s \left( \Theta_s^{\Delta t} - \Theta_s^{**} \right) \\
  (\alpha_g \rho_g E_g)^{\Delta t} &=& (\alpha_g \rho_g E_g)^{\Delta t} - \frac{3}{2} \alpha_s \rho_s \left( \Theta_s^{**} - \Theta_s^{0} \right).
\end{eqnarray}
The integration order of the above three steps are reversed with each call to $\mathcal{S}_{\Theta}^{\Delta t}$. On a second call, $\dot{\gamma}$ is integrated first, followed by $\phi_{\text{slip}}$, and then $\phi_{\text{visc}}$.

\section{One-Dimensional Test Problems}
A series of numerical experiments were performed to assess the accuracy and robustness of the method in one dimension.  The solutions were advanced in time using a third-order Runge-Kutta \cite{spiteri2003new} with a time-step size based on the maximum wave speed:
\begin{equation}
   \Delta t = \text{CFL}\frac{1}{\max(|u_g| + c_g, |u_s| + c_s)}
\end{equation}
where the CFL number was 0.5.  Solid volume fractions below $\alpha_{\text{min}} = 10^{-10}$ are set to zero as discussed in step 10 of Sec. 3.3.  Boundary conditions for the granular phase were either reflected for a symmetry condition or extrapolated for inflow and outflow.

\subsection{Advection of a Material Interface}
\begin{figure}
\centering
 \subfloat[$\rho_g$ and $\alpha_s$]{\includegraphics[width=0.45\textwidth]{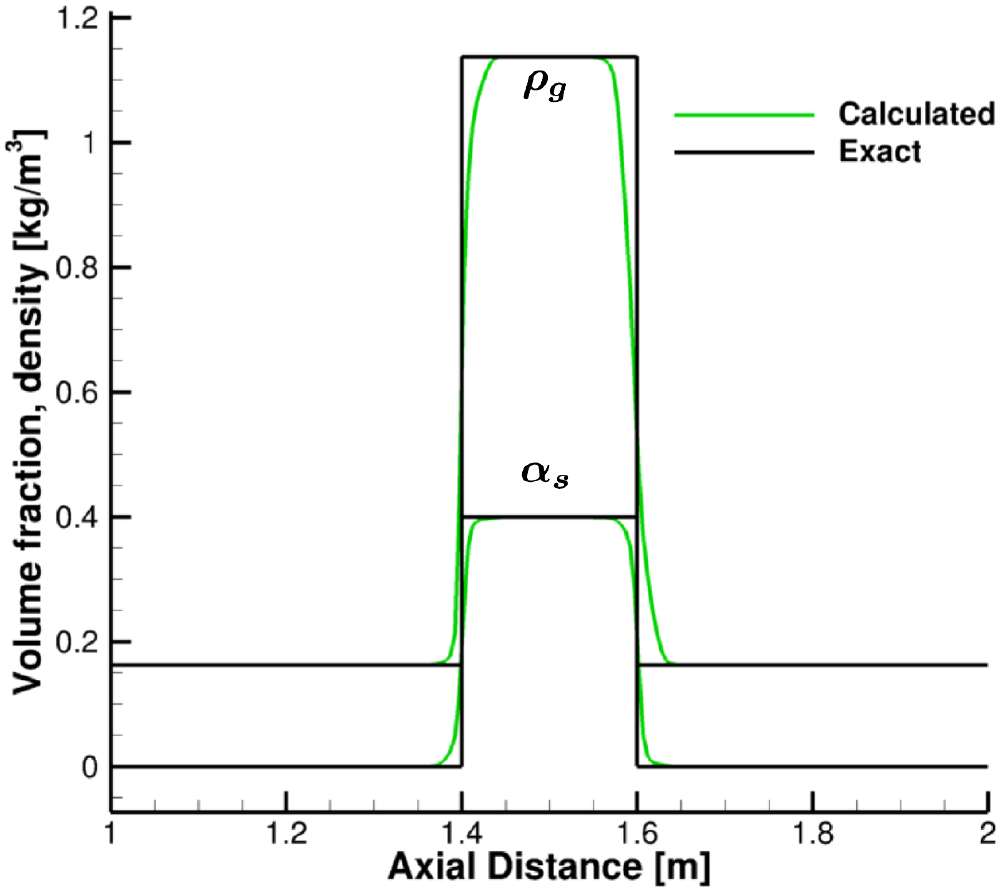}} \quad
 \subfloat[Pressure and temperature error]{\includegraphics[width=0.48\textwidth]{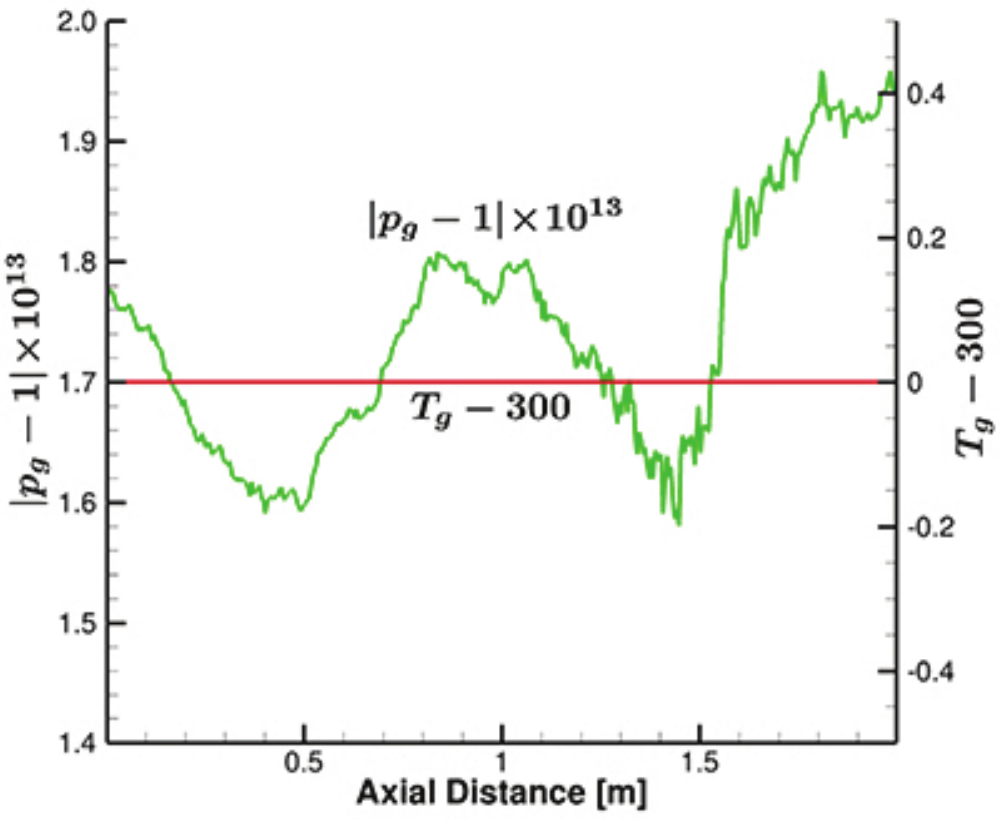}}
\caption{Advection of a granular and multicomponent gaseous interface with uniform temperature for a distance 1 m. (a) Comparison of gas-phase density and $\alpha_s$ to the exact solution and (b) gas-phase temperature and error in pressure.}
\label{fig:constTempAdvect}
\end{figure}

The first tests consist of advecting a material interface at uniform pressure and velocity.  The initial conditions are: $Y_{N_2} = 1$ and $\alpha_s = 0.4$ if $0.4 < x < 0.6$ and $Y_{He}=1$ and $\alpha_s=0$ otherwise.  The pressure in the domain was 1 atm and the temperature was 300 K.  The velocity was 100 m/s and the interfaces were advected for a distance of 1 m.  The domain was discretized with 200 grid points per meter.  The computed solution, shown in Fig.~\ref{fig:constTempAdvect}, indicates that temperature and pressure error are negligible.  

\begin{figure}
\centering
\includegraphics[width=0.48\textwidth]{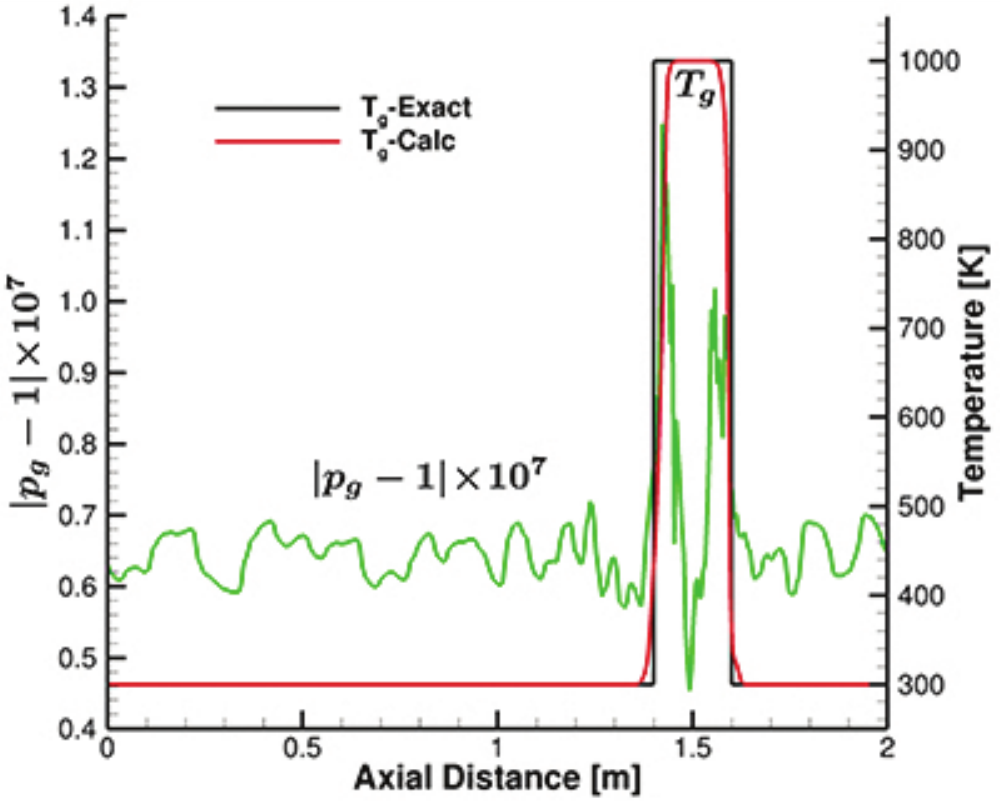}
\caption{Computed temperature and error in pressure after advecting of a granular and multicomponent gaseous interface with discontinuous temperature for a distance 1 m.}
\label{fig:variableTempAdvect}
\end{figure}

The problem is more challenging when a temperature discontinuity exists in the gas phase at the material interface.  The problem discussed above was repeated, but with the gas-phase temperature in the nitrogen bubble was increased to 1000 K, while the temperature in the helium is left at 300 K.  The pressure error and temperature are shown in Fig.~\ref{fig:variableTempAdvect}.  Even in this more challenging case with a multiphase, multispecies, and multitemperature contact surface, the computed pressure error remains small, with a maximum of $1.3 \times  10^{-7}$, which is mainly caused by small interpolation errors with the tabular approach [see Eqn.~\eqref{eqn:cpTabular}] used to calculate the gas-phase specific heat.  

\subsection{Transonic Flow Through a Granular Nozzle}
\begin{figure}
\centering
\includegraphics[width=0.5\textwidth]{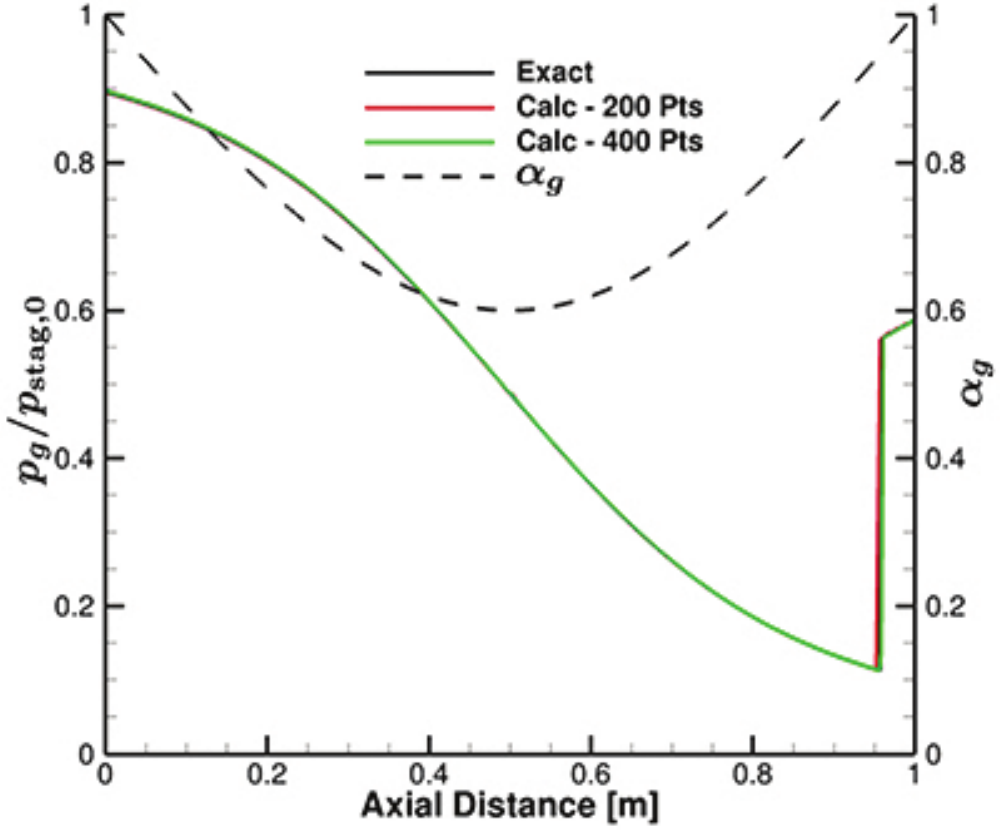}
\caption{Computed ratio of static to inlet stagnation pressure ($p_{\text{stag,0}}$) in a one-dimensional, transonic, granular ``nozzle''.  Note that the computed and exact solutions lie on top of each other.}
\label{fig:ArgonNozzle}
\end{figure}

In this test problem, a transonic nozzle was modeled by freezing the particles and neglecting all of the coupling sources except the gas-phase nozzling term. Under these conditions, the granular-phase equations are eliminated, and, if diffusion and viscous effects are neglected, the gas-phase equations reduce to one-dimensional nozzle flow.  The shape of the nozzle, determined by the gas-phase volume fraction, is $\alpha_g = 1 - 0.4 \sin(\pi x)$.  Argon gas was used to allow direct comparison with an exact solution.  The initial conditions were taken to be a uniform background of subsonic flow. Boundary conditions at both the inlet and outlet were non-reflecting.  The equations were integrated until a steady state was achieved.  The resulting stagnation pressure at the nozzle inlet was used to compute an exact solution.  A comparison between the exact and the computed solutions is shown in Fig.~\ref{fig:ArgonNozzle}.  The exact solution and computed solution on both grids are in excellent agreement and lie on top of each other. 

It is a more difficult problem when the change in volume fraction is discontinuous.  The gas-phase volume fraction is piecewise constant where $\alpha_g = 1$ if $x < 0.5$ and 0.6 otherwise.  The one-dimensional nozzle equations are invalid if the nozzle shape has a step change since the $p dA/dx$ term is non-differentiable and a perfect and piecewise constant solution is not expected for this test.  Yet, this test is important to determine if gas-phase pressure and velocity overshoots near a discontinuous interface grow with grid refinement. This calculation was advanced to steady state using the same procedure that was used to simulate the sinusoidal nozzle.

 \begin{figure}
 \centering
 \includegraphics[width=0.5\textwidth]{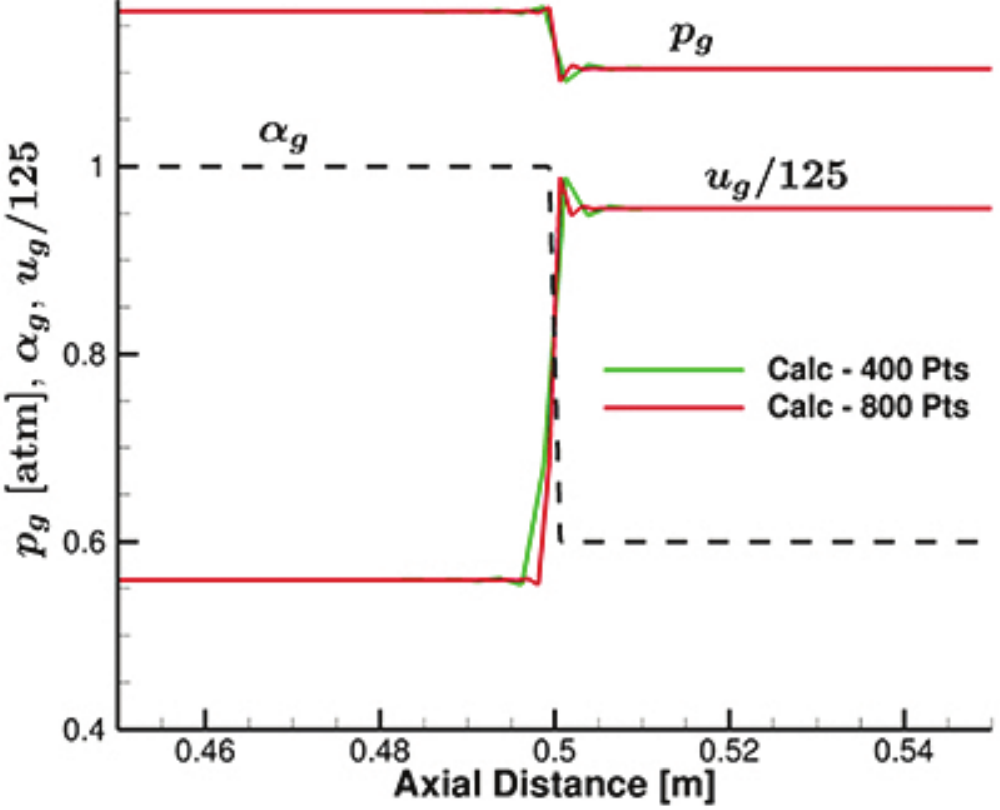}
 \caption{Computed solution of a one-dimensional granular nozzle with a step change in volume fraction.}
 \label{fig:Argon_Step}
 \end{figure}
The results of this calculation, summarized in Fig.~\ref{fig:Argon_Step}, show that the computed solution is, for the most part, piecewise constant. There are, however, small oscillations near the discontinuity in volume fraction.  These form even if an exact solution is given as the initial condition. The origin of these oscillations can be understood by noting that a computed discontinuity cannot be resolved over a single grid point, while, at the same time, the stationary granular interface is  artificially forced to lie within a single grid point. Such oscillations do not grow with grid refinement and would have little effect on full-scale simulations where intense drag in dense granular regions, numerical dissipation, and particle motion would smear the granular interface and significantly reduce the strength and time of the oscillations (see Sec. 6.5.)  Developing methods for eliminating such unphysical oscillations from an unphysical test problem is beyond the scope of this paper.

\subsection{Two-Phase Granular Riemann Problem}
\begin{figure} 
  \centering
  \subfloat[]{\includegraphics[width=0.49\textwidth]{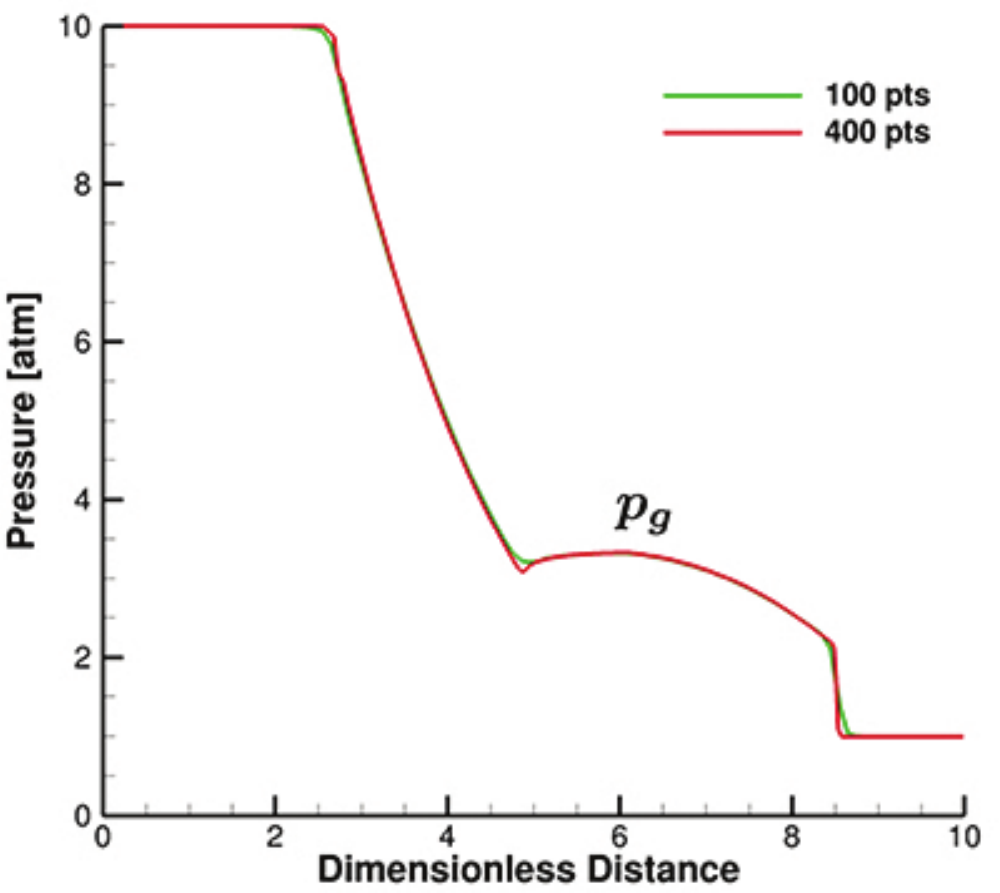}} \ 
  \subfloat[]{\includegraphics[width=0.49\textwidth]{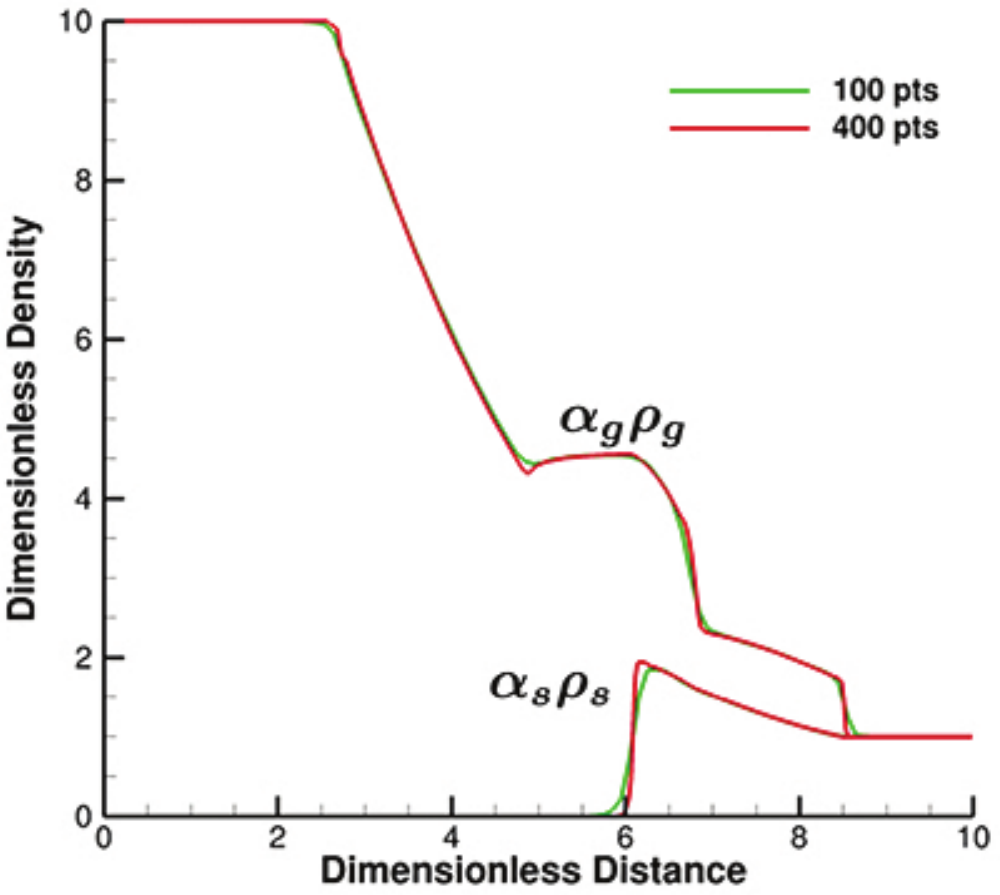}}
\caption{Computed (a) gas-phase pressure and (b) bulk densities of the gas and granular phases for the two-phase granular Riemann problem at $184$ $\mu$s.  The densities were scaled by the initial gas-phase density to the right of the interface and the length was scaled so that the domain ranges from 0 to 10 to match output used by Fedorov \textit{et al.} \cite{fedorov2007reflection}.}
\label{fig:SaitoShockTube} 
\end{figure}

There are not any exact solutions that can be used to verify multiphase shock-tube problems using dust-gas or kinetic theory-based granular models, even though the physics is well known for dilute particle suspensions \cite{miura1982dusty, saito2003numerical}.  Instead, calculations can be compared to the previous results \cite{saito2003numerical, fedorov2007reflection}, which used the following initial conditions
\begin{gather}
    \begin{matrix}
     p_g^L = 10 \ \text{atm},  & \quad  p_g^R = 1 \ \text{atm}, \\
     T^L = 270 \ \text{K},     & \quad  T^R = 270 \ \text{K}, \\
     Y_{g,\text{air}}^L = 1,   & \quad  Y_{g,\text{air}}^R = 1, \\
     \alpha_s^L = 0,             & \quad  \alpha_s^R = 5.172 \! \times \! 10^{-4}, \\
     \Theta_s^L = 0,             & \quad  \Theta_s^R = 0.
    \end{matrix}
\end{gather}
The density, diameter, and specific heat of the particles are $\rho_s = 2500$ kg/m$^3$, $d_s = 10 \ \mu$m, and $C_{V,s} = 718$ J/kg K, respectively. The domain is $0.257798$ m in length and the diaphragm is placed at $0.129$ m.  The solution at $184 \ \mu$s after the calculation is initiated is shown in Fig.~\ref{fig:SaitoShockTube} for two grid spacings.  The computed solutions compare well with those of Saito \textit{et al.} \cite{saito2003numerical} and Fedorov \textit{et al.} \cite{fedorov2007reflection}, and the discontinuities converge  with increasing refinement.   

\subsection{Multicomponent Granular Riemann Problem}
In this problem we solve the He-N$_2$ Riemann shown in Fig.~\ref{fig:HEN2Riemann}, but add particles to the right side of the domain similar to the previous problem.  The initial states for this test case are
\begin{gather}
    \begin{matrix}
     p_g^L = 10 \ \text{atm},  & \quad  p_g^R = 1 \ \text{atm}, \\
     T^L = 300 \ \text{K},     & \quad  T^R = 300 \ \text{K}, \\
     Y_{g,\text{He}}^L = 1,   & \quad  Y_{g,\text{He}}^R = 0, \\
     Y_{g,\text{N}_2}^L = 0,   & \quad  Y_{g,\text{N}_2}^R = 1, \\     
     \alpha_s^L = 0,             & \quad  \alpha_s^R = 4.555 \! \times \! 10^{-4}, \\
     \Theta_s^L = 0,             & \quad  \Theta_s^R = 0.
    \end{matrix}
\end{gather}

The density, specific heat, and restitution coefficient of the particles are $2500$ kg/m$^3$, $745$ J/kg K, and $0.999$, respectively. The domain is $1$ m in length and the diaphragm is placed at $ 0.5$ m. The volume fraction of the right state was chosen so that the bulk densities of the gas and granular phases were the same.   Calculations were done for two different particle diameters, 10 $\mu$m and 25 $\mu$m.  The equations were integrated to 400 $\mu$s of physical time and 400 grid points were used in each calculation.  

\begin{figure}
\centering
 \subfloat[Pressure]{\includegraphics[width=0.49\textwidth]{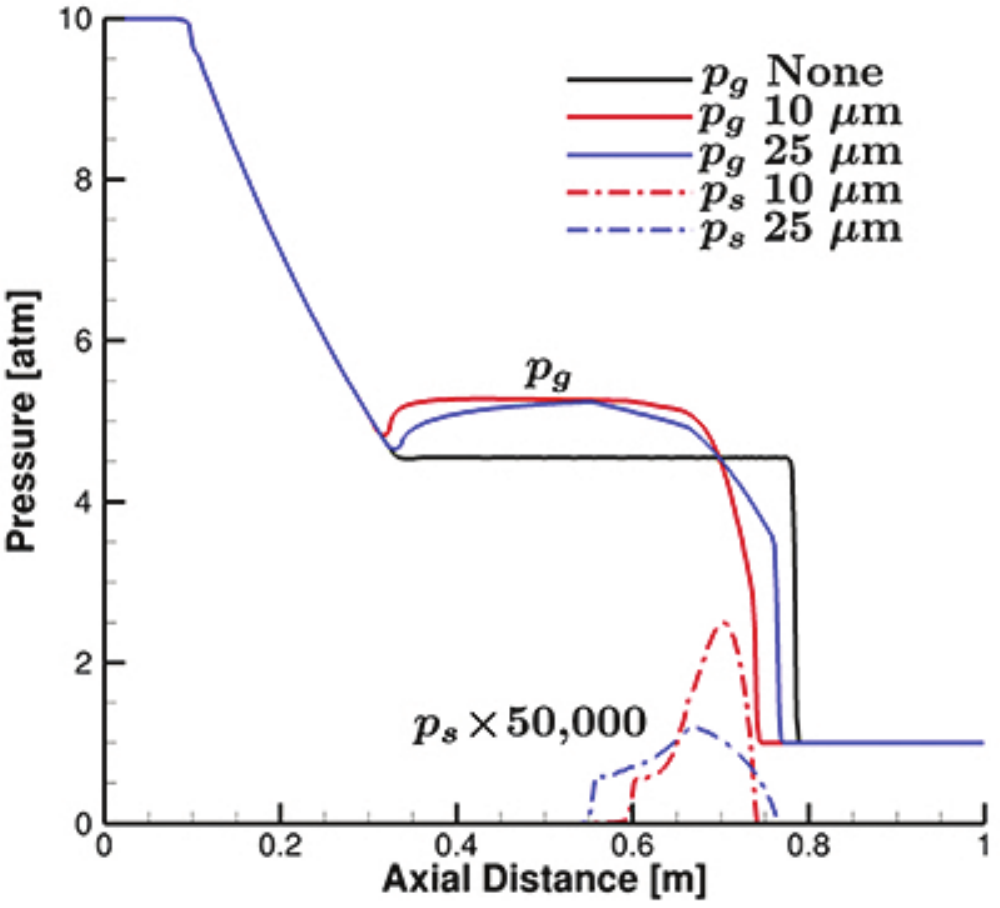}} \
 \subfloat[Velocity]{\includegraphics[width=0.49\textwidth]{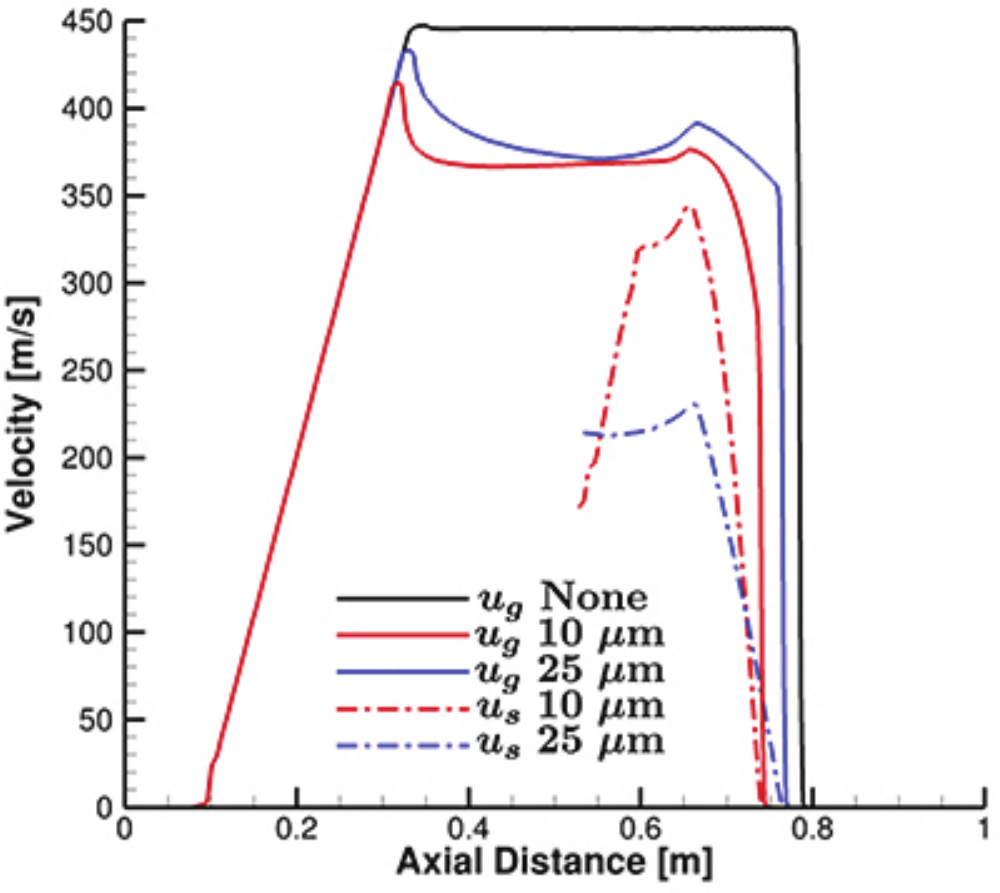}} \\
 \subfloat[Temperature]{\includegraphics[width=0.49\textwidth]{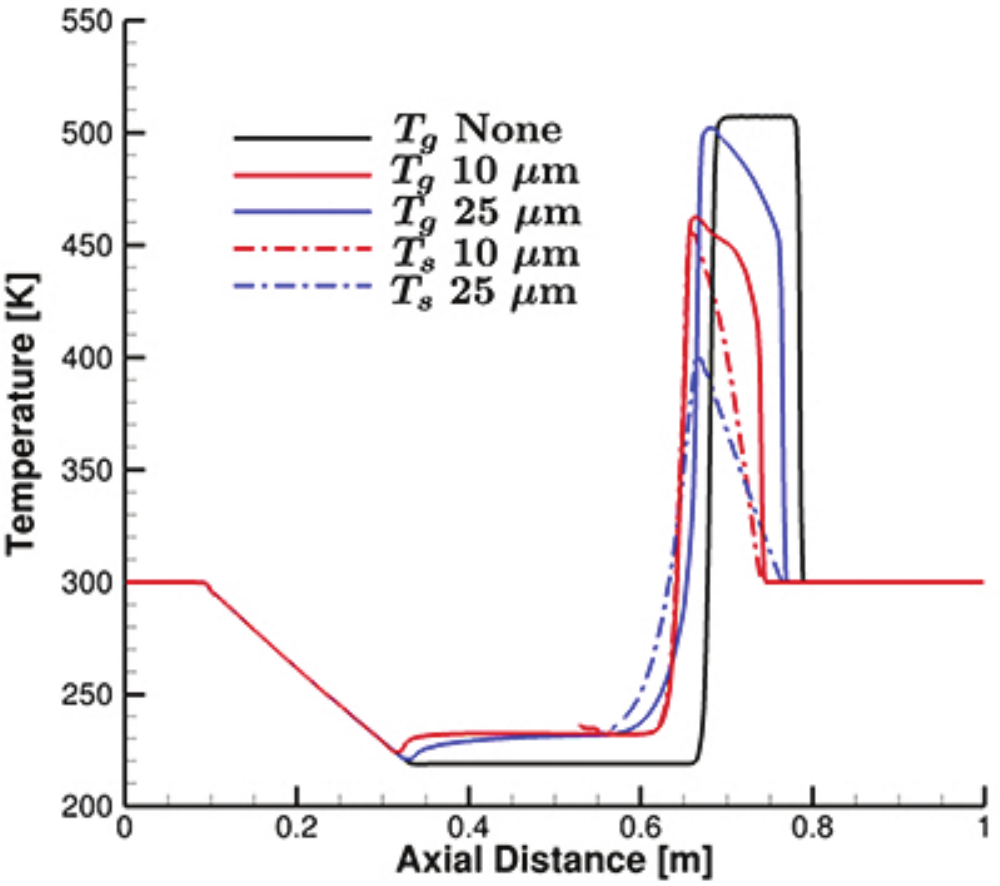}} \
 \subfloat[Bulk Density]{\includegraphics[width=0.49\textwidth]{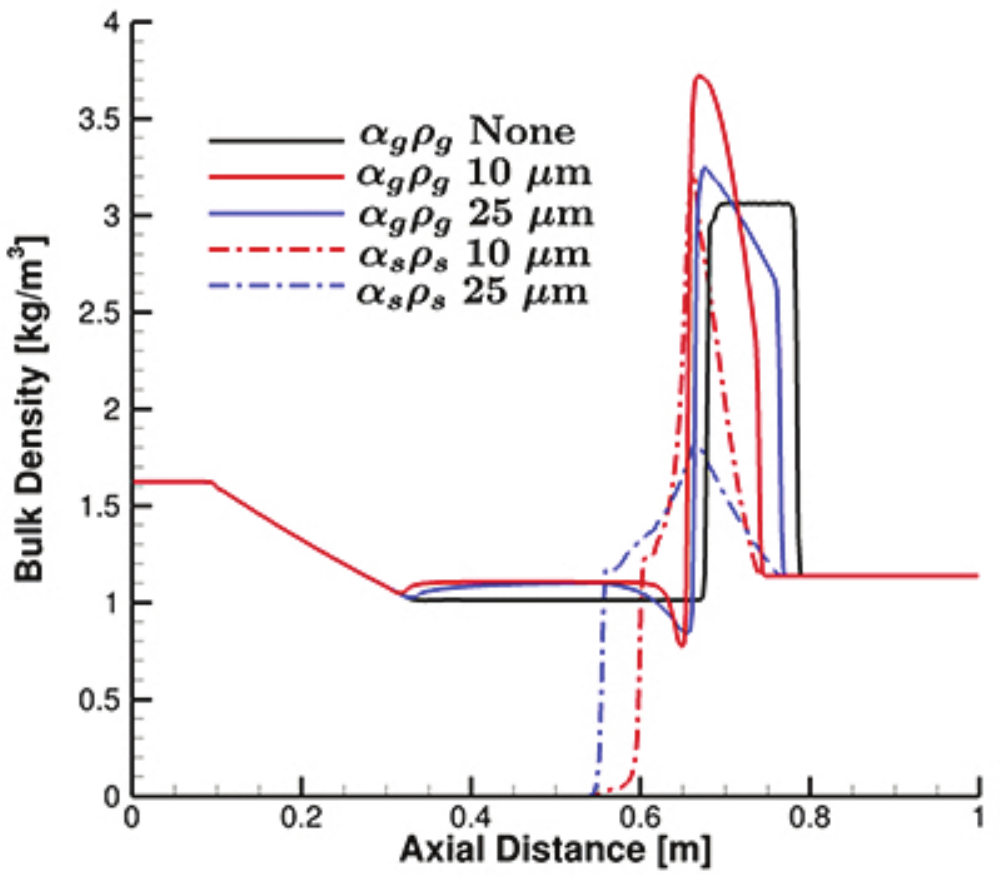}} 
\caption{Computed (a) gas and solids pressure, (b) velocity, (c) temperature, and (d) bulk density for multiphase shock tube problem with a He-N$_2$ interface at 400 $\mu$s.  The solid temperature and velocity are not shown if the particle volume fraction lies below 10$^{-8}$.}
\label{fig:LowVolFracShockTube}
\end{figure}

The multiphase shock tube results shown in Figs.~\ref{fig:SaitoShockTube} and \ref{fig:LowVolFracShockTube} indicate that the post-shock gas-phase pressure in multiphase shock tubes is actually higher if particles are present.  This is due to drag reducing the gas velocity and converting some of the gas-phase momentum into static pressure.  This raises the pressure of the post-shock gas where particles are present, which, in turn, generates compression waves that propagate towards the rarefaction.  These compression waves are responsible for the dips in pressure located on the tail of the rarefaction. The peak of solid bulk density and velocity at $\sim \! 0.65$ m is caused by a sharp change in the drag coefficient at the He-N$_2$ contact surface as a result of the large change in $\rho_g$.  This causes a pileup of particles on the nitrogen side of the gas-phase contact surface where the drag force is higher.   
The solids pressure is very small and has a negligible influence for the low volume fractions in this problem.

\subsection{Dense Granular Shock-Tube Problem}
\begin{figure}
 \centering
  \subfloat[Pressure]{\includegraphics[width=0.49\textwidth]{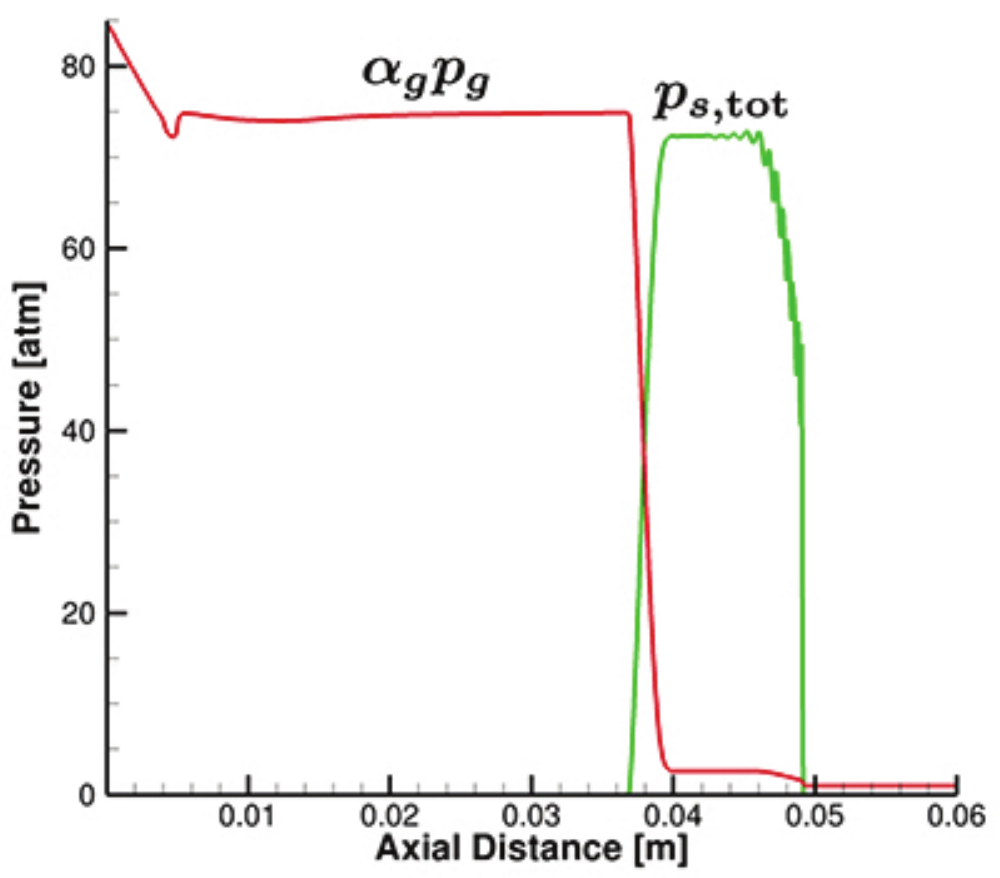}} \
  \subfloat[Velocity]{\includegraphics[width=0.49\textwidth]{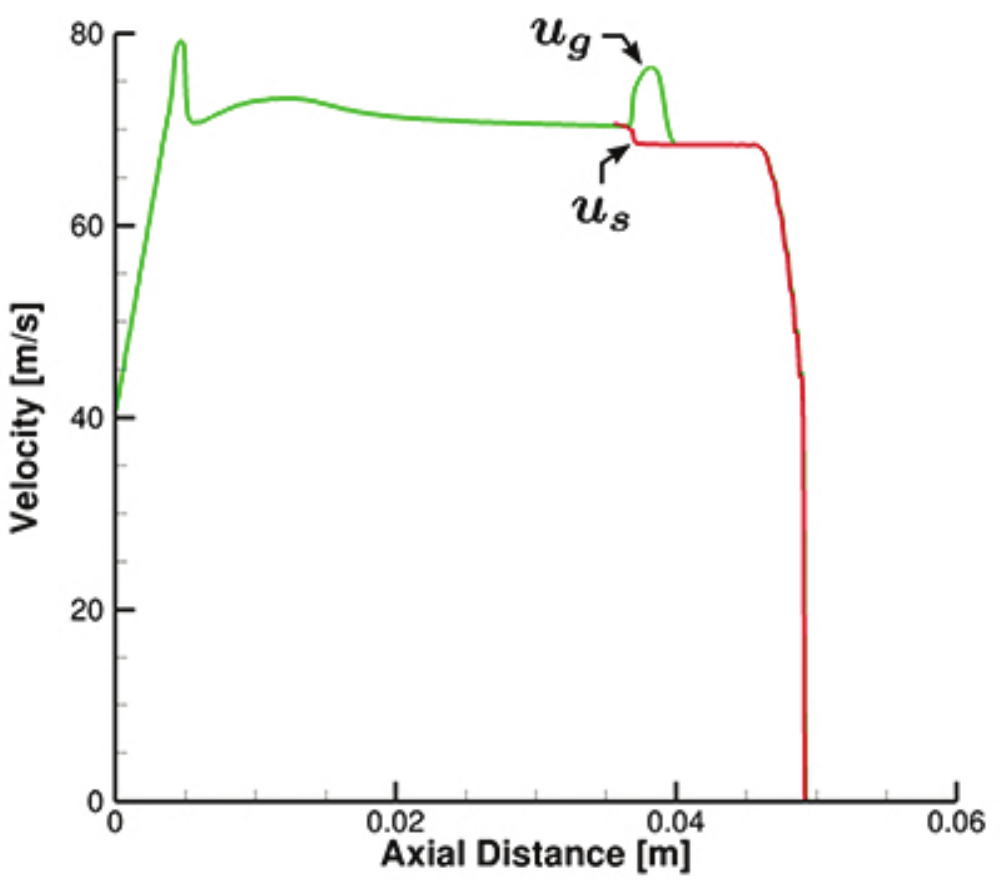}} \\
  \subfloat[Temperature]{\includegraphics[width=0.49\textwidth]{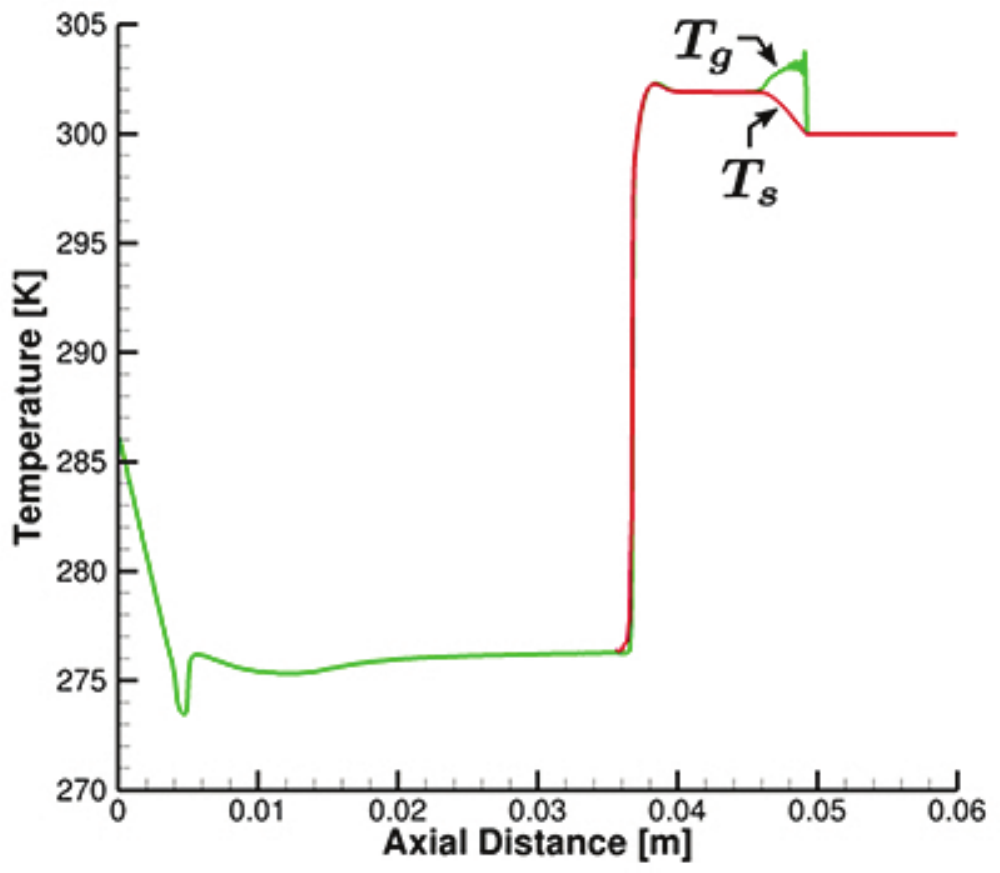}} \
  \subfloat[Bulk density]{\includegraphics[width=0.49\textwidth]{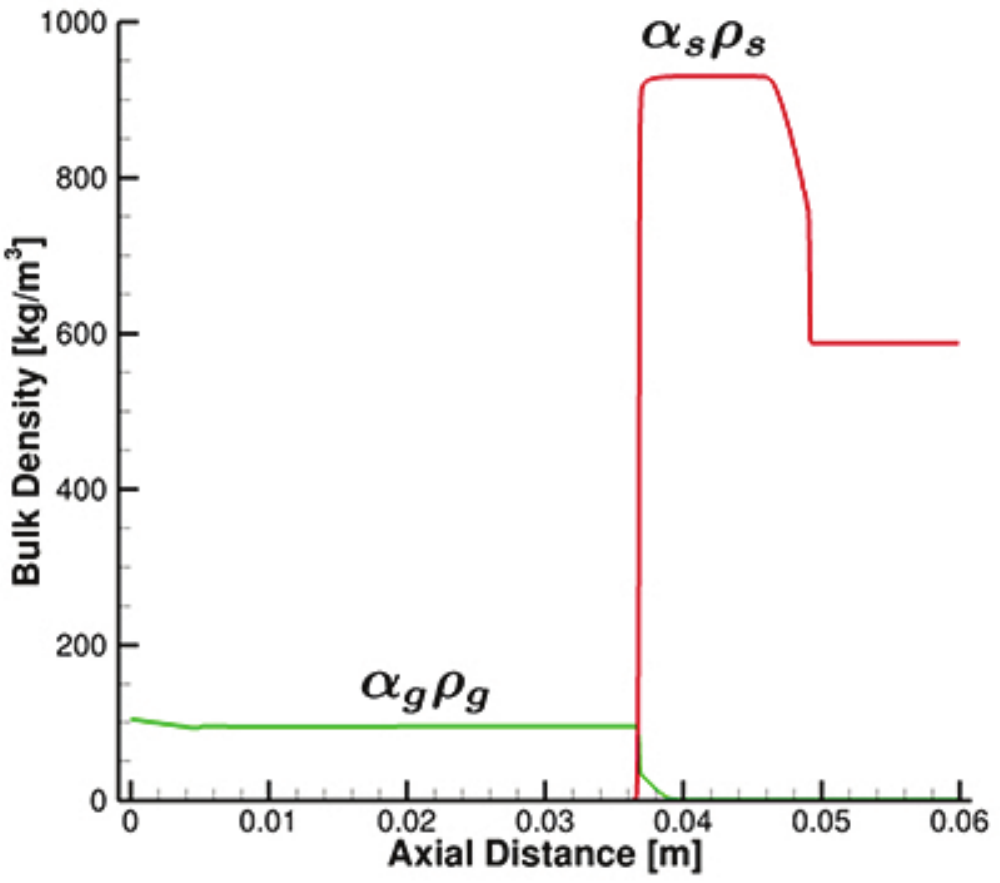}} 
 \caption{Computed (a) pressure, (b) velocity, (c) temperature, and (b) bulk density for the dense granular shock tube problem at 100 $\mu$s with $e=0.999$ and $\mathcal{D}=1$.}
 \label{fig:HighVolFracShockTube}
 \end{figure}

This problem tests the capabilities of the  method to compute high-pressure gases interacting with dense granular regions that approach the packing limit.  The initial states for this problem are
\begin{gather}
    \begin{matrix}
     p_g^L = 100 \ \text{atm},  & \quad  p_g^R = 1 \ \text{atm}, \\
     T^L = 300 \ \text{K},     & \quad  T^R = 300 \ \text{K}, \\
     Y_{g,\text{Air}}^L = 1,   & \quad  Y_{g,\text{Air}}^R = 1, \\
     \alpha_s^L = 0,             & \quad  \alpha_s^R = 0.4, \\
     \Theta_s^L = 0,             & \quad  \Theta_s^R = 0.
    \end{matrix}
\end{gather}
The domain measures 0.06 m in length and the diaphragm was placed at 0.03 m.  The solution was advanced to 100 $\mu$s on a grid with 1200 computational cells.   The particle diameter, density, and specific heat were 5 $\mu$m, 1470 kg/m$^3$, and 987 J/kg K, respectively. The  solution is shown in Fig.~\ref{fig:HighVolFracShockTube} with the hyperbolic dissipation parameter $\mathcal{D}$ set to 1 [see Eqns.~(\ref{eqn:granAusm}), (\ref{eqn:granF}), and (\ref{eqn:granG})] and a coefficient of restitution ($e$) of 0.999.  

\begin{figure}
  \centering
 \centering
  \subfloat[$e$=0.999]{\includegraphics[width=0.49\textwidth]{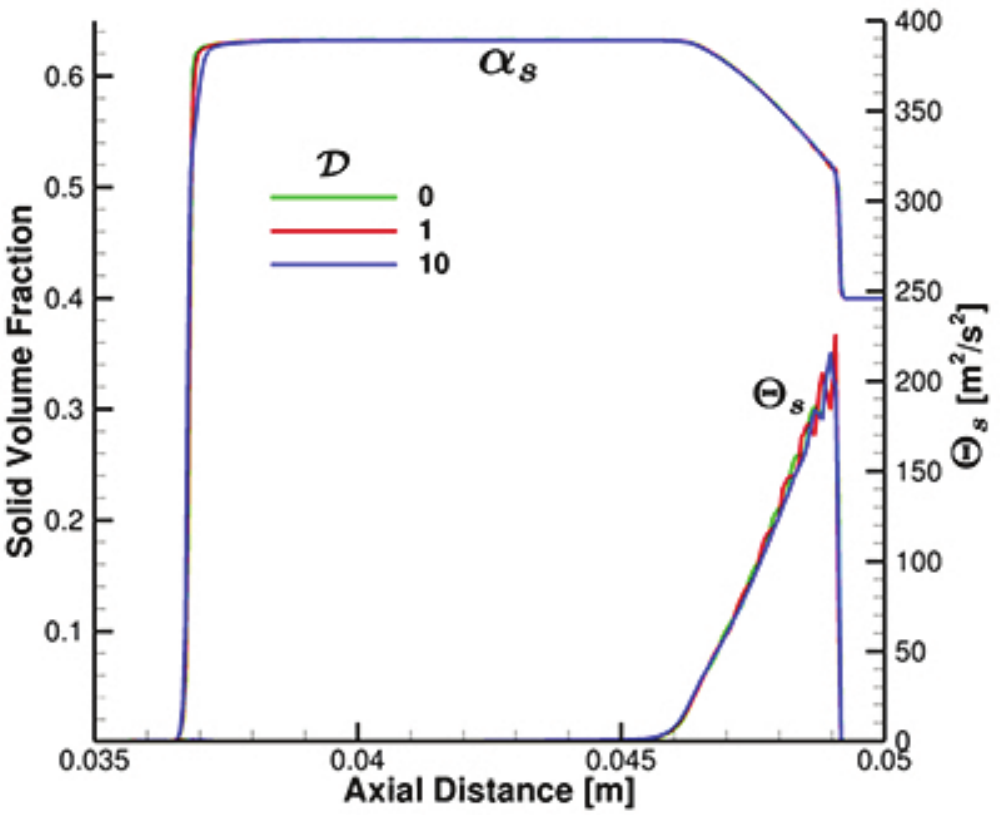}} \
  \subfloat[$e$=0.9]{\includegraphics[width=0.49\textwidth]{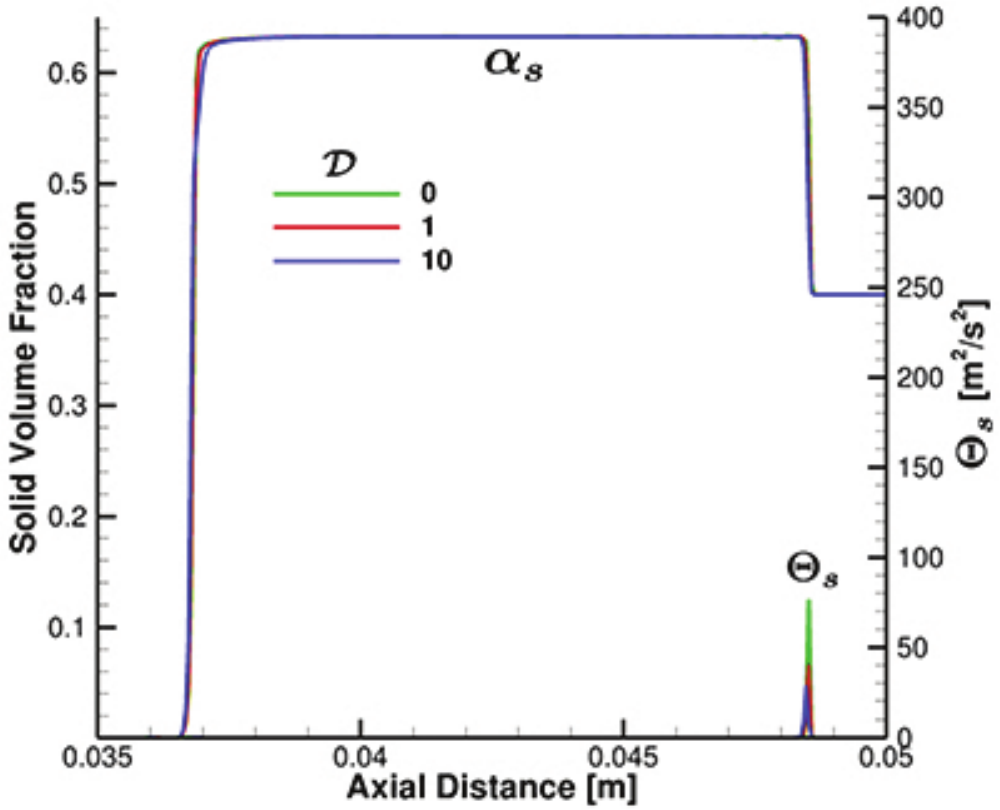}} 
 \caption{Effect $\mathcal{D}$ and $e$ on calculated profiles of $\alpha_s$ and $\Theta_s$ near a granular shock.}
 \label{fig:HighVolFracShockTube_alpha}
 \end{figure}

\begin{figure}
  \centering
 \centering
  \subfloat[$e$=0.999]{\includegraphics[width=0.49\textwidth]{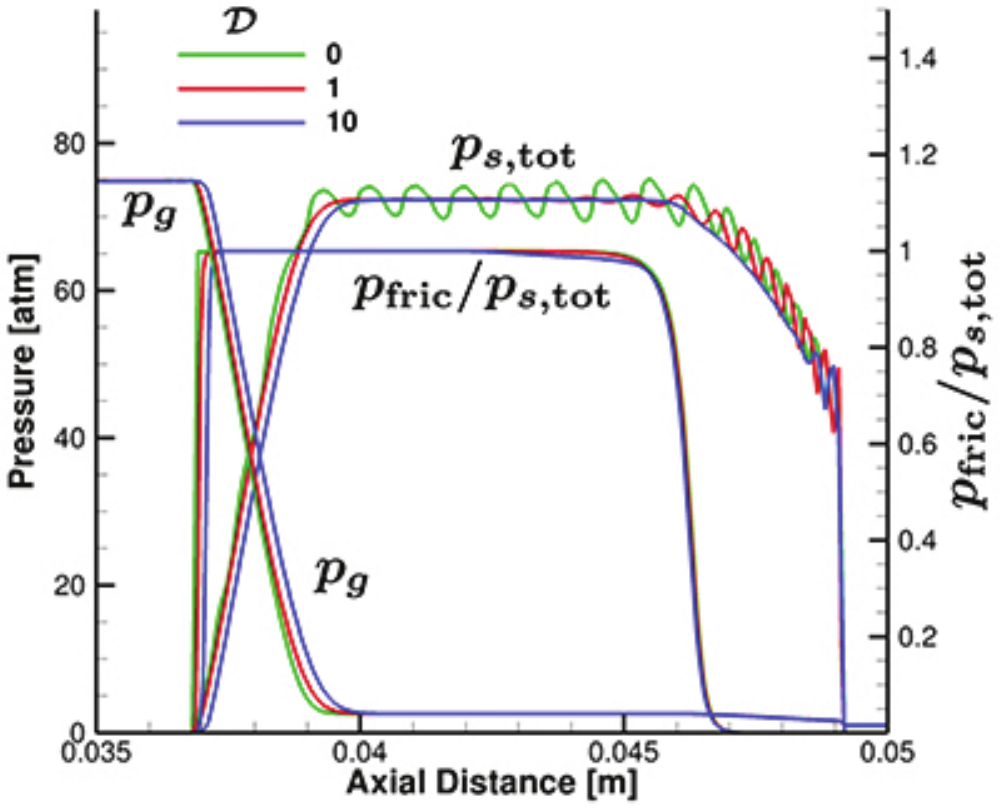}} \
  \subfloat[$e$=0.9]{\includegraphics[width=0.49\textwidth]{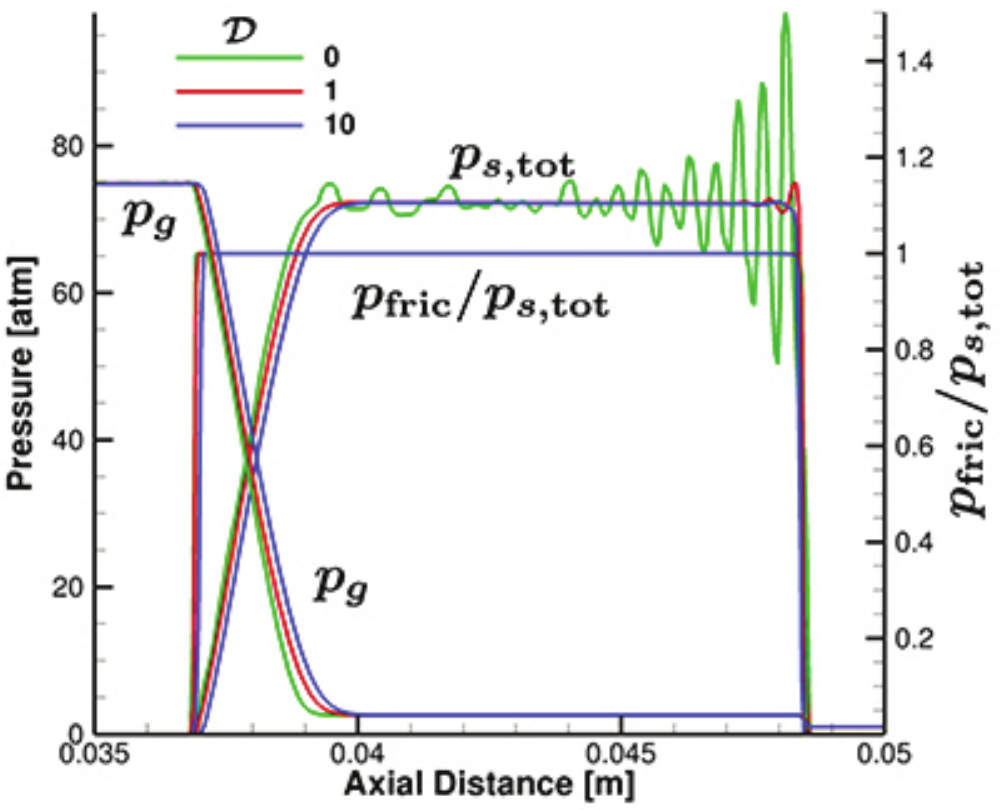}}
 \caption{Effect $\mathcal{D}$ and $e$ on calculated profiles of pressure and $p_{\text{fric}}/p_{s,\text{tot}}$ near a granular shock.}
 \label{fig:HighVolFracShockTube_press}
 \end{figure}

The structure of the granular shock as a function of $\mathcal{D}$ and $e$ are shown in Figs.~\ref{fig:HighVolFracShockTube_alpha} and \ref{fig:HighVolFracShockTube_press}.  
 Particle-particle interactions are dominant, unlike previous problems.  A strong granular shock is transmitted into the granular phase due to the combination of buoyancy forces ($\alpha_s \! \nabla \! p_g$) and particle drag from the high pressure gas flowing into the dense layer of particles.  The computed granular shock has a relaxation structure  similar to those reported by Kamenetsky \textit{et al.} \cite{kamenetsky2000evolution}.  Sources of granular energy, $E_s$, decrease rapidly away from the granular shock.  Without these sources, granular dissipation, $\dot{\gamma}$, decreases $\Theta_s$ and $p_s$.  Nevertheless, mechanical equilibrium needs to be maintained and the solid volume fraction rises to compensate for the reduction in $\Theta_s$ by increasing the friction pressure.  Eventually the granular temperature decreases to zero, and the total intergranular stress, $p_{s,\text{tot}}$, is solely from friction pressure, as shown in Fig.~\ref{fig:HighVolFracShockTube_press}.  The granular temperature decreases much faster when the coefficient of restitution is lowered to $e=0.9$, and then the granular shock structure is almost entirely due to friction pressure.  

Transmission of the granular shock also compresses the gas and produces a sharp rise in temperature and pressure.  Small gas-phase temperature oscillations on the order of 0.5 K are present near the granular shock.  These do not cause any numerical instabilities and quickly damp from heat transfer between the gas and granular phases.  The bump in gas-phase velocity near the solid contact in Fig.~\ref{fig:HighVolFracShockTube} is caused from a reduction in flow area when the gas flows into the particle bed.  Oscillations of pressure, velocity, and temperature near the granular interface are not present, as they were in the granular nozzle with a step change in flow area shown in Fig.~\ref{fig:Argon_Step}.      

The effectiveness of increasing $\mathcal{D}$ to control oscillations of intergranular stress in dense granular regions is shown in Fig.~\ref{fig:HighVolFracShockTube_press}.  Severe oscillations of friction pressure are produced when $\mathcal{D}$ is zero, but increasing $\mathcal{D}$ to 1 or 10 effectively removes the oscillations at the cost of some smearing of the granular interface.

\subsection{High-Pressure Outgassing}
\begin{figure}
\centering
 \subfloat[Pressure]{\includegraphics[width=0.49\textwidth]{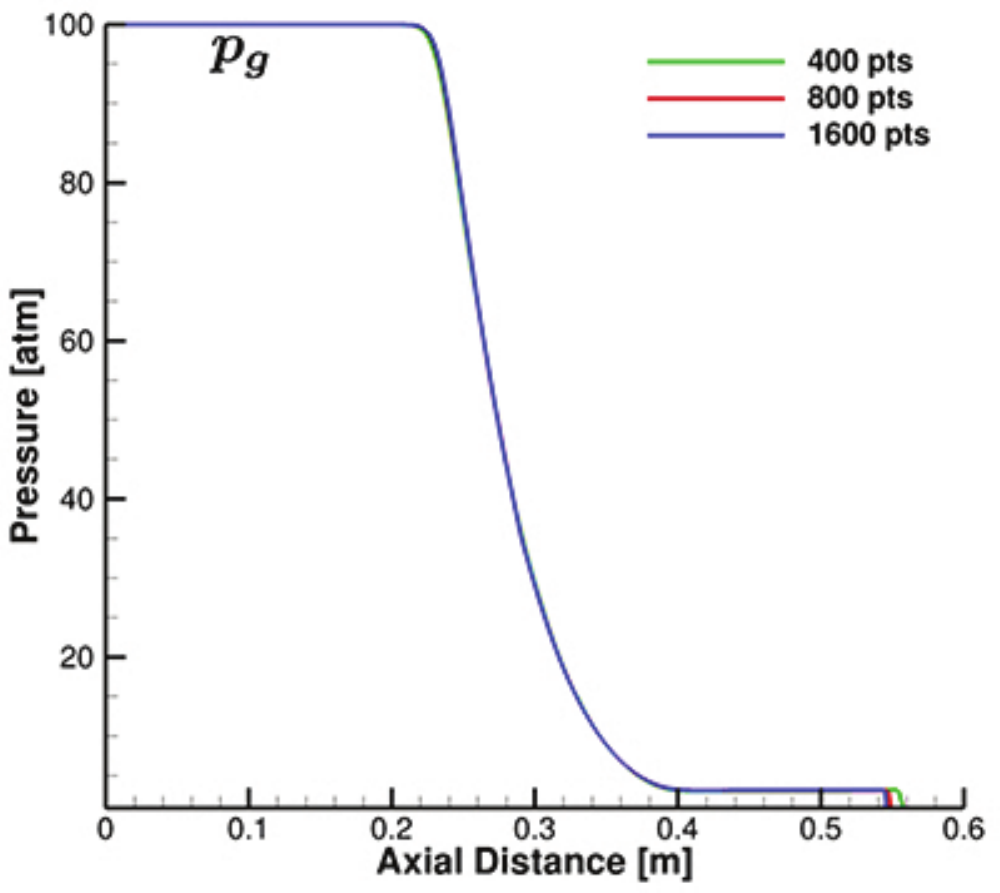}} \
 \subfloat[Velocity]{\includegraphics[width=0.49\textwidth]{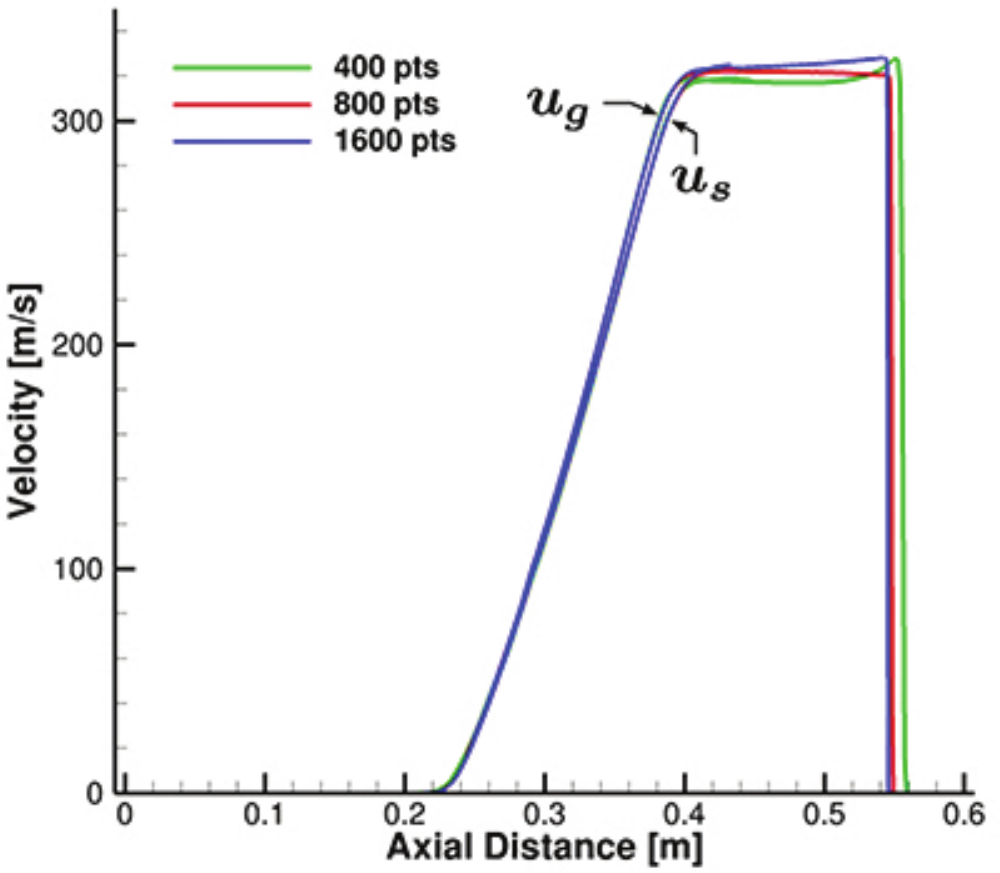}} \\
 \subfloat[Temperature]{\includegraphics[width=0.49\textwidth]{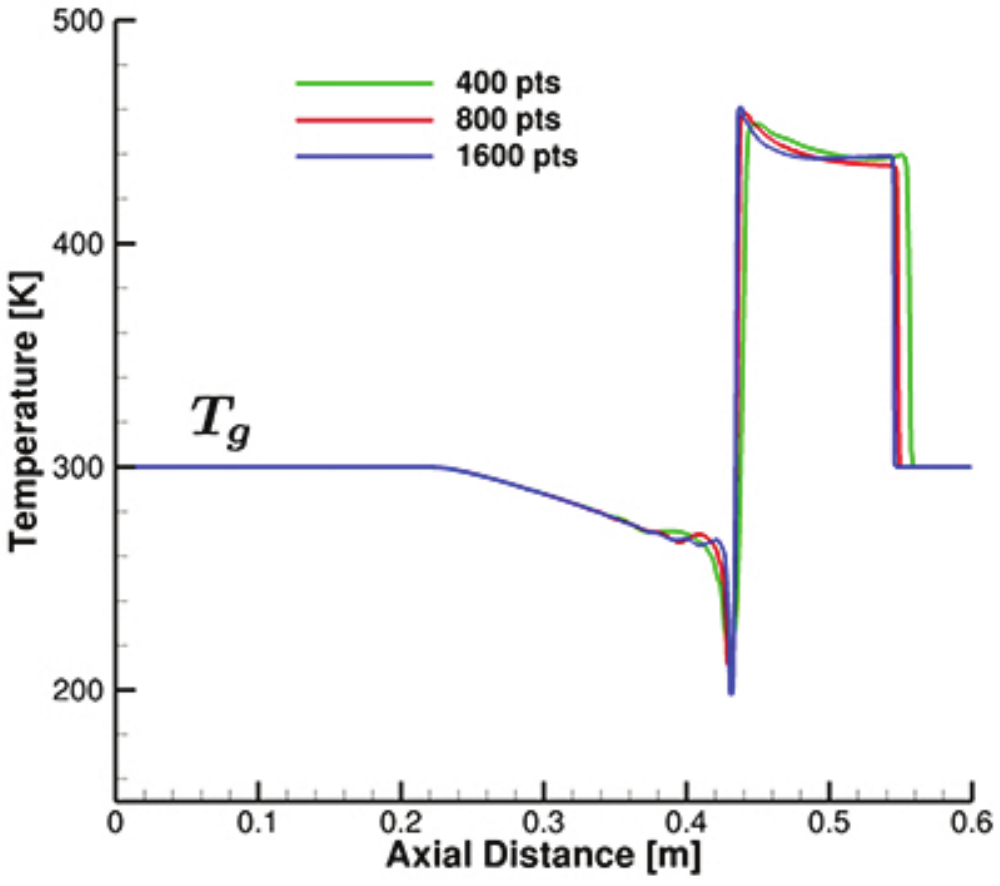}} \
 \subfloat[BulkDensity]{\includegraphics[width=0.49\textwidth]{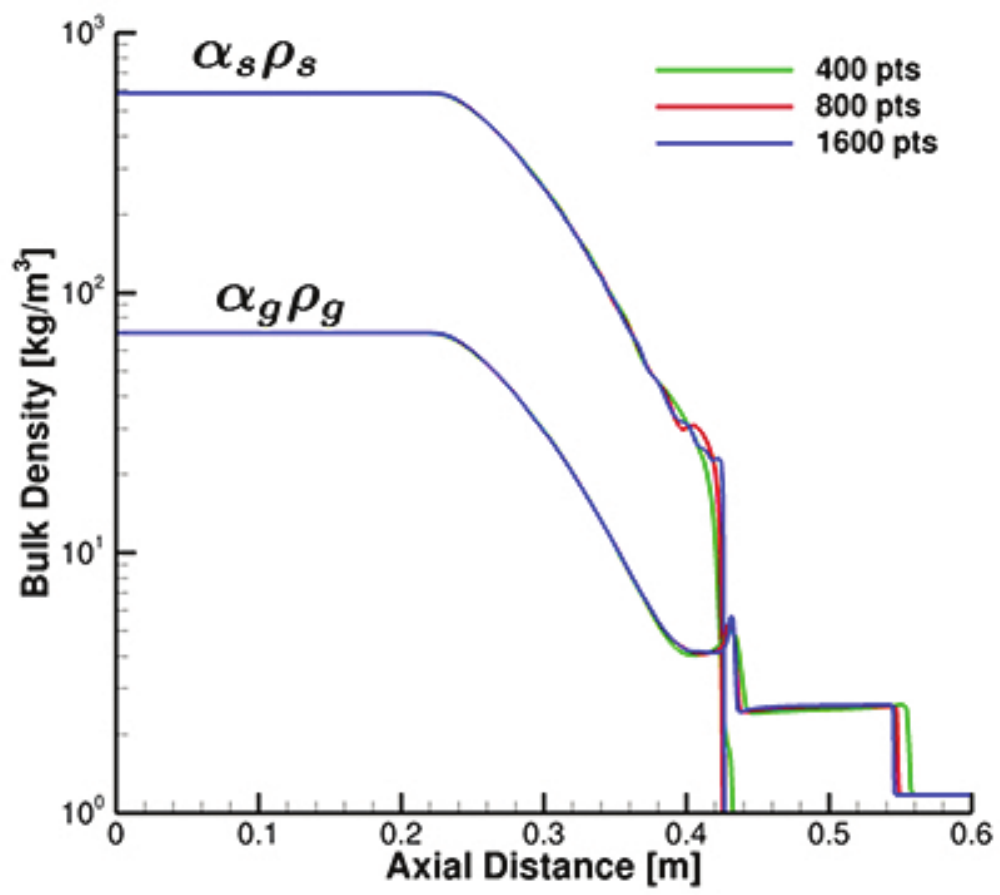}} 
\caption{Computed (a) gas-phase pressure, (b) velocity, (c) gas-phase temperature, and (d) bulk density  for high-pressure gaseous ejection from a dense layer of particles at a time of 400 $\mu$s. The solid velocity is not shown if the particle volume fraction lies below 10$^{-8}$.}
\label{fig:HighVolFracEjection}
\end{figure}

 Shock waves and expansion waves come and go throughout the course of a calculation, depending on the problem being solved.  Consider a high-pressure gas that is injected into a granular layer as a shock passes by.  This gas can subsequently be ejected when an expansion wave reduces pressure at the surface.  The outflow induced by the expansion wave will entrain some of the particles along with it. Whereas the previous problem simulated a high-pressure gas entering a region of dense particles, this problem simulates a high-pressure gas leaving a  region of dense particles.  The same initial conditions as the previous problem were used, with the exception that the particles are now located in the high-pressure region of the shock tube,
\begin{gather}
    \begin{matrix}
     p_g^L = 100 \ \text{atm},  & \quad  p_g^R = 1 \ \text{atm}, \\
     T^L = 300 \ \text{K},     & \quad  T^R = 300 \ \text{K}, \\
     Y_{g,\text{Air}}^L = 1,   & \quad  Y_{g,\text{Air}}^R = 1, \\
     \alpha_s^L = 0.4,             & \quad  \alpha_s^R = 0.0, \\
     \Theta_s^L = 0,             & \quad  \Theta_s^R = 0.
    \end{matrix}
\end{gather}
The domain is $0.6$ m long and the diaphragm was placed at $= 0.3$ m.   The solution at 400 $\mu$s, is shown in Fig.~\ref{fig:HighVolFracEjection} for three grid resolutions.

Drag and buoyancy forces induced by the escaping gas ejects particles from the dense granular region.  This reduces the particle volume fraction in a profile that is similar to the gaseous rarefaction wave propagating into the particle bed.  This reduction in particle volume fraction is not a granular expansion wave since the solids pressure is nearly zero in the entire domain.  Intense losses to the gas phase by entraining particles severely weaken the transmitted gaseous shock.  The dips in temperature near the granular interface are physical and result from particles heating the gas as the rarefaction wave propagates through the particle bed and a slight gap between the gas and granular contact surfaces.  Heat transfer increases the gas-phase temperature of gas flowing through the granular contact surface relative to gas that is behind the gas-phase contact but has not pass through the particle bed.  In addition $ u_g$ is slightly higher than $u_s$ at the granular contact surface and, as a result, the gas-phase and granular contacts remain separated by a short distance as the calculation progresses.  The effects of particle heat transfer and the small distance between the gas and granular contacts produce the sharp spike of gas-phase temperature in Fig.~\ref{fig:HighVolFracEjection}(c). 

The solution converges with sufficient refinement.  Slight differences between the grids are introduced by sharpening the initial granular contact surface. This has a small effect on the gas-phase post-shock state.  The small oscillations near the granular contact surface  are reduced with increasing grid refinement, unlike the results, shown in Fig.~\ref{fig:Argon_Step}. 

\section{Two-Dimensional Test Problems}

Multidimensional tests employ block-structured adaptive mesh refinement with the Paramesh library \cite{macneice2000paramesh}.  Refinement is based on smoothness of the mixture density $\rho_m = \alpha_g \rho_g + \alpha_s \rho_g$. Details of how Paramesh was used for this work can be found in \cite{Houim20118527}.  The time-step size is based on the maximum wave speed over the x- and y-directions with a CFL number of 0.5.

\begin{figure}
\centering
\includegraphics[width = 0.6\textwidth]{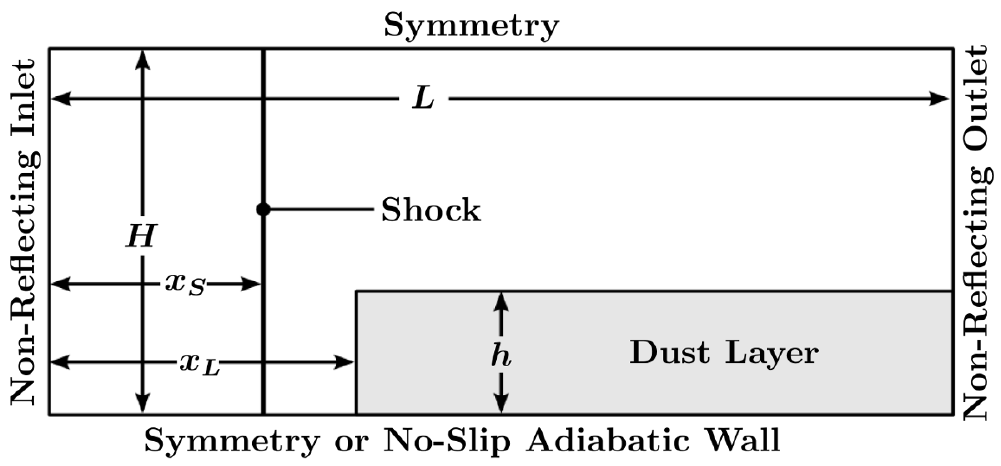}
\caption{Schematic diagram of the initial conditions used for multi-dimensional test problems.}
\label{fig:LayerICs}
\end{figure}

The initial conditions for all of the multidimensional test problems consist of a shock wave interacting with a layer of dust.  A schematic diagram of the geometrical setup and initial and boundary conditions is shown in Fig.~\ref{fig:LayerICs}.  Most of the calculations presented are inviscid, and the bottom boundary is a symmetry plane.  No-slip adiabatic conditions were used for viscous calculations for both gas and solid phases. Application of kinetic theory-based partial-slip boundary conditions for the granular phase \cite{GidaspowBook} could be used for more realistic problems, but is beyond the scope of this work.

\subsection{Interaction of a Shock Wave and a Low-Volume-Fraction Dust Pile}
\begin{figure}
\centering
 \includegraphics[width=0.80\textwidth]{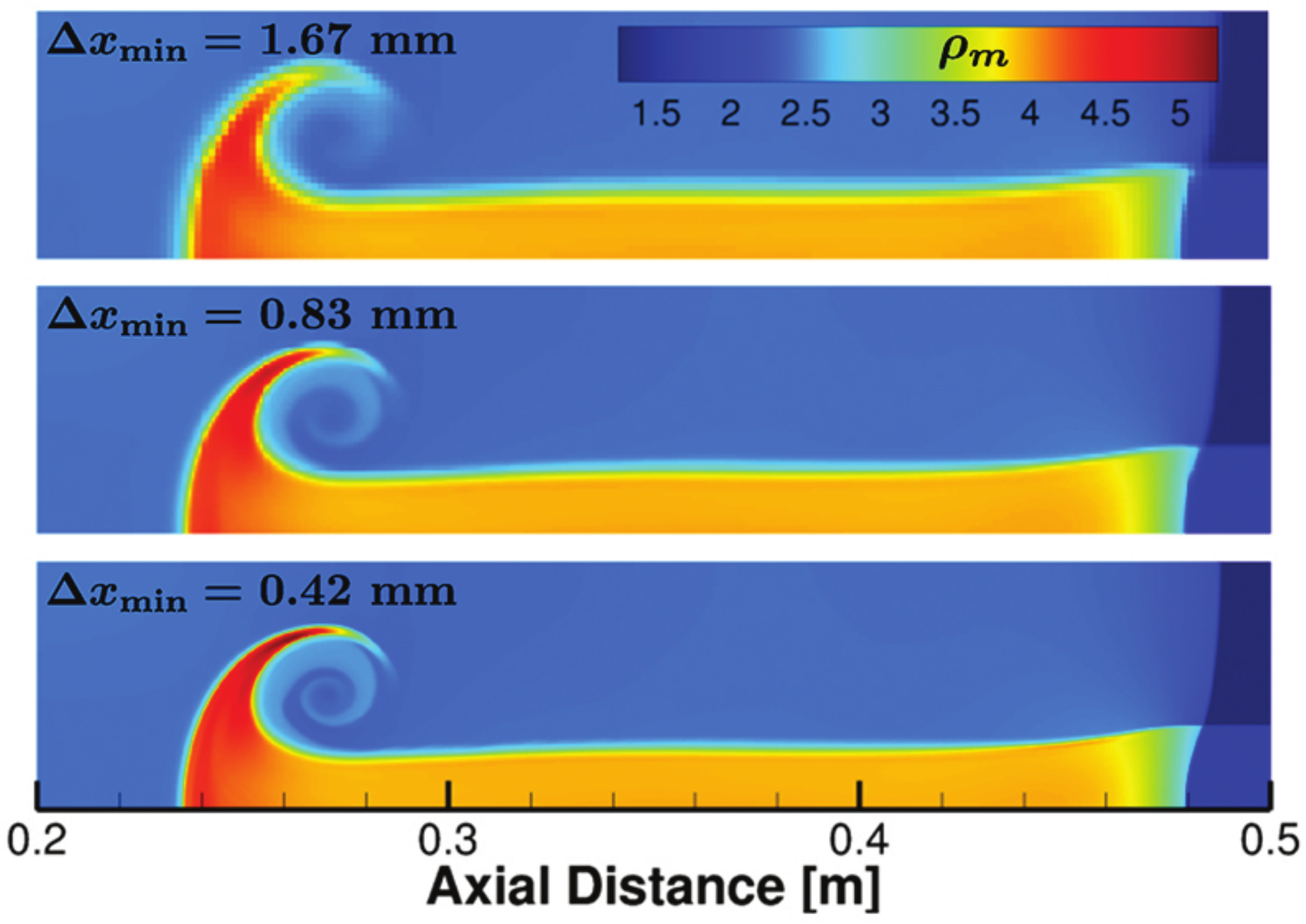}
\caption{Computed mixture density (in units of kg/m$^3$) for the low-volume fraction shock wave and dust layer interaction problem at 900 $\mu$s for three different grids.}
\label{fig:lowVolFraccomposite}
\end{figure}

This problem models the interaction of an inviscid shock wave with a loose dust layer, similar to that presented by Fedorov \textit{et al.} \cite{fedorov2010numerical}.  The heights of the channel and dust layer are $6$ and $2$ cm, respectively.  The left edge of the dust pile is placed at $2$ cm, and the shock is initially located at $1$ cm.  A shock of Mach number 1.6 propagates into air at 1 atm and 288 K.  The initial volume fraction of the dust layer is 0.04\%.  The particle diameter, density, and specific heat are $d_s=5$ $\mu$m, $\rho_s = 1470$ kg/m$^3$, and $C_{V,s}=987$ J/kg K, respectively.  Figure~\ref{fig:lowVolFraccomposite} shows the computed mixture density  at 900 $\mu$s for grid spacings of 1.67, 0.83, and 0.42 mm at the finest refinement level.  The computed solutions, including the particle and gas-phase vortices converge with increasing grid resolution.

The solution shown in Fig.~\ref{fig:lowVolFraccomposite} agrees qualitatively well with \cite{fedorov2010numerical} despite the use of different drag and heat transfer models.   A vortex is introduced in the gas-phase from a shear layer produced as particle drag locally decreases the fluid velocity in the granular layer.  
The gas-phase vortex rolls up some of the particles along with it.  The particle vortex does not exactly coincide with the gas-phase vortex due to particle inertia. 
Mechanical and thermal losses from entraining particles cause the shock near the top of the layer to curve and weaken with time.    Drag from high-pressure gas entering the layer from the y-direction compresses the dust layer while the front of the pile is rolling up.

\subsection{Compaction-Wave Angles}
\begin{figure}
\centering
 \subfloat[]{\includegraphics[width=0.35\textwidth]{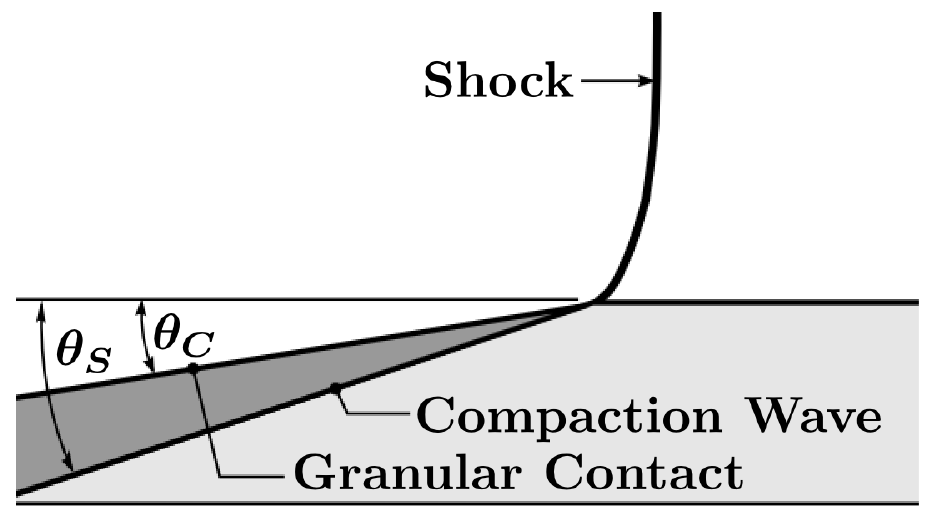}} \quad
 \subfloat[]{\includegraphics[width=0.55\textwidth]{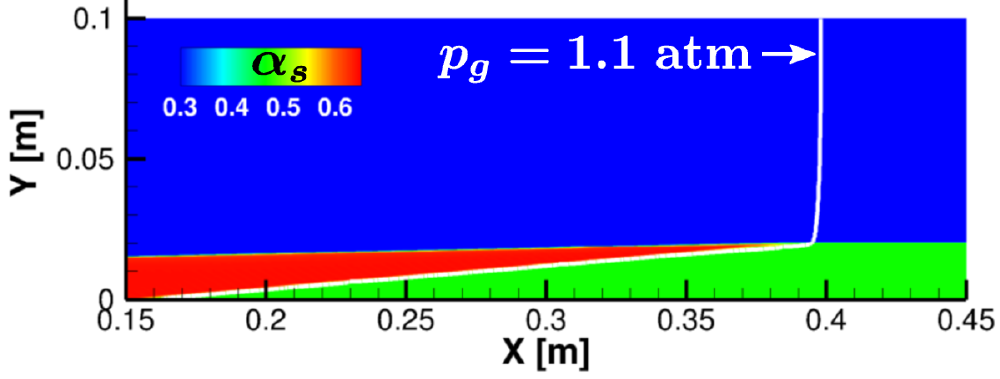}}
\caption{Angles of the transmitted compaction wave and granular contact from a shock wave propagating across the dust layer (a) definition of the angles and (b) computed solution at 100 $\mu$s and a grid spacing of $347$ $\mu$m at the finest level of refinement.}
\label{fig:compactionWave}
\end{figure}

There are not many tests that can be done for verification or validation of dense granular multiphase-flow simulations.  One experiment that can be compared to directly are the interactions of shock waves and dense dust layers described by Fan \textit{et al.} \cite{Fan2007LooseDust}.  These shock tube experiments measured the angle that a transmitted compaction wave and granular contact surface make with the horizontal when a shock wave passes over the dust layer, as shown in Fig~\ref{fig:compactionWave}(a).  For this test, the layer and domain heights are $2$ and $10$ cm, respectively.  The initial position of the shock is at $1$ cm, and the left edge of the dust layer is at 0 m.  The initial volume fraction was 47\%, and the density, diameter, coefficient of restitution, and specific heat of the particles was $\rho_s = 1100$ kg/m$^3$, $d_s=15$ $\mu$m, $e=0.9$, and $C_{V,s}=987$ J/kg K, respectively.  The velocity of the shock wave, which propagates into $1$ atm and $300$ K air, is 990 m/s.  The bottom boundary was changed to a non-reflecting condition to avoid reflections and increase the measurement distance for the transmitted compaction wave and granular contact angles.  The equations were integrated for 100 $\mu$s with $\Delta x=347$ $\mu$m at the finest level of refinement. 

The computed solid-volume fraction at  $100$ $\mu$s is shown in Fig.~\ref{fig:compactionWave}(b).  The computed transmission angles of the compaction wave and granular contact are $1.18^{\circ}$ and $4.78^{\circ}$, respectively, which are close to the measured values \cite{Fan2007LooseDust} of $1-2^{\circ}$ for the contact and $4^{\circ}$  for the compaction wave.    

\subsection{Interaction of a Shock Wave and a Dense Dust Pile}
In this problem the interaction between a strong Mach 3 shock and a dense layer of dust is examined.  The height of the domain is $10$ cm and the height of the dust layer is $6.67$ mm. The left edge of the dust pile is at $20$ cm, the shock is initially placed at $19$ cm, and the overall domain length is $50$ cm.  The Mach 3 shock propagates into air at 1 atm and 300 K.  The initial volume fraction of the pile is 47\%, which corresponds to typical volume fractions of settled layers of dust \cite{Fan2007LooseDust, BureauOfMines1988}.  Other parameters are identical to that of the low-volume-fraction shock and dust layer interaction problem discussed above.  The lift coefficient, $C_l$, was set to a value of 0.5. The grid spacing at the finest level of refinement was $\Delta x = 173$ $\mu$m.  

\begin{figure}
\centering
 \subfloat[$\log_{10}(\alpha_s)$]{\includegraphics[width=0.48\textwidth]{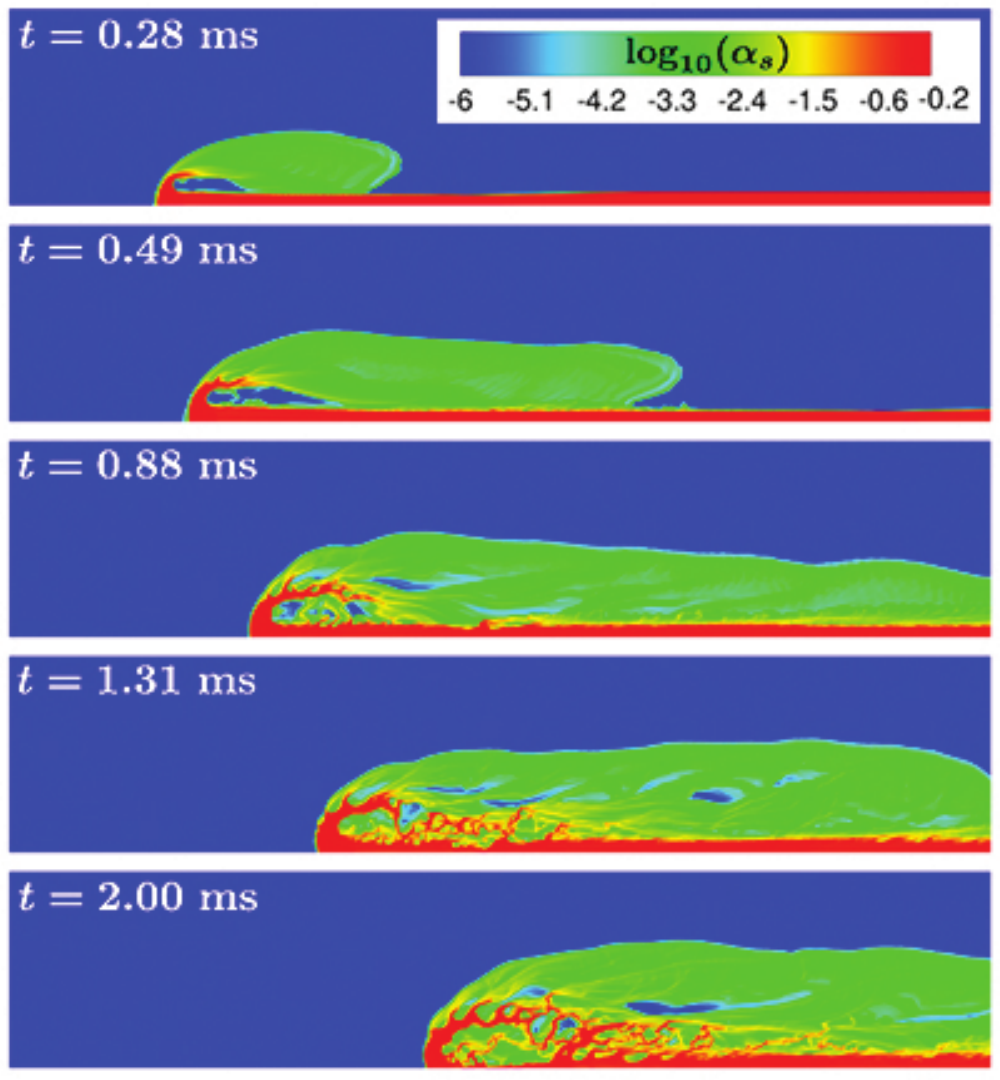}} \quad
 \subfloat[$|\nabla \! \rho_g|$]{\includegraphics[width=0.48\textwidth]{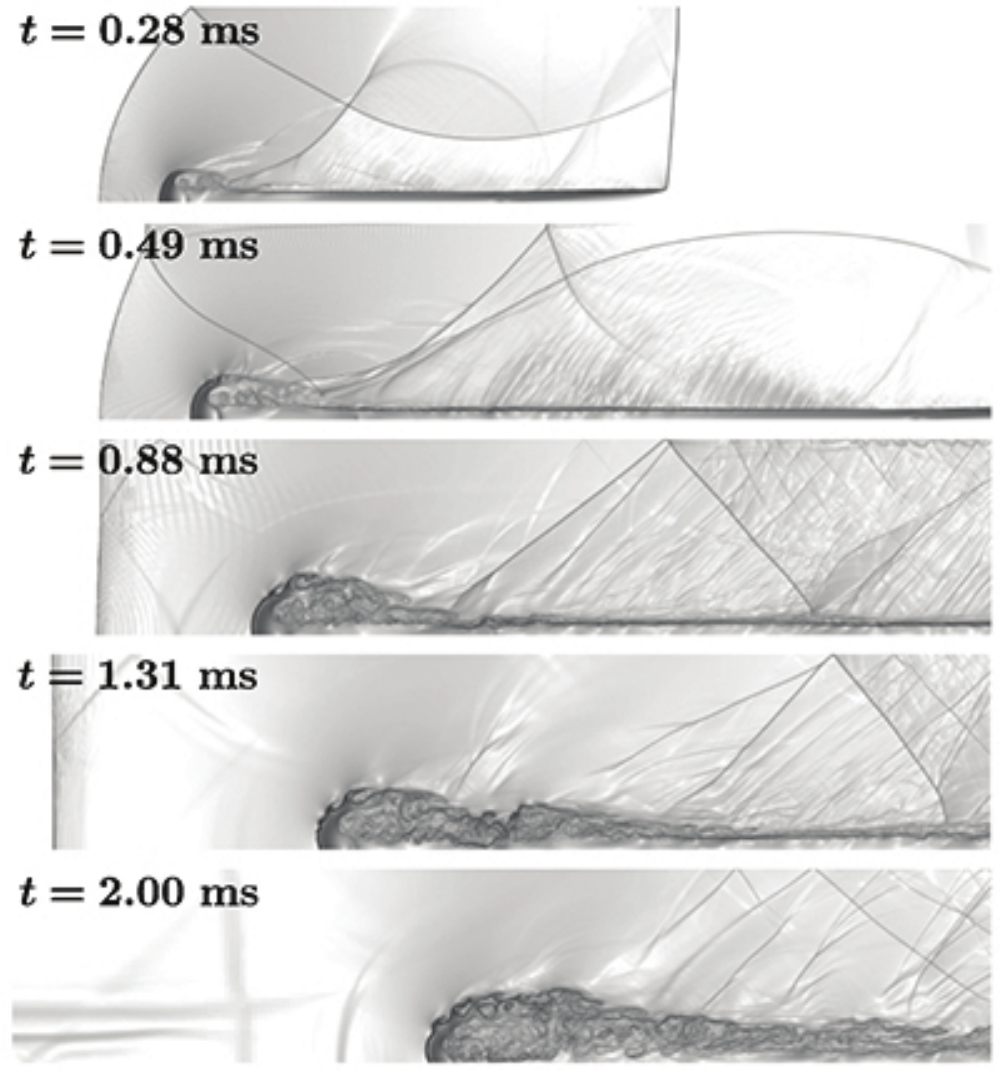}} 
\caption{Computed time sequence of (a) solid-volume fraction and (b) numerical Schlieren fields for a Mach 3 shock interacting with a dense layer of dust.}
\label{fig:highVAlpha}
\end{figure}

A sequence showing the solid-volume fraction and the numerical gas-phase Schlieren ($|\nabla \! \rho_g|$) is shown in Fig.~\ref{fig:highVAlpha} for $\mathcal{D}=1$ and $e=0.9$.  The entire domain is shown in each image.  
Initially, when the shock interacts with the dust layer, intense drag and buoyancy forces act on the particles.  This rapidly slows the gas and produces a strong reflection that resembles the interaction between a shock and a forward facing step.  The post-shock gases flowing up and around the front edge of the dust pile entrain some of the dust by drag and cause it to roll up.   The combination of lift forces and particle-particle interactions strip some of the particle from the surface of the rolled-up layer, producing a cloud with a volume fraction around 0.05\%.  Intense shear at the top of the layer and various shock reflections into the dust layer eventually destabilize the boundary layer as the shock moves downstream.  This perturbs the top of the dust pile and produces the Mach lines shown in the numerical Schlieren at $0.49$ ms.  Perturbing the surface of the layer intensifies turbulence and, as a result, throws more particles up into the post-shock gas.  By the end of the calculation, the layer of dust is highly distorted as a result of the turbulence.  

Starting at $\sim \! 0.49$ ms, lines of locally higher particle volume fraction start to form, which are sometimes called particle streamers.  At 2 ms, the volume fraction in these structures range from $0.5\%-20\%$.  Such structures have been observed in incompressible simulations of kinetic theory-based granular flow
\cite{Agrawal2001role, igci2008filtered}, computed using Eulerian-Lagrangian multiphase models that explicitly account for particle collisions
\cite{Helland2000210}, and have been experimentally observed in riser experiments \cite{Horio19942413}.  These structures result from two effects, the strong nonlinearity in the drag force as a function of volume fraction and inelastic collisions locally reducing the solids pressure.  Both of these effects produce dense clumps of particles that tend to move as a group and can be on the order 10 particle diameters across \cite{Agrawal2001role}.

\begin{figure}
\centering
 \subfloat[$\log_{10}(\alpha_s)$]{\includegraphics[width=0.48\textwidth]{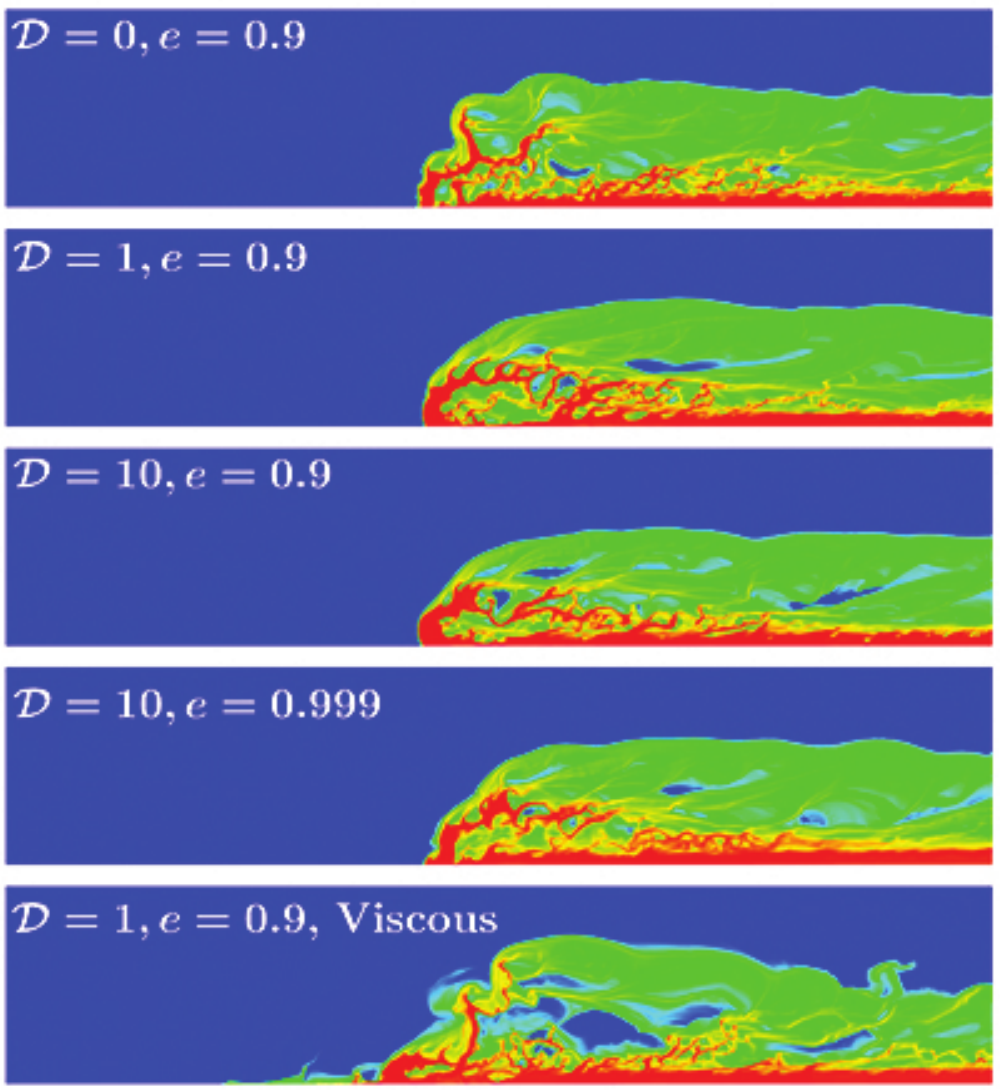}} \quad
 \subfloat[$|\nabla \! \rho_g|$]{\includegraphics[width=0.48\textwidth]{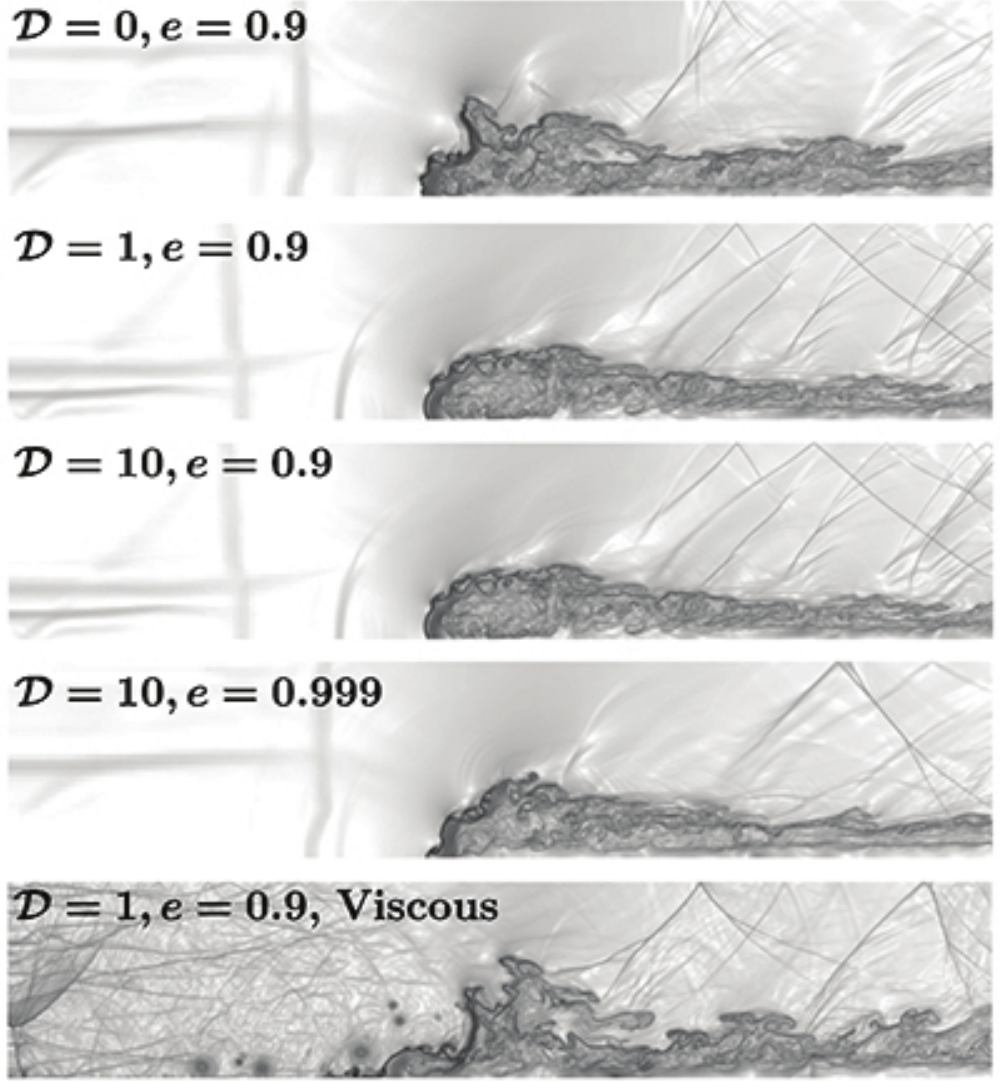}} 
\caption{Calculations showing the effect of $\mathcal{D}$, $e$, and viscosity for a Mach 3 shock interacting with a dense layer of dust at 2 ms on the computed (a) solid volume fraction and (b) numerical Schlieren fields for a Mach 3 shock interacting with a dense layer of dust.}
\label{fig:highCompare}
\end{figure}

The effect of the dissipation parameter, $\mathcal{D}$, the coefficient of restitution, $e$, and viscosity are shown in Fig.~\ref{fig:highCompare} at 2 ms.  The computed structures are qualitatively similar for all inviscid cases.  Differences are in the fine details.  High values of $\mathcal{D}$ and $e$ seem suppress surface instabilities in comparison to lower values and produce less turbulence.  Turbulence in the viscous case is actually higher due to upstream turbulence being produced from the initial reflected shock interacting with the boundary layer in front of the dust layer.  The upstream turbulence and the boundary layer change how the front of the dust pile becomes entrained.  The granular structures away from the edge of the dust pile are similar for inviscid and the viscous cases.
 
\section{Conclusions}
A numerical method was developed for solving kinetic theory-based granular multiphase models with volume fractions ranging from very dilute to very dense in highly compressible flows containing shock waves.  The numerical method assumes particle incompressibility, so that the particles can be thought of as blocking area in which gas flows.  The algorithm separates edges of a computational cell into gas and solid sections where gas and granular Riemann problems are solved independently.  Solutions from these individual flow problems are combined to assemble the fully coupled convective fluxes and nonconservative terms for both phases. A modified AUSM$^+$-up is used to compute the granular-phase fluxes with an additional tunable parameter to increase dissipation in dense granular regions.  This dissipation parameter works well in suppressing numerical oscillations when the packing limit is approached.  The technique converges as the grid is refined, even with very high density granular interfaces using low-dissipation numerical algorithms.  The method described advects dense granular material interfaces that coincide with multispecies gaseous contact surfaces without producing sharp spikes in pressure or temperature.  It also reproduces features from multiphase shock tube problems, granular shocks, transmission angles of compaction waves, and shock-wave interactions with dust layers.  This approach is relatively simple to implement into an existing code based on Godunov's approach and can be constructed from standard Riemann solvers for the gas-phase, a modified AUSM$^+$-up for the particle phase, and standard edge interpolation schemes.

\section*{Acknowledgements}

This work was supported by the National Research Council Postdoctoral Research Associateship Program and the University of Maryland through Minta Martin Endowment Funds in the
Department of Aerospace Engineering, and through the Glenn L. Martin Institute Chaired Professorship at the A. James Clark School of Engineering.
Computational facilities were provided by the
Laboratories for Computational Physics and
Fluid Dynamics at the U.S. Naval Research Laboratory  and the DoD High Performance
Computing Modernization Program.

\bibliographystyle{elsarticle-num}









\end{document}